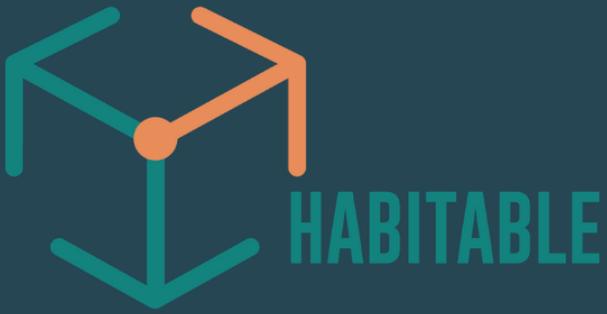

# Modeling the Senegalese Artisanal Fishing Migrations

Alassane Bah, UCAD

Timothée Brochier, IRD

D3.5

The **HABITABLE** project – Linking Climate Change, Habitability and Social Tipping Points: Scenarios for Climate Migration – aims at significantly advance our understanding of the **current interlinkages between climate change impacts and migration** and displacement patterns, and to **better anticipate their future evolution**.

Running for **4 years** (2020-2024), HABITABLE brings together **21 partners:** University of Liège, University of Vienna, Potsdam Institute for Climate Impact Research, University of Exeter, the IDMC, Lund University, Sapienza Università di Roma, adelphi, Université de Neuchâtel, Institut de Recherche pour le Développement, Council of Scientific and Industrial Research, UNESCO, University of Ghana, CARE France, University of Twente, Université Cheikh Anta Diop, Stockholm Environment Institute Asia, Raks Thai Foundation, Addis Ababa University, Institut National de la Statistique du Mali and Samuel Hall.

HABITABLE is the **largest research project on climate change and migration** to have ever been funded by the European Commission's Horizon 2020 programme.

Please visit www.habitableproject.org for more information about the project.

**Title:** Modeling the Senegalese artisanal fishing migrations

**Authors:**
Bah, Alassane, UCAD
Brochier, Timothée, IRD

**Reviewers:**

Jacob Schewe, PIK, Postdam, Germany
Diana Reckien, University of Twente, Neitherland

**Publication available on:**
This document will be made available on the project's website, once approved by the European Commission.

## Version history

| Version No. | Date | Information |
|---|---|---|
| 1 | Mars 2024 | Initial version submitted to the EC as deliverable |
|  |  |  |

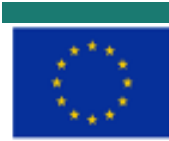

This project has received funding from the European Union's Horizon 2020 research and innovation programme under Grant Agreement No. 869395. The content reflects only the author's views. The European Union is not responsible for any use that may be made of the information it contains therein.

# Acknowledgements

The present work would not have been possible without the contributions of numerous partners and colleagues. In particular, we express here our gratitude to the colleagues affiliated to the Senegalese research institutions. Researchers and teacher-researchers from UCAD, in particular from ESEA and ESP, and the excellent LPAO-SF bring keystone contributions to the project. Colleagues from UGB, IRD, CIRAD, Sorbonne University, UAM and the research teams LOCEAN, UMMISCO, SENS, ESPACE-DEV, LEMAR, LOPS and the mixt international laboratory ECLAIRS also contributed. Senegalese's fishers contributed through the sharing of their knowledge. We acknowledge also the CRODT, part of ISRA.

# Acronyms

AI: Artificial Intelligence
CIRAD: Centre international de recherche en agriculture pour le développement
COPACE: Comité des Pêches pour l'Atlantique Centre-Est
CRODT: Centre de recherche océanographique de Dakar
ECLAIR2: Études intégrées du climat et de l'océan en Afrique de l'ouest et réponses aux changements climatiques au Sénégal
ECMWF: European Center for Medium term Weather Forecasting
ERA: ECMWF Re-Analysis
ESEA: École supérieure d'économie appliquée
ESP: École supérieure polytechnique
ESPACE-DEV: Observations spatiales, modèles et science impliquée
FAO: Food and Agriculture Organisation
Fish-MIP: The Fisheries and Marine Ecosystem Model Intercomparison Project
FU: Fishing Unit
GAMA: GIS Agent-based Modeling Architecture
GIS: Geographical Information System
IPCC: Intergovernmental Panel on Climate Change
IRD: Institut de Recherche pour le Développement
ISRA: Institut Sénégalais de Recherche Agricole
ITCZ: Inter-Tropical Convergence Zone
INN: Illegal, Unreported and unregulated (fisheries)
LEMAR: Laboratoire d'étude de l'environnement marin
LOCEAN: Laboratoire d'Océanographie Numérique
LOPS: Laboratoire d'observation physique et spatiale – Simeon Fongang
LPAO-SF: Laboratoire de physique de l'atmosphère et de l'Océan
OFT: One Factor at Time
SENS: Savoirs Environnements Sociétés
SSP: Shared Socioeconomic Pathways
SST: Sea surface temperature
UAM: University Amadou Mahtar Mbow
UCAD: University Cheikh Anta Diop
UMMISCO: Unité de modélisation Mathématique et Informatiques des systèmes complexes

# Index of Figures

## Appendix Figures

# Index of Tables

# Table of contents

# Abstract

The North-West African coast is enriched by the Canary current, which sustain a very productive marine ecosystem. The Senegalese artisanal fishing fleet, the largest in West Africa, benefit from this particularly productive ecosystem. It has survived the ages with remarkable adaptability, and has great flexibility allowing it to react quickly to changes, in particular by changing fishing gear and performing migrations. However, since the 1980s, the increasing fishing effort led to a progressive fish depletion, increasing fisher's migration distances to access new fishing grounds. Since 2007 many fishers even started to navigate to Canary archipelago in order to find a more lucrative job in Europe, carrying candidate to emigration in their canoes. This phenomenon further increased since 2022 due to a new drop in fishery yields, consecutive to the development of fishmeal factories along the coast that amplified overfishing. Climate change may also impact fish habitat, and by consequence the distribution of fishing grounds. The question addressed in this research was how climate change, fishing effort and socio-economic parameters interact and determine the artisanal fishery dynamics. An interdisciplinary approach allowed us to collect data and qualitative information on climate, fishing effort and socio-economic parameters. This served as a basis to build a multi-agent model of the mobility of Senegalese artisanal fishing. We implemented a first version of the model and presented some preliminary simulations with contrasted fishing effort and climate scenario. The results suggested that first, climate change should have only a slight impact on artisanal fishing, even in the most extreme climate scenario considered. Second, if fishing effort was maintained at current levels, we found a collapse of the fishery with massive fishers migrations whatever the climate scenario. Third, with reduced fishing effort, a sustainable fishery equilibrium emerges in which Senegal's artisanal fishery catches ~250,000 tons of fish a year mostly in Senegal, approaching the 2000s catches records. This sustainable equilibrium maintained with the two-climate change scenario tested. Fishers migrations provide clues of the fish populations state and have implications for the sustainable exploitation of fishing resources. Senegalese artisanal fishers' migrations impact the regional distribution of the fishing effort, therefore must be taken into account in regional development and planning policies for this sector, particularly in a context of increasing infrastructure and spatial management measures (*e.g.* marine protected areas). This work lays the foundations of a computer simulation tool for decision support.

# Keywords

# Foreword

UMMISCO is an international research unit gathering researchers and professors-researchers from five countries including France and Senegal. The Senegalese team of UMMISCO gather researchers from IRD (Institut de recherche pour le development, France) and UCAD (Université Cheikh Anta Diop, Dakar, Sénégal). Research on the Senegalese small scale fisheries at UMMISCO started in 2014. It focuses on socio-economic models for studying how coastal fisheries are impacted by management measures as artificial reefs, marine protected areas and international agreements on fisher's mobility. This research direction was later boosted by its inclusion in the Habitable project.

Modeling a phenomenon as complex as artisanal fishers migrations is a challenge that calls on different types of knowledge. We constituted an interdisciplinary research group around UMMISCO modelers in order to support research efforts in physical oceanography, marine biology and socio-economy on the issue. Members of this group included 2 oceanographers, 3 climate change scientist, 2 fisheries scientist, one socio-economist, one mathematician and one informatician from ESP-UCAD, UGB, Sorbonne and IRD. This interdisciplinary group collectively established a conceptual framework and a roadmap, following the evolution of the research during meetings in 2020 and 2021 in Senegal. During these meetings, key parameters were identified, and serves as a basis for master students research co-directed by the modeling team and the disciplinary professors. Five master theses dealing with fisher's migrations were defended in informatics, socio-economics and marine biology. Apart from the master students, two post-docs in mathematical modelling and climate change were hired for few months. CRODT-ISRA, the Senegalese recherche center on fisheries participated to the project as a third part partner.

# 1 – Introduction

## 1.1 Fisheries in North-West Africa

North-West Africa coast lies among the world most productive areas in terms of fish. This is due to the fact that surface water is enriched by deep waters upwelled near the coast under the action of the trade winds from Morocco to (seasonally) Guinea Bissau. When nutrients rich waters reach the surface, phytoplankton blooms occur and sustain a large trophic net, including many fish populations. While long distance fisheries from Europe and Asia have long been exploited the area, from Morocco to Guinea artisanal fisheries mainly developed in Senegal.

### 1.1.1 Senegalese Artisanal Fishers Migrations

Artisanal fishery is a major economic sector in Senegal, involving more than 60 000 direct jobs, 14 000 canoes and contributing to the regional food security. The sector was described as very adaptive and resilient to changes in the ecosystems, including modern fishing gear as encircling net. However, as the artisanal fleet grew and got more efficient, industrial fishing both legal and illegal also increased, and the very productive marine ecosystem progressively suffered overfishing (Belhabib *et al.*, 2014). Fish target species reduced in size and abundance and overall abundance progressively reduced to 50% since 1980.

Senegalese artisanal fisheries are very compartmented in terms of gender. While crew on fishing trips are exclusively composed of men, seashell collection in tidal areas and artisanal fish processing, are exclusively a women business. While the usual definition of fishing unit in Senegal refer to the operational canoe, its crew and fishing gear, a more exhaustive definition should include the "land team", mainly women. Historically, the two cannot be separated, but the development of industrial fish processing progressively reduced the space of women, while fish scarcity forced men to spend more and more time at sea. Hereafter, by the term "fishers' migrations" we refer to the change of landing sites of the fishing units along the coast, which may occur for durations ranging from weeks to years. When migrations are long, it is common that the whole family move together and inhabit the new landing site; alternatively, fishermen can have a family in each landing site (Deme et al., 2021). Hereafter we use the neutral term of "fishers" to refer to the people related to the artisanal fishing units, both men and women (Branch & Kleiber, 2017).

Failler et al. (2020) and Deme et al. (2021) draw attention to the significant contribution of Senegalese migrant artisanal fisheries to the total landings recorded where they are stationed, in Senegal and at the scale of the sub-region, contributing up to 70% of small pelagic fish yearly landings. It is the fact of large pirogues fishing with several kilometers long purse seine which constantly migrate to follow fish schools and disembark at landing sites where there is demand (Figure 1). It appears that migrant fishers are largely contributing to fishmeal factories supply, selecting their landing sites according to the presence of these factories. The authors point out the importance of taking this mobility into account in the management of the sector, particularly in the context of the proliferation of fishmeal factories which are causing demand to explode (Deme et al., 2023). For this, a detailed understanding of the mobility processes of artisanal fishermen is necessary.

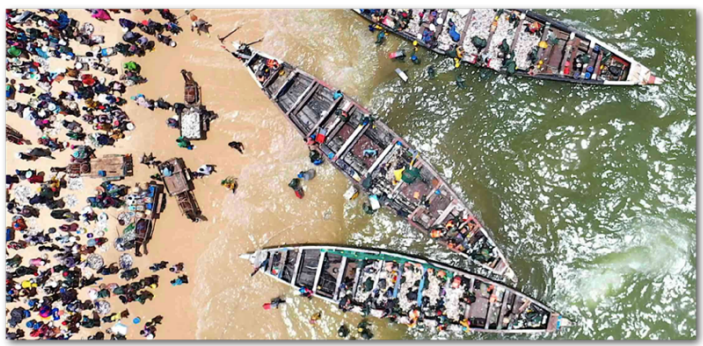

*Figure 1: Arrival of large pirogues, possibly migrant, in the Kayar landing site, Senegal. Photo: UMMISCO*

Senegalese artisanal fishers' migration is a cultural, economic and social phenomenon. Chauveau et al. (2000) presented a historical and modern description of these migrations, documented by numerous research studies in the fields of social sciences and fisheries (Figure 2). This historical perspective show that at the beginning of the 20th century, migrations were motivated by the search of new markets that developed in new urban centers and exportation places. As local fishing pressure increased around cities, fishermen migrated along the coast to reach less exploited fishing grounds. The motorization of the canoes, during the 1960s, initially caused a reduction in alongshore migrations because fishers could access offshore fish. This was followed by the decolonization and the establishment of fishing licenses which caused a reduction of the long-distance industrial pressure, also resulting in a reduction of artisanal fishers' migrations as fish abundance increased. However, since the 1980s, the progressive fish depletion caused a continuous increase of migration distances, and the decrease of fishing sector incomes. Combined with the agricultural crisis, and regional geopolitical instabilities, it leads to an increase of candidate for migration to Europe in order to find security and a more lucrative job. Since 2007 many experimented fishers then rent out their services as migrant transporters, either coming back to Senegal to repeat the operation or staying themselves in Europe. The phenomenon was recently amplified due to a new drop in fish yields since 2019, consecutive to the development of fishmeal factories along the coast that further amplified overfishing (Brochier et al., 2023; Enríquez-de-Salamanca, 2023).

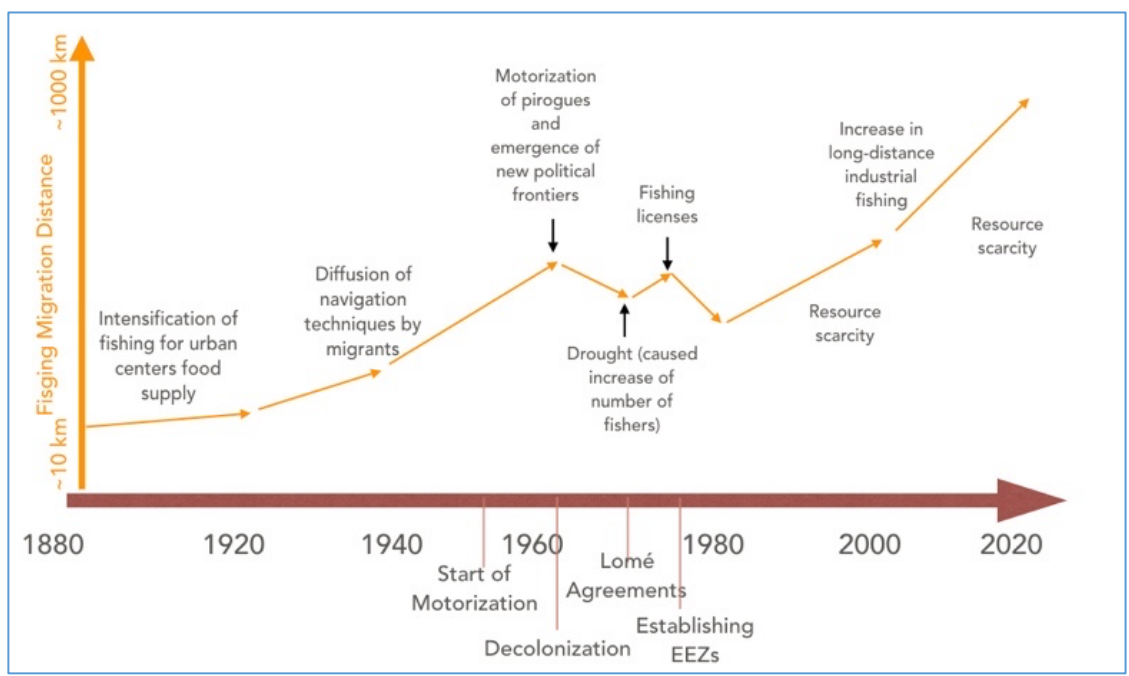

*Figure 2: Socio-Economic and ecological tipping points that impacted fisher's migrations in north West Africa. Only two events caused a reduction temporal of fishing migration distance. Conceptual scheme constructed from the information in Chauveau et al. (2000).*

### 1.1.2 Impacts of Climate Change

Climate change already impacted North-West Africa, and future projections suggest changes on precipitations that may negatively impact agriculture (Sultan et al. 2013). However, the way in which climate change might impact the marine upwelling ecosystem remains uncertain (Sylla et al., 2019). In the 1970s, severe drought negatively impacted agricultures, but the marine ecosystem was not impacted. In fact, in period of drought the trade winds are likely more frequent, which favor the upwelling enrichment of the marine ecosystem, and thus fish productivity. This contrast in climate impacts between agriculture and fisheries resulted in human migrations from inland to the coast, artisanal fisheries acting as a refuge sector. Such mechanism could also be at play at present time, as agricultural production reduce in Senegal under the effect of multi-factors effect of climate change, including rainfall reduction (Faye, 2022). However, in the current overexploited state of the marine ecosystem, climatic refugees may not be able to find a place in artisanal fisheries.

At seasonal scale, climate strongly impact fish habitat distribution in North-West Africa, and thus impact fish migrations. Fish migrations could be simulated by biophysical models (e.g. Brochier *et al.*, 2018) and are mainly determined by the seasonal fluctuations of the ITCZ (Inter-Tropical Convergence Zone). These seasonal fluctuations alternates in Senegal an upwelling ecosystem from December to June with a warm coastal water ecosystem from July to November. Indeed, seasonal variability of the upwelling-favorable winds causes large fluctuation of sea surface temperature, plankton, and finally fish habitat. Thus, many fish species migrate according to these seasonal fluctuations, which may be modulated by climate change. Also, as fish abundance reduced, seasonal fish migration patterns became more important, in particular for the small pelagic fish fishery. For example, in Senegal some fish species that were historically present year-round became available only a part of the year. Thus, climate change may impact fisheries by modulating these patterns of fish habitats.

The IPCC highlighted the complexity of the link between climate change and migration patterns, due to the interaction of social, economic and environmental factors. The overall aim of the project HABITABLE was to clarify this complexity within the conceptual framework of "habitability". Here, we understand "habitability" as the potential of northwest African coast to sustain the Senegalese artisanal fishers under climate change scenarios. But since a probable impact of climate change would be to increase the number of fishers, the question turns into how much fishing effort can the marine ecosystem sustain in the context of climate change. We tackle this question by building a tool to forecast the effect of climate change on Senegalese artisanal fisheries. This task implied interdisciplinary research in order to integrate the effect of climate change with socio-economic and biological drivers of artisanal fishery dynamics.

## 1.2 Interdisciplinary research strategy

The artisanal fishers' migration is a complex phenomenon resulting from the socio-ecosystem dynamics. Descriptions of the fisher's migration were obtained from literature study, expert elicitation and fisher's focus groups which revealed the societal and political context of these migrations. On the other hand, remotes and in situ sensors and government sectorial reports provided information about the physical, ecological and economic environment of the fishers. This evidenced processes and parameters at play that were described in the frame of sociology, economics and geography studies in one hand, and marine biology and fishery science on the other hand. This information was gathered in an ontology of the Senegalese artisanal fisheries which fed the onset of a conceptual model formulated either as equation-based model or as a multi-agent, informatic model. Within this research procedure several activities occurred at the same time, or in an iterative way (Figure 3).

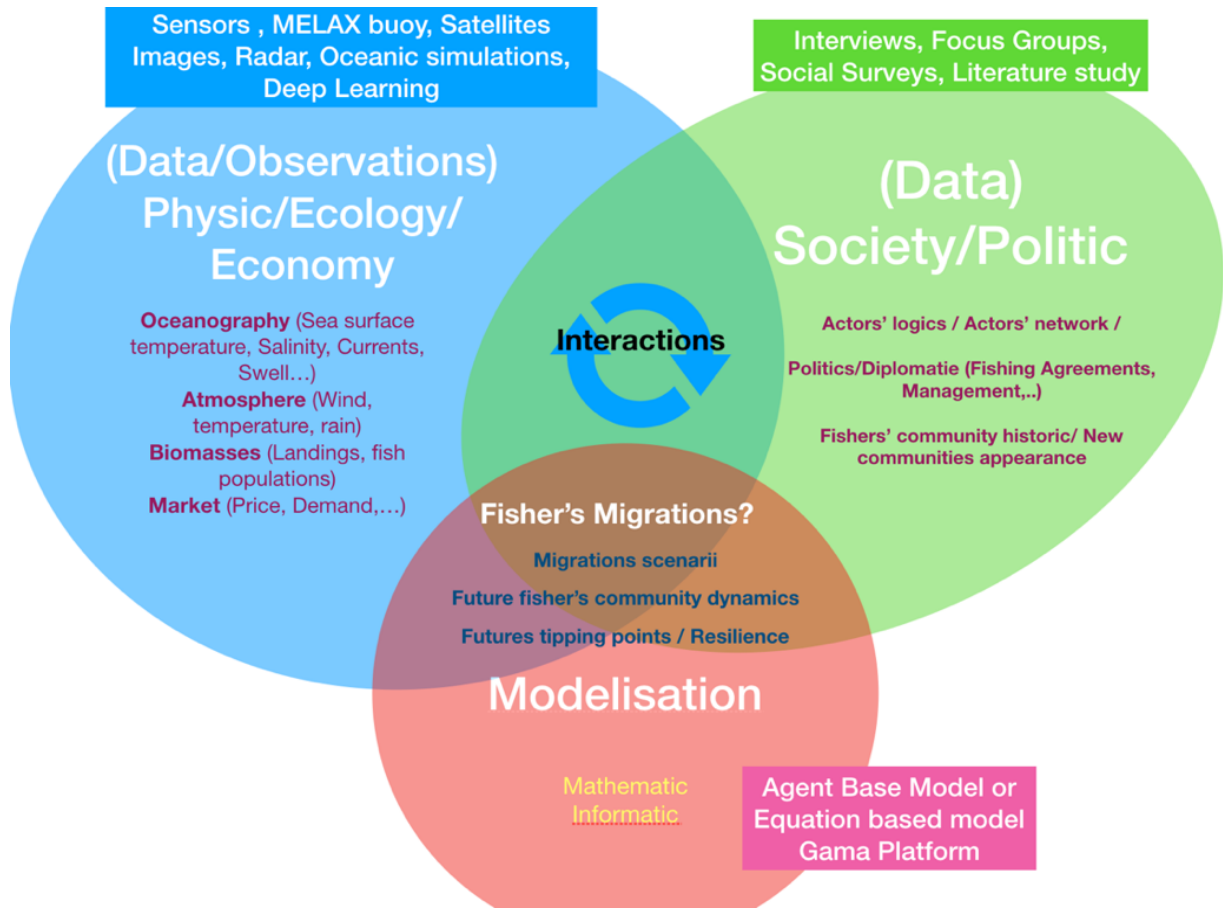

*Figure 3: Schematic view of the interdisciplinary research strategy. Processes and parameter to be identified are in the circles, while methods to approach them are in the squares*

### 1.2.1 Social and Politic Parameters

The literature contained many information on Senegalese artisanal fishers' migrations, and specifically the main migration routes were described according to the different fisher's community (Figure 4). Expert elicitation and fisher's focus groups allowed us to complete this information with a detailed picture of the fisher's decision to do migrations, the preparation of the migration, the organization of the time during the migration, and the operations when coming back from migrations. These information's were collected with expert and stakeholders from 11 dedicated focus groups with different fisher's communities. Numerous interviews were also conducted along the Senegalese coast, which were analyzed in three theses from UCAD students (Fall, 2021; Diallo, 2022; Ndiaye, 2023). These approaches provided qualitative data about actor's logics and networks, and revealed complex interaction between fish overexploitation, fish migration patterns, demand from the market and the role of politics and diplomacy through management systems and international fishing agreements. Fishers migrations occurs at any time of the year, motivated by fish scarcity resulting from a combination of fish seasonal migrations and overexploitation. In the same season, some fishers can migrate north while others migrate south, depending on the target species. Fish-traders are keystone actors of these migrations. Their job consists in buying fish to fishers and selling it on the market. Fish-traders have contact along the coast that tell them when a target species appears, and if a market exists for this species at this time, then the fish-trader will finance fishers to migrate to this place, potentially in the frame of international agreements.

### 1.2.2 Physical, Ecological and Economic Parameters

One of the salient characteristics of the upwelling systems is the strong variability of the biological production induced by the physical parameters, in particular the variability in wind strength and direction. Understanding the mechanisms driving this variability is crucial to anticipate the effect of climate change. Previous research in physical and fishery oceanography have detailed how wind and upwelling variability impacted fish species distribution and migrations by combining knowledge on fish species biology and physical oceanography data as sea surface temperature (e.g. Binet et al., 1988; Brochier et al., 2018 and references herein).

The ecological context and economic state of the artisanal fishery was described by annual fish landings, number and distribution of fishing units, and landing sites infrastructures. Annual catch for each landing sites in Senegal is partly available from published time series, from yearly internal reports and FAO aggregated datasets for Senegal. FAO/COPACE reports also provided parameters for the exploited fish population at the sub-regional scale. The CRODT monitored the artisanal fleet distribution through a bi-annual census of fishing units in the main landing sites from 1980 (Figure 6). The censing extended to most landing sites in 2005, but until 2014, the censing could not occur in southern part of Senegal (Casamance) because of political instabilities. At this time, infrastructures at landing sites were very scarce so that it is likely that the landings in this area were mainly for local consumption and artisanal fish processing. Since 2015, Casamance landing sites were also censed and a growing number of fishing units was observed as local infrastructures developed. The censing occurred at the two contrasted seasons (warm and cold) in order to capture the effect of seasonal migrations. These data are part of the protocol for the artisanal fish landings monitoring, but it appeared that the bi-annual census was not enough to provide a good picture of fishing units distribution at a given time, because many fishing units actually moved among landing sites several time a year. However, for logistic and financial limitations it was not possible to count the fishing units at the appropriate time scale, which would be ideally the week. In order to contribute to overpass this limitation, we supported research to develop an automated fishing units census based on aerial images from drone and /or satellites and using artificial intelligence (AI) trained with experts to detect different categories of Senegalese artisanal fishing units (Figure 5).

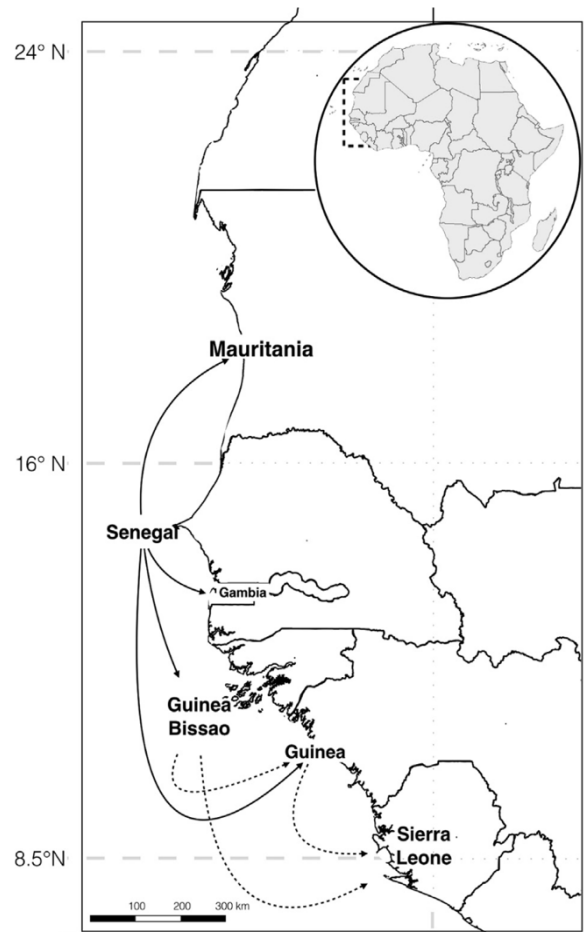

*Figure 4: Main artisanal fishing migration routes (adapted from Binet et al.2013).*

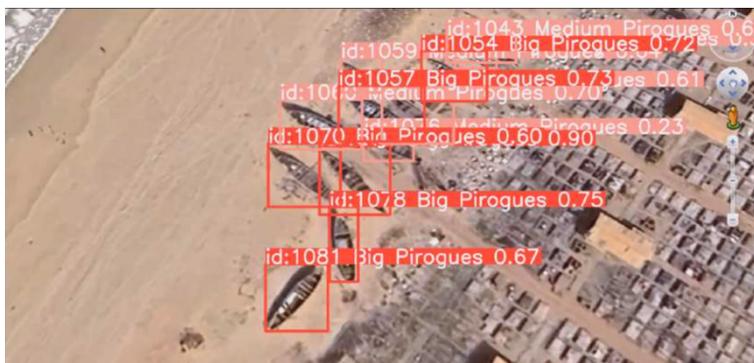

Figure 5: Example of Senegalese fishing unit's detection and counting on a landing site from satellite image (Olaluwoye et al., accepted).

Detailed reports of the fishing infrastructures and processing capacity at each landing sites were not available except for fishmeal factories, for which NGOs reporting also existed. Indeed, it appeared that fishmeal factories daily processing capacities represent the main part of the processing capacities of the landing sites where they are installed. Expert elicitation produced (much lower) estimations of processing capacities of artisanal transformation, export and fresh local market.

In order to study the impact of climate change, we needed to reconstruct climate change scenario adapted to our specific question, the impact on fisheries. The reference climate situation was the "ERA5" re-analysis (historical scenario), the closest to the real climate. To construct future scenarios, we analyzed 13 IPCC models parameterized with atmospheric $CO_2$ concentrations corresponding to the projections for the period 2064-2099 with the IPCC SSP2-4.5 and SSP5-8.5 scenarios respectively. SSP2-4.5 simulates a greenhouse gas emissions trajectory whose socio-economic assumptions are compatible with a continuation of current socio-economic trends. This emissions trajectory induces a radiative forcing of 4.5W/m2 in 2100, i.e. a warming level of approximately 2.7°C compared to the pre-industrial period. The SSP5-8.5 scenario corresponds to a higher climatic drift, where the hazards evolve most unfavorably and therefore where the level of risk would be the greatest. From the analysis of the future climate model, we found that in the most extreme scenario, the SST increased by 3 degree in the area. However, in these climate change global models, the small-scale features such as the seasonal coastal upwelling off Senegal were not well represented. Thus, in order to keep the SST structure typical to a coastal area, which may continue to exist, we constructed an "optimistic future scenario" and a" pessimistic future scenario" that were the reference scenario (ERA5) with a uniform SST correction of + 1.5°C and + 3°C (Table *1*). SST increases in future climate scenarios shifted the fish habitat northward (see figures in Appendix C).

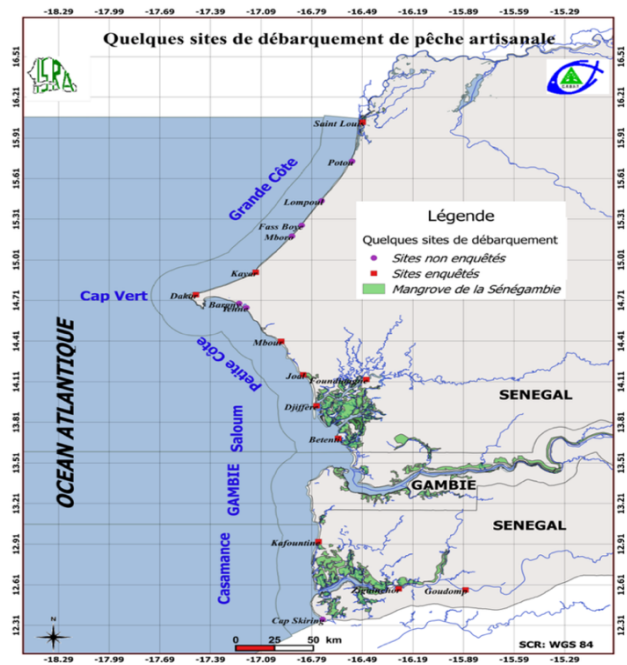

*Figure 6: Artisanal fisheries landing sites where the number of fishing units are monitored bianualy, representing 80% of the total number of fishing units (CRODT).*

### 1.2.3 Model Construction

Artisanal fisher's migration models in the literature were reviewed, and synthetized by a post-doc researcher at UCAD. This study (unpublished) showed the existence of mathematical approaches and fishing cost optimization models, but point to the lack of exploratory models for inferring the effect of climate change. Fishery science usually considers either the impact of fisheries on fish populations or the impact of climate on fish. Attempt to integrate both interaction in a single model led to complex models useful for research but difficult to interpret for decision making (Travers et al., 2007). The Fisheries and Marine Ecosystem Model Intercomparison Project (Fish-MIP) tackle these limitations by investigating multiple complex models, but still conclude that the level of uncertainty still prevent their use to support country-level adaptation policies (Tittensor et al., 2021). Thus, rather than adapting a pre-existing model, we started a new model, with reduced complexity, based on the specificities of Senegalese artisanal fisheries. Parameters and processes identified by the interdisciplinary research were used to construct the conceptual model.

The multi-agent approach allowed us to easily integrate spatial data with processes from different nature, described in different disciplines, into a single model which can then be explored numerically. To avoid the technical difficulties linked to fish population closures, i.e. density dependence, fish population dynamics were based on a simple mathematical model imbedded in the multi-agent, informatic model. Model implementation was iterative, including step by step climate, fish migration, infrastructures and fishing unit parameters (see lower complexity models in Brochier and Bah, 2020; Deugue, 2020). The objective of the model was to be able to identify future climate and socio-economic tipping points for the Senegalese artisanal fisher's community. In this approach, tipping points corresponded to threshold values of the model parameters separating a sustainable fishery from a collapsed fishery. Iterative model implementations highlighted the existence of missing information, which motivated new field research, and in turn model evolution.

The "ODD" protocol (Grimm, 2010) was designed to provide a description of agent-based models allowing code rewriting. We present the conceptual model in section 2 and its implementation and running in Gama, an agent-based simulation platform written in Java (https://gama-platform.org/), in section 3. Some preliminary simulation results were discussed and future research directions are suggested in section 4.

# 2 – Description of the Model

The model description follows the ODD (Overview, Design concepts, Details) protocol for describing individual- and agent-based models (Grimm et al. 2010). The objective of the protocol is to present first the high level strata, simple, of the model, while entering in the complexity in subsections.

## 2.1. Purpose

The purpose of the model proposed here was to lay the foundations of a computer simulation tool for artisanal fishery management decision support including the climate and socio-economic parameters that may impact the fishery. The impact of climate change on fisheries is not straightforward, thus the purpose here was to make it emerge from the impact of spatially explicit climate change scenarios on the distribution of fish habitat. In turn, the distribution of fish habitat impacted artisanal fishing trips distance and duration. Artisanal fisheries in low governance areas, such as developing countries, are weakly impacted by regulation, thus the purpose was to relate the fishing effort to market demand, itself controlled by processing capacities infrastructures. Social behaviors and financial mechanism constrained the onset of fishing operations and migrations, the purpose of the model was to assess the impact of these processes through migrations probability and duration parameters.

In the configuration presented here, the model explicitly represents the particularities of the Senegalese artisanal fishery fleet, which includes different size categories of canoes, fishing gear and target species. The purpose was to assess the effect of climate change and fishing pressure in north-west Africa on the sustainability of the artisanal fishery in Senegal and the corresponding migration rates. It helps to disentangle the effect of climatic and fishery parameters, and thus to identify how and where to focus management actions in order to restore sustainability of the artisanal fishery. The model was designed to be evolutive and adaptable to other geographical and socio-economic configurations.

## 2.2 Entities, State Variables, and Scales

The socio-ecosystem of artisanal fishing is made up of three subsystems, the fishery itself, the climate, and the economic and social context (Figure 7). In this model we focused on the fisheries subsystem, while the "climate" and "socio-economic" systems were represented in a simplified manner. The entities of the model were the fishing units, the landing sites, the environment of the fishing grounds and the target species (Figure 8). These four entities were defined on the scale of the geographical area going from Morocco to Guinea, the area of action of Senegalese artisanal fishery (Binet et al., 1998). The time horizon considered was several years to describe seasonal movements while taking into account the knock-on effects that can influence from one year to the next. The spatial dynamics of the resource have a significant influence on fishing trips, which were explicitly modeled. A spatial resolution of 10 kilometers and a time step of 1 hour were in line with the average velocity of the canoes (10 km/h).

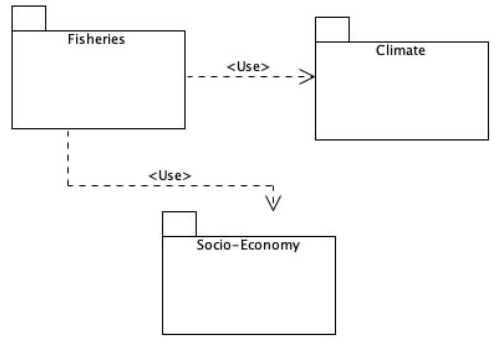

*Figure 7: General model of the socio-ecosystem of fisheries*

### 2.2.1 Fishing Unit

In the model, fishing units were considered as individuals, or agents. In the sense of Charles-Dominique and Mbaye (2000), these are the units which are constituted for each fishing trips. They belong to a family business. In the model, fishing units come in different types, and each type is "armed" differently depending on the type of fishing for which it is intended. Following the typology in Chavance and Morand (2020), we considered three types of fishing units corresponding to different boat sizes and fishing practices. These practices are line or longline for small ones; the gillnet for intermediate sizes and the ring seine for larger boats. Medium and large canoes can be equipped with coolers, which allows trips over several days, until 10 days in the model, while ensuring the conservation of catches.

Each category of fishing unit was associated with target species either demersal or pelagic. Demersal species, which live close to the bottom, were the main targets of small fishing units while large fishing units targeted in priority pelagic species, which live close to the surface, but could also access demersal species as second choice. Medium-sized fishing units randomly targeted demersal and pelagic fish. The variable engine power and fishing gear size among fishing unit categories was reflected in the model by contrasted catchability, i.e. the proportion of fish biomass present in the fishing ground (biomass patch in the model) catch per hour when fishing (Table 6).

Fishing trips can be of different durations and target more or less distant fishing sites. In the model, each category of fishing unit was associated with a type of mobility. Small fishing units can only perform daily outings to target fishing sites within a 50 km radius. Medium-sized fishing units can also perform "tides", which are outings that can last up to 10 days at sea within a radius of 500 km (Table 6). Finally, larger size fishing units can also perform "campaigns", i.e. changes of landing site, possibly abroad, for a duration up to 10 months. When a fishing unit does not find any fish, there was a probability that of starting a campaign, that is, it will change its landing site.

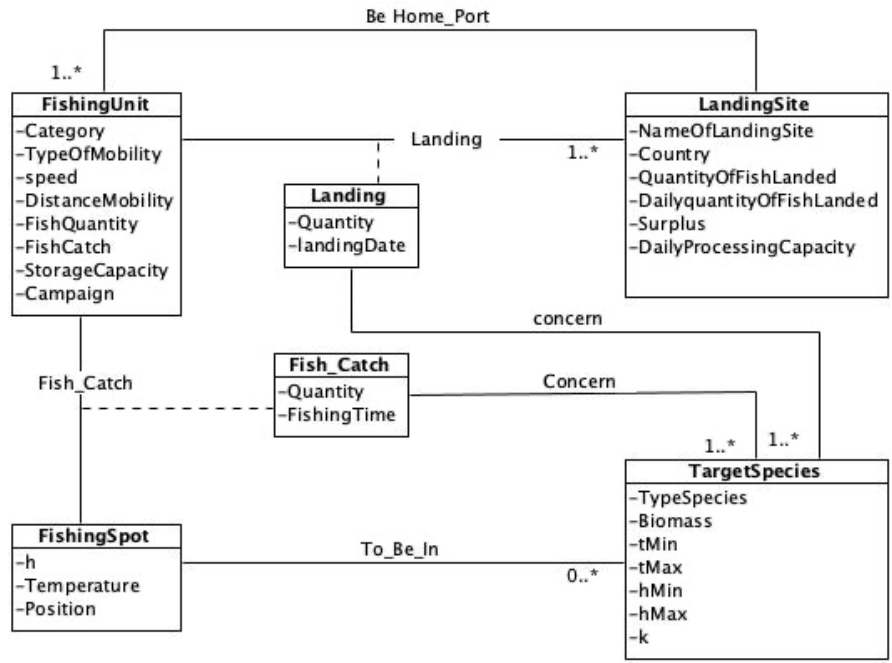

*Figure 8: Class diagram of the "Fisheries" subsystem.*

### 2.2.2 Target Species

The target species of artisanal fishing are those that are accessible on the continental shelf, dominated by 17 main species that usually constitute ~80% of the landings (Ndiaye et al., in prep). We group them into 4 model-species based on their pelagic or demersal habitat, the bathymetric limits of their distribution and their thermal preference. These model-species each represent a sum of species which constitute the pelagic and demersal fish associated with warm waters, referred to as "Guinean fish affinity", and the one associated with cold waters referred to as "Saharan fish affinity". According to the literature, we hypothesize the uniqueness of the populations of these 4 model-species at the sub regional scale, *i.e.* the reproduction rate depends on the total biomass.

### 2.2.3 Environment

Space was represented realistically, in the form of geo-referenced grid with a 10 km resolution, covering the entire area of action of the Senegalese artisanal fishing from guinea to south Morocco. Each grid cell corresponding to the sea contained local bathymetry and monthly sea surface temperature. Fish biomass patches were distributed in this grid, and thus each grid cell that own a fish patch is potentially a fishing site. Bathymetry and water surface temperature determined the habitat of each model-species. The bathymetry, or bottom topography, was variable in space but constant in time, while the sea surface temperature (SST) changes over time, with a resolution of 1 month. SST was the climatic parameter in each grid cell of the model, which can be modulated according to climatic scenarios and which constrained the fish distribution. Target species temperature and bathymetry ranges allowed us to simulate monthly variability of habitat for different climate scenarios (see figures in Appendix C).

### 2.2.4 Landing Sites

Artisanal fishing landing sites are places along the coast where fishers can beach their canoe and directly sell their catch. These are located near to road connections, and can have specific infrastructure dedicated to fish processing. Landing sites were distributed along the coast and had different daily processing capacities according to the infrastructures. In the model, if a landing site received more fish in one day than its daily processing capacity, it ceased to be an attractive area for fishermen seeking to market their catch. The daily processing capacity parameter stand for the presence or absence of infrastructures as fishing wharf, ice factories, processing units, fish meal factories and connection to the road, and the number of canoes that can be received.

## 2.3 Process overview and scheduling

### 2.3.1 Fishing Units

After initializing the distribution of fish and fishing units, the simulation began with an increment of the days of the year scrolling the date, and therefore the corresponding physical environment, i.e. the SST distribution, according to the climate scenario used. For each day, processes relating to the spatial dynamics of fishermen and target species are calculated every hour. For fishing units, the day of simulation includes 4 processes: identification of a fishing

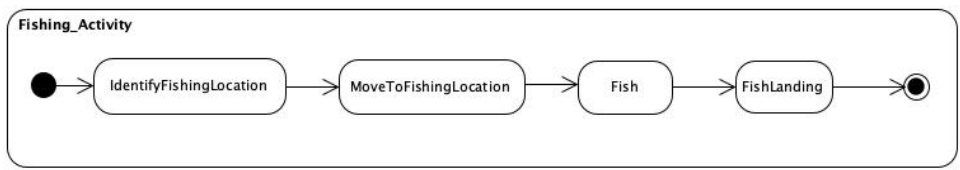

*Figure 9: General view of model actor's logic of a fishing trip. Each box correspond to a method implemented whih deal with all particular cases. Details are listed in algorithm 1. If no fishing ground can be identified in vicinity of the home landing site, then the fishing unit migrate to another*

site, movement towards this site, fishing action and fish landing (Figure 9). However, fishing trips can have a variable duration, and therefore at each time step could find a fishing unit could be at different stages of this logic, which is handled by Algorithm 1 (below):

---

**Algorithm 1:** Fishing Unit Outings

---
**begin**
    **for** *each fishing unit* **do**
        **if** *Fishing unit is at sea* **then**
            **if** *There is still time to fish* **then**
                ⌊ Continue fishing
            **if** *It is time to come back* **then**
                | Move back to the landing site
            **else**
        **if** *Fishing unit is at land* **then**
            **if** *Daily Outing* **then**
                List potential fishing grounds in a 50km radius Move toward one of
                potential (random) fishing ground Fish for a time such that total time
                of the outing does not exceed 12 hours Move back to landing site
            **if** *Tide Outing* **then**
                List potential fishing grounds in a 500km radius Move toward one of
                potential (random) fishing ground Fish for a time such that total time
                of the outing does not exceed 10 days Move back to landing site
            **if** *Campaign* **then**
                ⌊ Move the fishing unit to an other landing site nearer to the fishing area.
        **else**

---

### 2.3.2 Fish Population

Fish population dynamics included 3 processes: habitat search, reproduction and movement. The habitat of each species was calculated each new month of simulation, but population growth was only calculated twice per year of simulation, as described in Algorithm 2:

---

**Algorithm 2:** Fish Species population dynamics

---
**Data:** *Tmin, Tmax, bathymin, bathymax, K, k, r*
**begin**
    **for** *each fish species* **do**
        1 - Search for habitats: cells with depth and temperature in the range of the
        species preferences
        2 - Compute total population biomass growth using (B, K,r, t)
        **for** *each fish species individual* **do**
        ⌊ Fish move algorithm

---

Finally, the mobility of each fish biomass patch was computed each simulation day for pelagic species, and each simulation month for demersal species following Algorithm 3:

---

**Algorithm 3:** Fish move algorithm

---
**Data:** *Tmin, Tmax, bathymin, bathymax, K, k, r*
**if** *fish species is demersal* **then**
    **if** *a new month started* **then**
        /*move each month */ **for** *each super individual* **do**
        ⌊ New random position within the habitat

**if** *fish species is pelagic* **then**
    **if** *A new day started* **then**
        /*move each 24h... */ **for** *each super individual* **do**
        ⌊ Move toward an adjacent cell where habitat is ok

---

## 2.4 Design concepts

### 2.4.1 Basic Principles, Emergence and Adaptation

The main hypothesis of the model was that artisanal fishing fleet migrations continuously adapt to fish biomass distribution and landing sites processing capacities. Fish biomass resulted from the balance between growth and fishing mortality, and its distribution was constrained by climate through its effect on fish habitat. By contrast, landing site processing capacities directly resulted from fishery management decisions.

The distribution of fishing units along the coast emerged from the interaction between climatic, biological, social and economic factors. The total biomass of fish catches each year and landed in each landing site resulted from the correspondence of landing sites processing capacities with the distribution of fish biomass. The stabilization of fish catches depends on the balance between fishing mortality and fish growth. Fishing mortality emerge from the number of fishing units, the distance between fishing grounds to landing sites and the processing capacities of the landing sites.

Fish agents adapted their position depending on variations in sea temperature, in order to remain in their grid cells where SST and bathymetry correspond to their habitat. These rules aim to reproduce the observed migrations of fish as a function of temperature, which differ depending on the species. The fishing units had several adaptive traits, to reflect the adaptive capacity of Senegalese artisanal fishing. Fishing "tactics" and "strategy", following a simplified version of the Pech et al. (2001) model, aimed to maximize the profitability of the activity. The fishing strategy consisted of adapting the preparation of the fishing unit according to environmental data and the socio-economic context. Fishing tactic consists of adapting the action of fishing to the abundance of fish encountered, such as changes in fishing site, or change of fishing gear, and in the choice of the landing place according to distances and price.

### 2.4.2 Entities Objectives, Sensing, Interactions and Stochasticity

Fishing units' strategies and tactics were dictated by the economic criterion, *i.e.* fish marketing. The objective of each fishing unit was to find a fishing site and find a suitable landing site. This amounts to maximizing the income of fishers by reducing travel costs and maximizing catches and sales prices. Note that the social criterion also determined the back migration of fishing units, which depended on its home port.

Fishers continuously exchange information, by phone even at sea when near the coast, to find out fish grounds and the prices in the different landing sites. In the model the price was not explicit but fishermen adjusted their landing place in the event of saturation of the daily processing capacity of the landing sites. Implicitly, fishers were also considered to communicate information about fish patches, so in absence of available fish patch within their radius they didn't go fishing or they migrate to another landing site.

Fishing units targeting small pelagic fish spend all the time scanning the sea around them to detect fish schools. Thus, in the model, these fishing units detected fish patches they encounter on their route. By contrast, fishing units targeting demersal species always move to their predefined destination, and adjust their position if they get low catch there.

Fish patches entities had no other objectives than remaining within their preferred habitat. In the model, fish sense the SST and bathymetry and move to stay within their habitat as SST change according to seasons.

Models' stochasticity occurs at two levels, the fish biomass distribution and the fishing units' choices. For the fish biomass distribution, stochasticity comes from the initial random fish patches distribution and their random walk within their habitat. For fishing units, the choice of fishing ground within the radius of action, and in case of campaign migration the choice of the new landing site, also involves stochasticity.

### 2.4.3 Observation

Here "observation" refers to the variables that were archived during each simulation and plotted for results discussion. The variables archived after each day of simulation were the distribution of canoes in the landing sites, the quantity of landings at each landing site, and the biomass of the four fish-model species. This allowed to comment, for each scenario tested, on the evolution of the total catch and distribution among the landing sites, and eventually on the fisher's migrations between landing sites.

## 2.5 Initialization

### 2.5.1 Fishing Unit's Initial Distribution

At the first-time step of the model, the fishing units were distributed among the landing sites as specified in the simulation's configuration file. The initial landing site remained an attribute of the fishing units, this was the place where it came back after a migration. Also, for each fishing unit (FU) the quantities of fish caught and the number of days at sea were initially set to zero.

### 2.5.2 Fish Biomass Initial Distribution

The habitat of the four model-species were computed at the first-time step. The initial biomass of each model species, a parameter of the simulation, was distributed in patches randomly within the habitat, with the maximum biomass per patch "k". The order of magnitude of the total biomass of each model-species was based on evaluation reports of the state of fishing stocks (FAO, 2015).

### 2.5.3 Environment Initialization

At the first-time step, the physical grid containing bathymetry and SST at each grid cell was read, corresponding to the first day of January of the first year of simulation of the selected climate scenario.

### 2.5.4 Landing Sites Initialization

The landing sites were each located in the model's map at their geographical position, and their daily landing quantity were set to 0. Note that this quantity was reinitiated, *i.e.* set to 0, every 24 hours of simulation time.

## 2.6 Input data

Input data determines the scenarios explored with each simulation. These data were gathered in the configuration files listed in appendix Table *8*.

### 2.6.1 The Climate Scenario (Marine environment).

Sea surface temperature (SST) from global reanalysis of the past climate ("ERA5") were extracted and interpolated to a 10km-resolution grid from 10°South to 25°North and from 10°West to 19°West, with a monthly temporal resolution. Each sea grid point therefore represents 100 km$^2$ with an average depth and monthly SST. Average depth or bathymetry in grid cell was extracted from the "Etopo" global database. Climate scenarios specifically built as inputs were listed in Table *1* and marine environment attributes were listed in *Table 2*.

*Table 1: List of climate scenarios used.*

| Climate scenario | Period covered/Scenario | Configuration File Name |
|---|---|---|
| Historical reanalysis | 1979 – 2014 | Climat_era_1979_Lolli |
| Optimistic future scenario | 1979 – 2014 + 1.5°C | Climat_era_1v5deg_Lolli |
| Pessimistic future scenario | 1979 – 2014 + 3°C | Climat_era_3deg_Lolli |

*Table 2: Variables of the marine environment defined at each grid point.*

| Variable | Values |
|---|---|
| Surface temperature (°C) | 18-30°C depending on season |
| Depth (meters) | 0-3000 m |
| ZEE* | Country Name |
| Marine protected area* | Yes No |
| *: not used in the simulations presented here | |

### 2.6.2 Landing Sites Parameters

We considered 15 main artisanal fishing landing sites, which represent the sum of smaller landing sites, distributed in Mauritania, Senegal, Gambia, Guinea Bissau and Guinea (Table *3*). We considered 3 contrasting scenarios of processing capacity distribution among the landing sites: (1) the distribution of processing capacity estimated in 2020 (notably including numerous fish meal factories in Mauritania), (2) the same total processing capacity but distributed homogeneously among landing sites and (3) homogeneous but 10 times smaller processing capacity (Table *8*).

*Table 3: List of northwest Africa artisanal fishing landing sites considered in the model. Estimated daily processing capacity gather infrastructure from other neighbor fish landings and were based on 2020's reports.*

| Landing Site | Latitude | Longitude | Country | Processing Capacity |
|---|---|---|---|---|
| Nouadhibou | 21.00 | -17.00 | Mauritania | 15400 tons/day |
| Tiouilit | 18.82 | -16.16 | Mauritania | 700 tons/day |
| Nouakchott | 18.10 | -16.02 | Mauritania | 9400 tons/day |
| **Saint-Louis** | 16.02 | -16.51 | Senegal | 450 tons/day |
| **Fass Boye** | 15.25 | -16.85 | Senegal | 100 tons/day |
| **Kayar** | 14.92 | -17.12 | Senegal | 450 tons/day |
| **Dakar** | 14.76 | -17.48 | Senegal | 750 tons/day |
| **Mbour** | 14.41 | -16.97 | Senegal | 750 tons/day |
| **Joal** | 14.17 | -16.85 | Senegal | 400 tons/day |
| **Tanji** | 13.36 | -16.8 | Gambia | 350 tons/day |
| **Gunjur** | 13.15 | -16.78 | Gambia | 650 tons/day |
| **Kafountine** | 12.92 | -16.75 | Senegal | 350 tons/day |
| **Cap Skiring** | 12.38 | -16.74 | Senegal | 30 tons/day |
| **Bissau** | 11.80 | -15.58 | Guinee Bissau | 10 tons/day |
| **Conakry** | 9.510 | -13.71 | Guinea | 10 tons/day |

### 2.6.3 Target Species Parameters

The parameters for logistic population growth (the carrying capacity "K", the yearly reproduction rate "r") and the maximal local biomass density "k" were estimated from grey literature (Table 4). For each model species, 4 habitat limits parameters defined a temperature and bathymetry range. The bathymetry range represented for the variable offshore distribution of the species represented.

*Table 4: Attributes of "fish" agents (estimated values).*

| **Model-Species** | **species represented** | **Temperature range** | **Bathymetry range** | **K (Tons)** | **K (tons/km2)** | **r (year⁻¹)** |
|---|---|---|---|---|---|---|
| **Coastal demersal "Guinean affinity"** | *Arius spp., Sepia officinalis, Pomadasys jubelini, Dentex macrophtalmus* | 24-29°C | 0-100m | 300 000 | 30 | 0.5 |
| **Coastal demersal "Saharan affinity"** | *Octopus vulgaris, Pagrus spp, Pagelus bellotii, Galeoides decadactylus, Epinephelus aenus, Cynoglossus spp., Pseudotholitus spp.* | 18-25°C | 0-100m | 500 000 | 50 | 0.5 |
| **Coastal pelagic "Guinean affinity"** | *Sardinella maderensis, Ethmalosa fimbriata, Caranx rhonchus, Trachurus trecae* | 24-29°C | 0-100m | 1M | 100 | 1.5 |
| **Coastal pelagic "Saharan affinity"** | *Sardinella aurita, Scomber colias* | 18-25°C | 0-300m | 3M | 100 | 1.5 |

### 2.6.4 Characteristics of the Fishing Fleet

The three categories of fishing units in the model correspond to the technical characteristics in Table *5*. In the model, the three fishing unit categories differs in their storage capacity, catchability and migration behavior, but can share similar home ports (Table *6*). We considered 3 contrasted fishing fleet scenarios corresponding to (1) the number of canoes from the 2014 census, (2) a number of canoes 2 times less and (3) 2 times more than the 2014's situation. The probability and the maximum duration of campaign for each fishing unit were set such that larger FU performed longer migrations (campaign).

*Table 5: Technical characteristics of the three main categories of Senegalese artisanal fishing units (adapted from Chavance and Morand, 2020).*

|  | Fishing Unit Category I (line or longline) | Fishing Unit Category II (gillnet) | Fishing Unit Category III (Rotating Seine) |
|---|---|---|---|
| **Boat size (m)** | 9.5 | 12,6 | 21.5 |
| **Average number of crew** | 3,7 | 4,3 | 11,3 |
| **Outing rate (%>2j)** | 57.4 | 55.7 | 96 |
| **Good fishing period** | 1->12 | 11->07 | 10->07 |
| **Engine power** | >=15ch | >=15ch | >=40ch |
| **Cost of purchase (canoe)** | 952,000 xof | 1 443 000 xof | 1,907,000 – 11,274 |
| **Average cost of day fishing**\*\* | 100,000 xof | 100,000 xof | 200,000 xof * |
| **Average cost of tidal fishing**\*\* | 150,000 xof | 150,000 xof | 300,000 xof * |
| **Average turnover** | 190,000 xof | 220,000 xof | 800,000 xof |
| **Squad in Senegal (2020)** | 4888 | 4955 | 758 |
| *2 canoes must go out together, one carries the net, the other the fish \*\*Excluding salaries and depreciation | | | |

*Table 6: Attributes of the three categories of fishing units.*

|  | Cat I | Cat II | Cat III |
|---|---|---|---|
| Storage capacity | 500 kg | 5000 kg | 30 000 kg |
| Fishing Radius | 50 km | 100 km | 1000 km |
| Maximum time at sea | 1 day | 15 days | 15 days |
| Catchability | $10^{-4}$ | $10^{-3}$ | $10^{-2}$ |
| Campaign Probability | 0.1 | 0.2 | 0.3 |
| Campaign Maximum Duration | 4 months | 8 months | 12 months |
| Homeport | Initial landing zone | | |

## 2.7 Submodels

### 2.7.1 Fishing

The biomass captured (the catch) at each time step by a fishing unit that reach its fishing ground was given by a relationship that mimic the Holling III functional response for low biomass levels (eq.1):

$$C_t = \frac{q b_t^2}{b_t + b_{crit}} \qquad \text{(eq. 1)}$$

Where $C_t$ = biomass catch at time step t; q = catchability (per time step); $b_t$ = biomass of the fished fish patch at time t and $b_{crit}$ = local biomass threshold for which the proportion of fish catch was divided by 2 (Figure 10). At each time step, the biomass catch is added to the fishing unit total catch of the current fishing trip, and subtracted from the fish patch biomass. Fishing stops either at the end of the allowed time at sea or when the biomass captured equals or exceeds the storage capacity of the canoe. If the local patch is depleted fishing can continue in the neighbor patches.

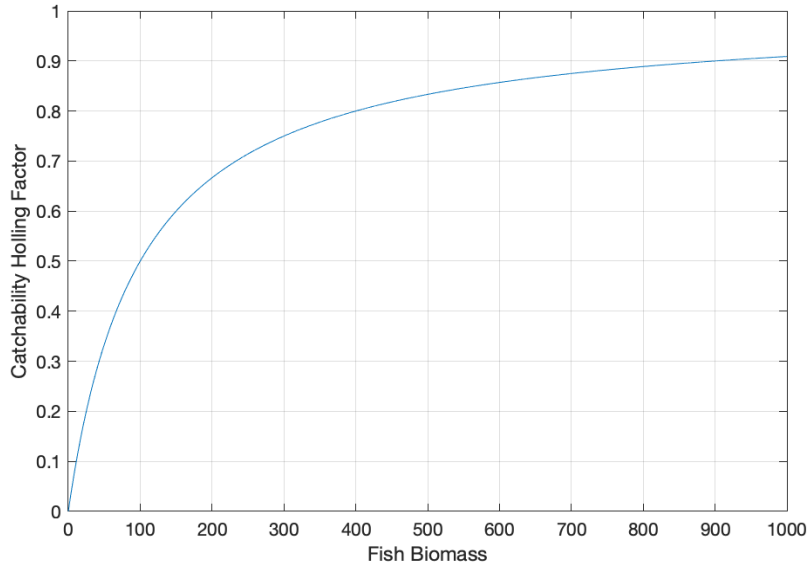

*Figure 10: Catchability correction factor depending on fish biomass, Holling type III like, applied from simulation 7. Qcor = B/(B+100)*

### 2.7.2 Fish Population Dynamics

The biomass dynamics of each model-species was represented using a logistic growth model and was split in biomass patch of the size of grid cells (10 km$^2$) distributed within the available habitat. The maximum biomass per patch was a parameter of each model-species.

The growth of the fish population was represented by a logistic law applied at the population level at specific times of the year, while fishing mortality was applied continuously in patches where fishing occurs. The amount of biomass catch was described in previous section. The population growth procedure is done for each model species "s" times a year (eq. 2):

$$\begin{cases} B = \sum_{i=1}^{n} b_i \\ G = \frac{r}{s} B \left(1 - \frac{B}{K}\right) \quad \text{(eq. 2)} \\ N = \frac{G}{k} \end{cases}$$

Where B is the total model-species biomass, computed as the sum of the remaining biomass in the patches, $b_i$ is the biomass in the patch i and n the number of patches. "s" is the number of reproduction event per year. Thus, reproduction event occurs each 12/s months. G is the biomass growth, and N the number of new patches to be distributed randomly within the species habitat.

The total biomass of each model species was equal to the sum of the biomasses 'patch. When a fishing Unit fish in a patch, the biomass was reduced and when the biomass patch falls below 100 tons (1 ton per km2), the patches were deleted. This stands for fish patch senescence which was not explicit in the model. Since pelagic species are very mobile, in the model they performed continuous random walk within the habitat, updating their position at a daily rate. By contrast demersal species are usually more sedentary, thus in the model they also performed a random walk within the habitat but only one time per month.

# 3 Simulations

The model was implemented in Gama. We called the obtained simulator "Lolli" that we used for the sensitivity analysis of the model. Using this simulator, a first analysis all of the model parameters were explored, in order to know the sensitivity of the model to each of them. Secondly, preliminary climate and fish parameters scenarios were explored to disentangle the effect of these factors. At this stage, historical simulations were not yet performed but the preliminary simulations allowed us to understand the dynamics emerging from the model assumptions.

## 3.1 Presentation of the "Lolli" Simulator

The model was implemented on the platform "*Gama*". The code was available uploaded on "github", an online version manager. It will be publicly released after complete validation procedure and first results publication.

The simulation interface (Figure 11) has a set of input parameters that allow the modeler to test various configurations of the model by changing one factor at a time, often referred to as the "One Factor at a Time" (OFT) approach. These parameters are essential for understanding how different initial conditions and input variables can influence simulation results. For example, they could include values such as fish population parameters, number of canoes, migration behavior, or other characteristics that impact the behavior of agents in the model.

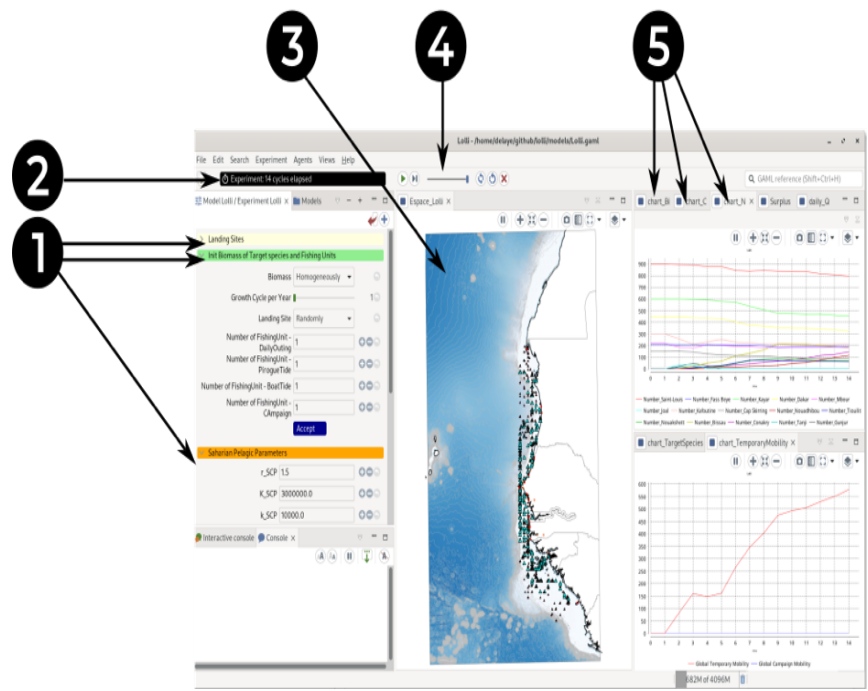

*Figure 11: Presentation of the "lolly" simulator graphical interface. 1- Simulation Parameters, 2-Visualization of Time, 3- Spatial Visualization, 4- Real-Time Interaction, 5- Monitoring Variables*

**Visualization of Time** is provided by default by GAMA to all models. It allows users to follow the temporal evolution of interactions between agents and understand how dynamics unfold over time. For example, this could include graphs showing how the fish population changes over fishing seasons. **Spatial Visualization** allows observers to spot and identify different elements, such as types of canoes, which are represented by colored triangles. Using this spatial visualization, users can track, verify and identify the spatial behaviors of agents. For example, it could show how canoes move across the lake, where they group together, or how they interact with schools of fish in specific areas of the lake. **Real-Time Interaction** GAMA offers a tool to interact with the progress of the simulation in real time. This default functionality allows users to make decisions and apply actions at certain points in the simulation. For example, a user might decide to introduce a new fishing regulation at some point in the simulation to see how it affects fishermen's behavior and the fish population. **Monitoring Variables.** Observers have the possibility to follow a set of individual or aggregated variables. For example, this could include tracking fisher productivity based on their position, or fish biomass changes in response to fishing activities.

## 3.2 Sensitivity Analysis

The most influential parameters on the output were identified using Saltelli (2010) sensitivity analysis method. This was done by running the model for each combination of 13 parameters listed in Table 7, and calculating the variance of the cumulative catch biomass after 6 months of simulation, a time allowing each process to intervenes in the cumulative sum of catch, except the fish population growth that would take a lot calculation time. The variance of this output was then decomposed into individual contributions from each input parameter. Note that this sensitivity analysis did not include the processing capacity parameters, as this parameter was later introduced in the model.

The most influential parameters for the total fish catch were the catchability parameter of fishing unit (FU) category 3, i.e. the largest fishing units considered (Figure 12). However, this can change as for the demersal biomass in Guinee Bissau which was more sensitive to catchability parameter of fishing unit category 2, probably due to the fact that campaign for FU category 3 did not start in Guinee Bissau during the sensitivity test. In general, the sensitivity results cannot be interpreted much, but the results still show a stability of the algorithm that may allow to study the effect of the different parameters in specific scenarios. Results of influential parameters for the fish catch per country were summarized in figures provided in Appendix B.

*Table 7: List of parameters and their range used for the Saltelli analysis of the model.*

| Input name | Range | Explanation | Source |
|---|---|---|---|
| **Climate Variability** | -2°C to + 2°C | This parameter simulates a uniform increase or decrease in sea surface temperature | IPCC |
| **Campaign Probability for fishing units' cat. 1,2,3 (3 parameters)** | 0 - 1 | After each fishing trip, if catches are null then there is a probability that the fishers migrate to another landing site. Is defined for each fishing units' categories (1,2,3) | Fishermen interviews |
| **Campaign Duration for fishing units' cat. 1,2,3 (3 parameters)** | 1-10 months | Max duration of the campaign. Each time a fishing unit start a new campaign, the duration is fixed in advance randomly between 1 month and the max duration fixed here. | Fishermen interviews |
| **Catchability (q) for fishing units' cat. 1,2,3 (3 parameters)** | 10-6 to 10-5 (cat 1) 10-5 to 10-4 (cat 2) 10-4 to 10-3 (cat 3) | Share of local fish biomass catch per hour of fishing. Is defined for each fishing unit's categories (1,2,3). | Fishermen interviews |
| **Canoe storage for fishing units' cat. 1,2,3 (3 parameters)** | 250 – 750 kg (cat 1) 500 - 1500 kg, (cat 2) 1 – 100 tons (cat 3) | Storage capacity on board. | Fishermen interviews |

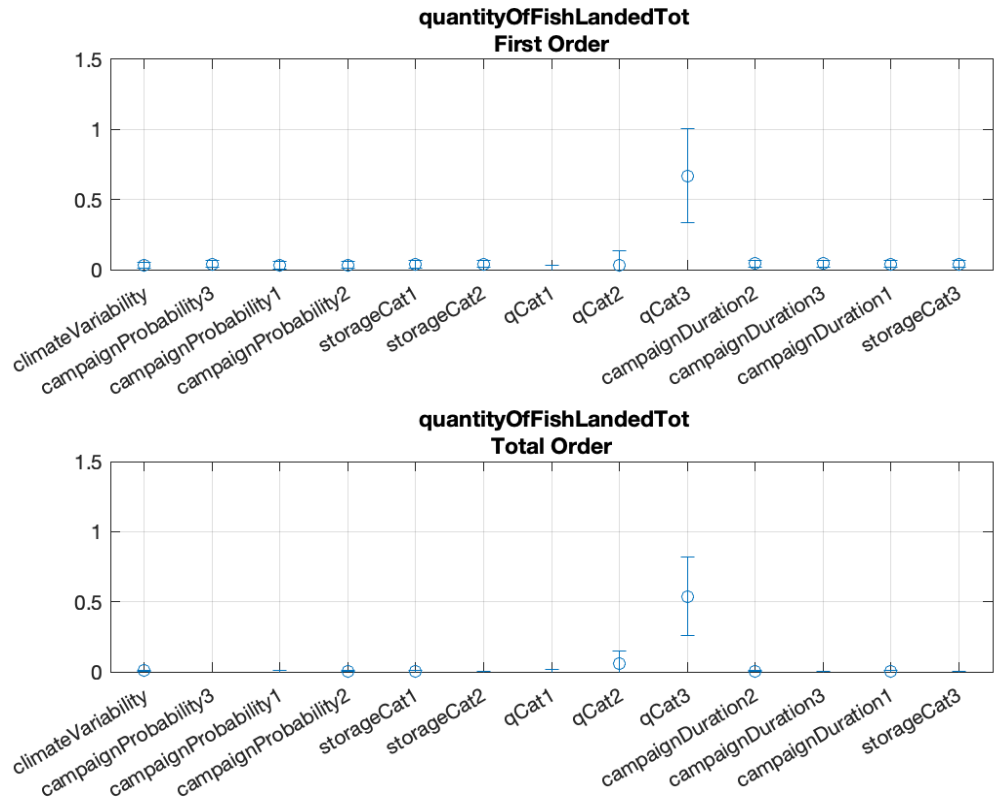

*Figure 12: Saltelli analysis of the effect of model parameters on cumulative total catch.*

## 3.3 Preliminary Simulations and First Scenarios Exploration

The objective in this section was to explore different climatic and socio-economic scenarios for artisanal fishing (Table *8*). However, before reaching this objective, preliminary simulations were needed to calibrate the catchability parameter. Indeed, catchability was the most arbitrary parameter in the model, since it defined the rate at which fishers catch fish, a very complex process that integrate many things as the experience of fishers, the number of fishers per canoe, the type and size of fishing gear, the fish biomass density, fish behavior, the state of the sea, the depth, and many others parameters. For this reason, in fishery models this parameter is usually used as a calibration parameter, *i.e.* its value is estimated such that model predictions of fish catch in average state are comparable to available data. Also, catchability may not be linear because the time fishers spend to find fish at very low density may increase, as described in the Holling III functional response. As it was coded in the model, the meaning of the catchability parameter is the fraction of fish biomass in a fished patch that is catch per time step (hour) per fishing unit. The FU's storage capacity always limit the maximum quantity of fish catch per fishing trip, but it appeared that given the large number of fishing unit it didn't prevent overexploitation.

*Table 8: List of scenarios and corresponding configuration files for the one factor at time sensitivity study*

| Model Parameter | Scenario Name | Filename | Description |
|---|---|---|---|
| Climate Scenario | SST +0°C (Reference) | Climat_era_1979_Lolli.tif | Reference climate file. Regional SST from climate reanalysis era 1979-2005 |
| | SST +1.5°C | Climat_era_1v5deg_Lolli.tif | Reference SST + 1.5 °C |
| | SST +3° | Climat_era_3deg_Lolli.tif | Reference SST + 3 °C |
| Coastal Infrastructures | 2020's Capacity (Reference) | DailyProc-Cap_2020s_Lolli.csv | Reference processing capacity of landing sites, based on 2020 data (dominated by fishing meal factories in Mauritania). |
| | Homogeneous Capacity | DailyProcCap_HOMOGE-NEOUS.csv | Homogeneous distribution of the reference processing capacities among all landing sites |
| | Reduced Homogeneous Capacity | DailyProcCap_HOMOGE-NEOUS_x01.csv | Homogeneous distribution of the reference processing capacities divided by 10 among all landing sites |
| Artisanal Fishing Fleet | 2014's Fleet | Fis-gingUnits_2014s_10p_Lolli.csv | Reference distribution of Fishing units according to the 2014 UEMOA survey (13000 fishing units), with 1 model fishing unit corresponding to 10 real ones. |
| | Double Fleet | FU_2014s_x2_Lolli.csv | Reference Fishing Unit distribution x2 |
| | Half Fleet | FU_2014s_x05_Lolli.csv | Reference Fishing Unit distribution /2 |

Thus, in preliminary simulations we explore the effect of the catchability parameter and how it interacted with climate change scenarios (Table 9). These simulations also served to test model algorithms, time of execution, parameters calibration and stability of the results. Most figures cited in this section can be found in Appendix D, only some contrasted results figures were included at the end of this section.

*Table 9: Preliminary simulations for calibration of the catchability parameter with corresponding configuration files for the "one factor at time" sensitivity tests.*

| Sim # | Climate scenario | Infrastructure scenario | Fishing Fleet scenario | Catchability parameter |
|---|---|---|---|---|
| 1 | SST +0°C | 2020's Capacity | 2014's Fleet | Qcat1 = 0.0005; Qcat2 = 0.005; Qcat3 = 0.05 $b_{crit}$ =0 (i.e. no Holling III behavior) |
| 2 | SST +0°C | 2020's Capacity | 2014's Fleet | Idem (repetition Sim#1) |
| 3 | SST +0°C | 2020's Capacity | 2014's Fleet | Idem (repetition Sim#1) |
| 4 | SST +0°C | 2020's Capacity | 2014's Fleet | Change in the code: Cat3 access 30% of demersal species |
| 5 | SST +0°C | 2020's Capacity | 2014's Fleet | As simu 4 but with reduced catchability (divided by 5 compared to Simu 1,2,3,4 simulations) |
| 6 | SST +0°C | 2020's Capacity | 2014's Fleet | As simu 4 but with further reduced catchability (divided by 50 compared to Simu 1,2,3,4) Qcat1 = 0.00001; Qcat2 = 0.0001; Qcat3 = 0.001 |
| 7 | SST +0°C | 2020's Capacity | 2014's Fleet | As simu 4 but with further reduced catchability Qcat1 = 0.000001; Qcat2 = 0.00001; Qcat3 = 0.0001; Holling type III with $b_{crit}$ =100 |
| 8 | SST +0°C | 2020's Capacity | 2014's Fleet | Idem (repetition of Sim #7) |
| 9 | SST + 1.5 °C | 2020's Capacity | 2014's Fleet | Same catchability as Sim #7, with optimistic climate scenario |
| 10 | SST + 3 °C | 2020's Capacity | 2014's Fleet | Same catchability as Sim #7, with pessimistic climate scenario |
| 8.1 | SST +0°C | 2020's Capacity | 2014's Fleet | Catchability increased to Qcat1 = 0.0001; Qcat2 = 0.001; Qcat3 = 0.01 |
| 9.1 | SST + 1.5 °C | 2020's Capacity | 2014's Fleet | Same catchability as Sim #8.1 |
| 10.1 | SST + 3 °C | 2020's Capacity | 2014's Fleet | Same catchability as Sim #8.1 |

At the beginning of simulation 1, many short-term migrations occurred (Appendix Figure 8), corresponding to fishing units that must change their landing site because of the lack of infrastructure (processing capacity) at their initial landing site. This was due to the fact that catchability parameter was too high, so that FU could systematically fill their storage capacity when fishing and then saturated the landing sites, with about 2 000 000 tons catches during the first year of simulation, mostly in Senegal and Mauritania, but rapidly dropping (Appendix Figure 11). This large catch value at the beginning of the simulation can be interpreted as the model "spin-up", i.e. the simulation time during which the model equilibrates starting from a "freezed" initial situation in which all fishing unit are at land and fish biomass are at an arbitrary level. After the first year of simulation, short term migrations were replaced by long term migration of almost all fishing units. Long-term migrations occur when fishers cannot find fish within their radius of search (which was 50, 100 and 1000km respectively for Cat 1,2 and 3). This was coherent with the pelagic fish biomass rapid drop and extinction (Appendix Figure 9) as these species were targeted by Category 3 fishing units. However, demersal species biomass increased and stabilized at their maximum value, the carrying capacity which was set to 300 000 tons for Guinean demersal and 500 000 tons for Saharian demersal fish (Appendix Figure 9). This was due to the fact that fishers depleted the fishing area around the landing sites, and by evolutionary process the fish biomass growth distributed outside the reach of fishers, although sporadically entering in the fishing area which allowed to maintain some catch (Appendix Figure 11). It was possible in this version of the model because only fishing units of Category 1 and 2 could access demersal species. During the first year of simulation most Senegalese fishing units migrated abroad (Appendix Figure 10), then the movements were correlated with the timing of fish reproduction (1 time per year in this configuration). After five years the distribution stabilized with ~750 fishing units in and 550 fishing units out Senegal. The catch increased again from year 2 to 4 in all landing sites (Appendix Figure 11) but started to drop at 5 years. The repetition of simulation 1 showed a strong convergence of the results, despite some variability due to the model's stochasticity (Appendix Figure 12).

In simulation 4, the value of the catchability parameter was not changed, but the code was modified such that demersal species could be partially exploited by Category 3 fishing units. Indeed, even if Category 3 fishing units normally target pelagic species, in case of fish scarcity they can decide to target demersal species, a behavior that was described in the focus groups. As a result, all the catches and fish biomass rapidly drop to 0 in simulation 4, including demersal species. The cumulative catch per fishing unit category rapidly stabilize, indicating the fishery collapse, with the higher catch for cat. 3 (Appendix Figure 13 and Appendix Figure 14). In simulation 5 despite a reduced catchability for all categories of fishing units, all fish biomass rapidly drops (Appendix Figure 15). In simulation 6 we further reduced the catchability for all categories of fishing units, but the fish biomass still rapidly drops (Appendix Figure 16). In all these simulations in which there is a rapid drop of fish biomass, we see that most fishing units migrates to Mauritania before the fisheries collapse (Appendix Figure 17). Once the fisheries collapsed, the fishing units went back to Senegal and continued to move sporadically to other countries in search of (no more existing) fish.

In simulation 7, we included a Holling III functional response which strongly reduce catchability when fish patch drops under 100 tons per patch (Figure 10), and we further reduced the max catchability so that the total fishing pressure might be ~100 times smaller than the one expected. As a result, the fish biomass stabilized and the catch also were stable, around 350 000 tons per year in Senegal plus ~50 000 tons out of Senegal, mostly in Mauritania (Appendix Figure 18 and Appendix Figure 19). Repeating simulation 7 provided contrasted results with ~10% higher catch (Appendix Figure 20), highlighting a significant impact of stochasticity compared to simulations 1-3. The same configuration but using climate change scenario SST + 1,5°C (simu 9, Appendix Figure 21) and SST + 3 °C (simu 10, Appendix Figure 22) generated similar results, with a catch variability below the effect of stochasticity observed between simulations 7 and 8. It means that at such a low fishing pressure, the climate changes scenarios did not impact the fisheries in the model.

In simulations 8.1, 9.1 and 10.1 we applied an increased catchability (x100 compared to sim 8, 9 10). Although high, such catchability value were more realistic, allowing fishers to fill their canoe in few hours in the case of very high fish biomass density, as it was reported in old fisher's interview. As a result, and despite the Holling III function that reduce catchability at low fish density biomass, fish biomass were depleted after 5 years in sim 8.1 (Appendix Figure 23 and Appendix Figure 24), in 6 years in simulation 9.1 and seven years in simulation 10.1 (Appendix Figure 25, Appendix Figure 26, Appendix Figure 27, Appendix Figure 28). The first year catch reach ~1,8 M tons in simulation 8.1 (current climate), ~1,65 M tons in simulation 9.1 (SST + 1,5°C) and ~1,7 M tons in simulation 10.1 (SST + 3°C), but here again these differences may not be significative, falling under the stochasticity effect. However, the better resilience of the fisheries with the SST+3°C climate scenario might be due to the fact that Guinean pelagic fish habitat was constrained offshore and thus less accessible to artisanal fisheries, which slow down the overexploitation effect.

To conclude on these preliminary simulations, the "actual fishing pressure scenario" (simu 8.1, 9.1, 10.1) resulted in fishery collapse and numerous long term migrations for the three climate scenarios tested, including the reference climate scenario (present) (see recapitulative Figure 13 and Figure 14). The fishing effort considered in this "actual fishing pressure scenario" was based on considering a fixed number of fishers and a constant investment in fishing effort, i.e. fishers are continuously prospecting for new fish ground by increasing their migration distance in the case of fish depletion. This is not realistic as fishers actually leave the fishery or reduce their fishing effort if the catch doesn't justify anymore the cost of fishing trips; however, this prediction still provides information that fishers have to leave the fishery on the short term. On the other hand, the "lower than actual fishing intensity" fishing scenario (simu 8, 9, 10) showed that for any climate scenario, the fishery stabilize around 250 000 tons per year, only in Senegal (see recapitulative Figure 15 and Figure 16). This scenario showed that according to the model, climate change alone would not be responsible for the fishery collapse, thus the only tipping point for fishery collapse have to be identified on fishing effort.

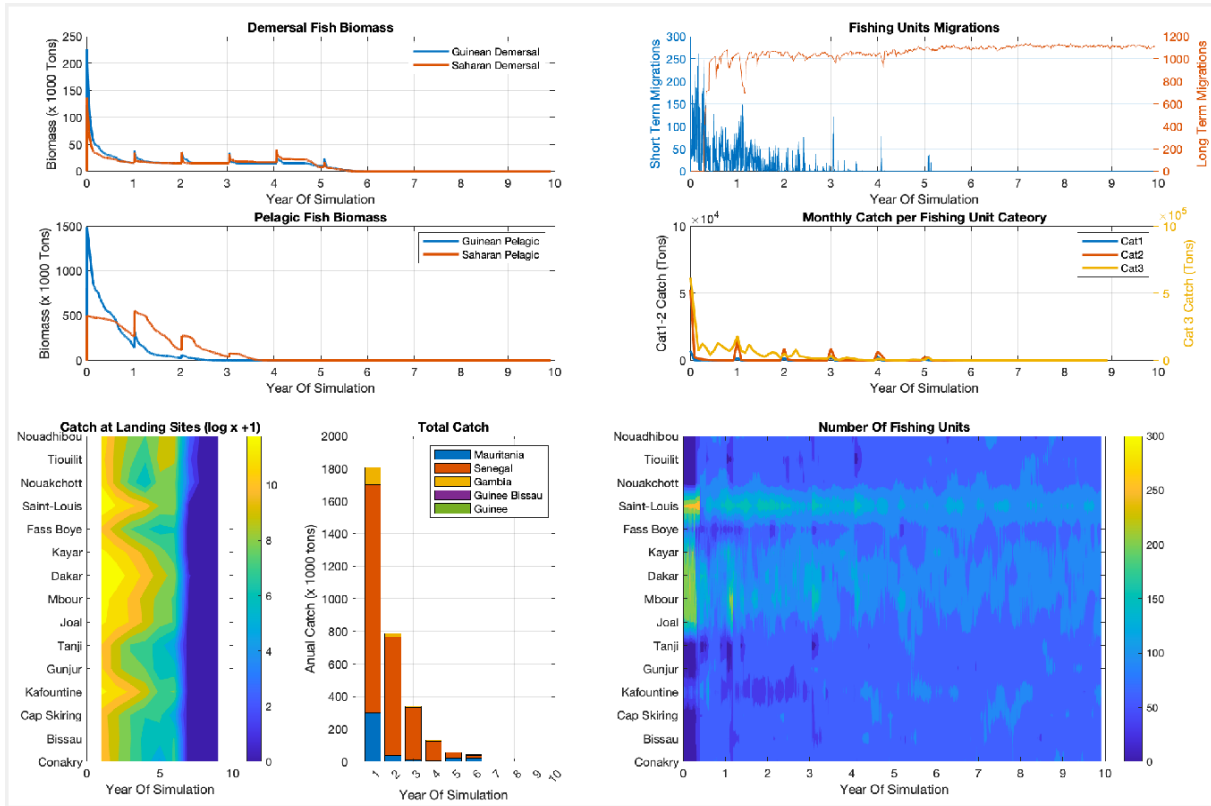

*Figure 13: Results from simulation 8.1, with actual fishing efficiency, reference climate scenario.*

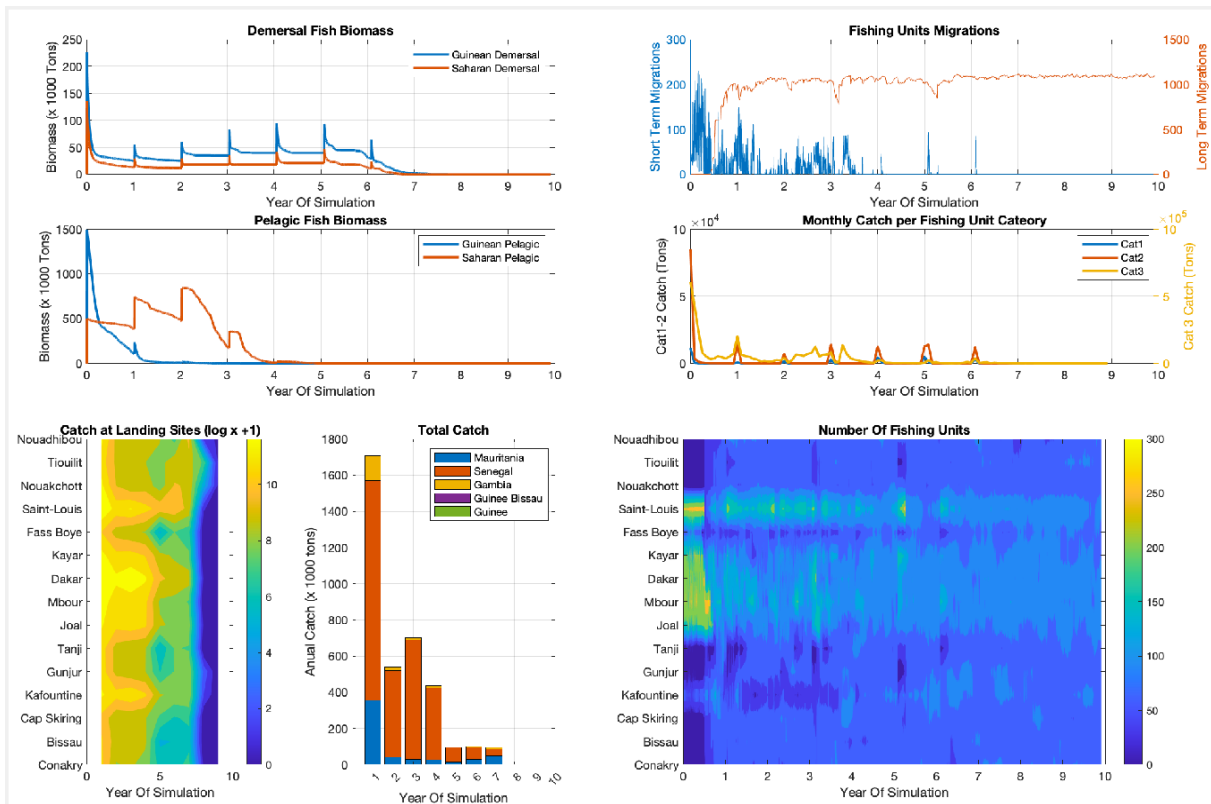

*Figure 14 Results from simulation 10.1, with actual fishing efficiency, SST+3°C climate scenario.*

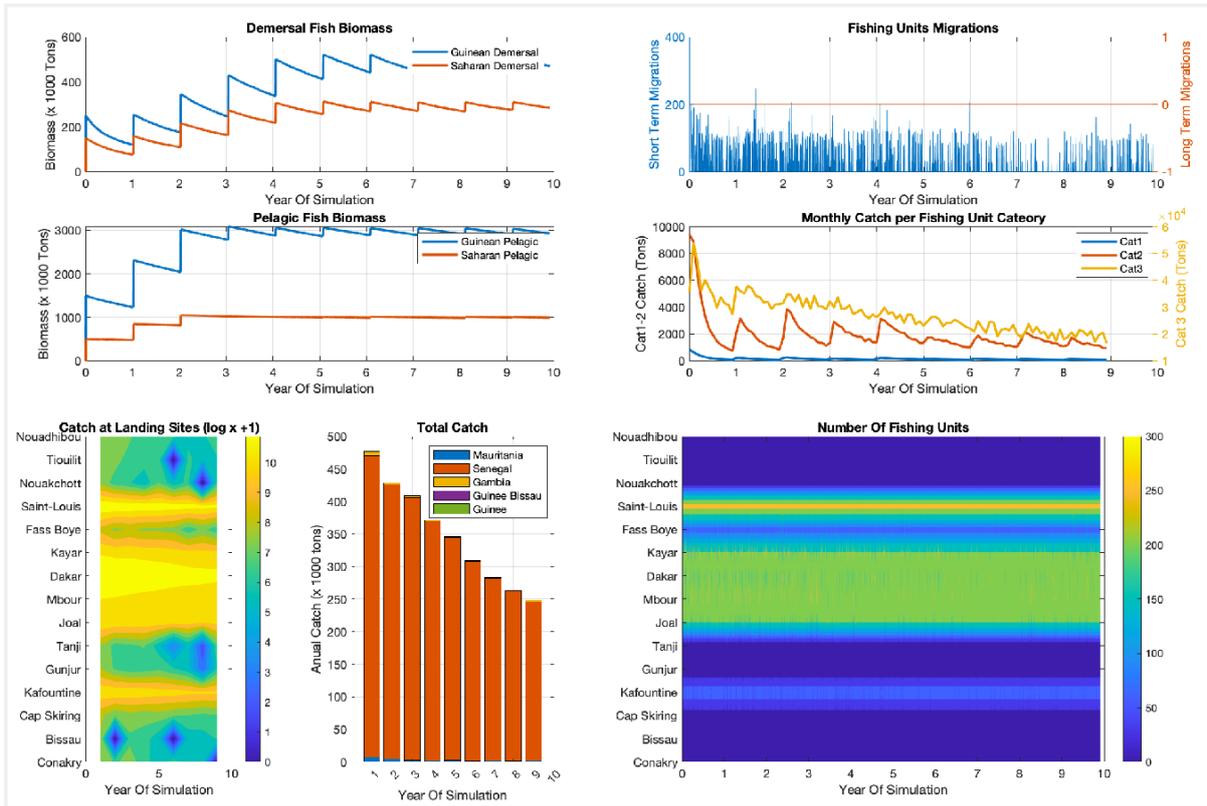

*Figure 15: Results from simulation 8, with a 100 time slower than actual fish catch, reference climate scenario.*

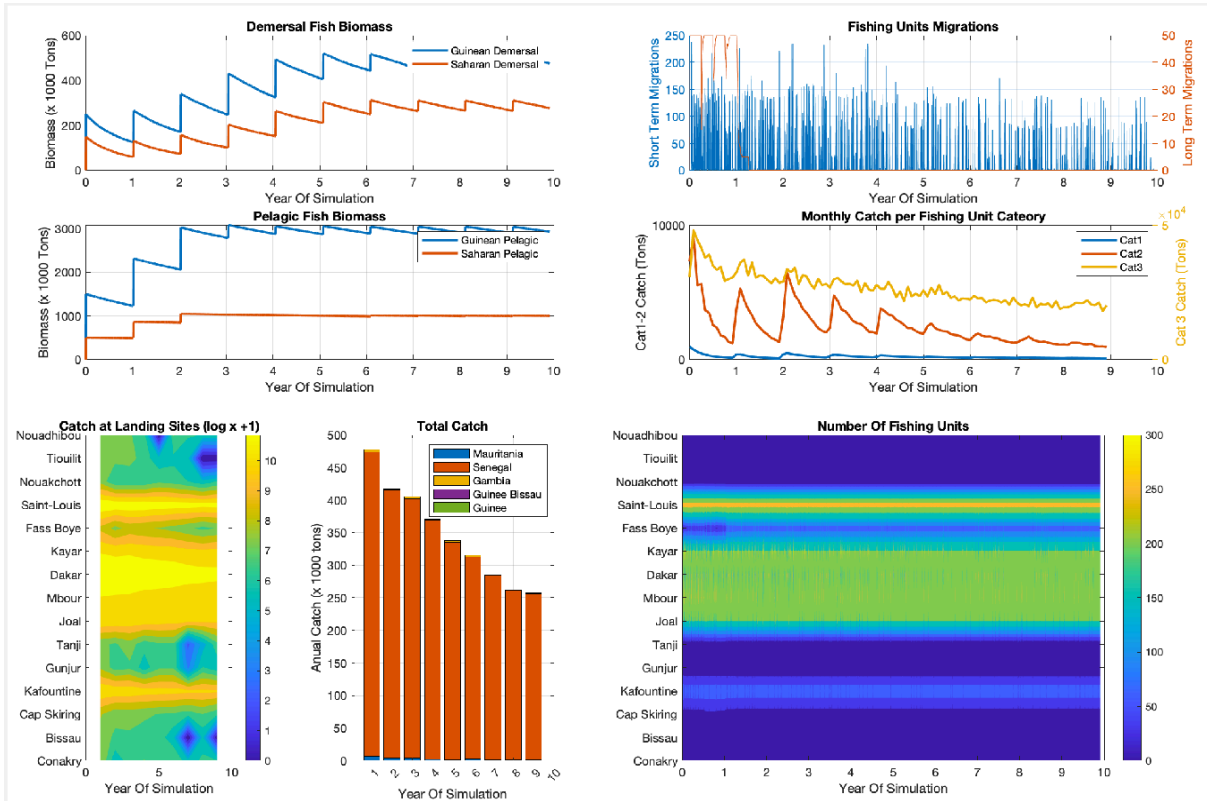

*Figure 16: Results from simulation 10, with a 100 time slower than actual fish catch, SST+3°C climate scenario.*

# 4 - Discussion

In a first section we discussed the choices made for the conceptual model. Secondly, we commented on preliminary simulations, and then we highlighted the limits of this approach before concluding with some perspectives.

## 4.1 Conceptual Model

We built a conceptual model for on the spatiotemporal dynamics of artisanal fisheries based on data collected on climate, oceanography, fish habitat, fish behavior, fisher's behavior, fishing units' characteristics, landing sites distribution and infrastructures. The conceptual model described how climate change may impact the contrasted habitats of four typical fish species defined by their thermal and bathymetric preferences, as well as their population parameters of logistic biomass growth. Three categories of fishing units were considered, defined by their fishing capacity, catch storage capacity and propensity for migration. Three types of migrations were described, corresponding to contrasting fishing distances and migration times to other landing sites. Landing sites were defined by their position along the coast and their catch processing capacity. In the event of saturation, these landing sites are no longer attractive to fishermen. Of the many processes and parameters described in the data, only a few were actually included in this conceptual model. Indeed, available information's obtained from preliminary interdisciplinary research described a very complex socio-system. In order to keep the model in the optimal level of complexity ("Medawar zone", Grimm et al., 2005), attention was paid to reduce the explicit modeling of the low levels or side processes as much as possible while keeping explicit the processes identified as most determinant for fishers' migrations. This approach may also favorize the model hypotheses understanding and thus results acceptance by actor's, and thus increase the model usefulness as a tool for managers decision help. The presentation of the conceptual model using the "ODD" protocol (Grimm et al., 2010) also contribute to make it a generic tool that could be adapted to other artisanal fisheries in the world, and let open the inclusion of specific development as more detailed economic processes or fish population dynamics.

Interestingly, it appeared that the expected results from model exploration differ among the experts that contributed in specific aspect of the conceptual model. This illustrated the non-trivial dynamics that emerge from the model, despite the very simplified representation of the processes at play. In particular, the model put in perspective three aspect of the fishery overcapacity that are the fishing action efficiency itself parametrized by the catchability parameter, the number and duration of fishing trips, and the processing capacities at the landing sites. The conceptual model we proposed also allow to explore the interplay between these aspects. Nevertheless, the effect of industrial fishery was not taken into account in this conceptual model that may include both national and long-distance vessel, in the frame of international agreements or not (IUU fishing). This could be interpreted, for example by considering part of the catch of the artisanal fisheries to be actually catch by the industrial fishery. The proportion of total catch that may be imputable to industrial fishery including IUU) may reach 50% in Senegal (Belhabib et al., 2014).

## 4.2 "Lolly" Simulator Preliminary Results

The model was implemented in Gama, an individual modeling platform developed at UM-MISCO that enabled us to integrate all the processes listed in the conceptual model. We called the obtained informatic simulation tool "lolly", which is the name of one of the fishing seasons in *Lebu*, the main historic fisher community in Dakar. We used the computing capabilities of UMMISCO in Dakar to explore this model, in particular the effect of fishing unit efficiency and climate scenarios.

The first simulations highlighted the relationship between the fishing unit's radius of action and the emergence of overexploited areas for demersal species. In these first simulations, rotating seine fishing units (category 3) were the only ones which, due to their large size, accessed all potential fishing sites, but they only targeted pelagic species. With the fishing capacities used, this resulted in the disappearance of pelagic species, but demersal species remained outside the range of the category 2 and 3 fishing units targeting them. Thus, in these simulations, demersal species kind of learned to maintain themselves at a distance from landing sites. The stabilization of this situation was due to an effect of the logistic growth model, since when the population reaches its maximum level with biomass patches beyond the reach of artisanal fishing, growth became null and there were no new patches randomly distributed in the habitat, as described by the growth algorithm used. This artifact, which nonetheless reflects a tangible phenomenon, was mitigated in subsequent simulations when we considered that rotating seine fishing units (Cat 3) could incidentally catch demersal species, from simulation 4 onwards. Category 3 fishing units therefore had access to all target species over the entire continental shelf zone. In fact, these category 3 fishing units shared this characteristic with large industrial distant-water fishing vessels, except that artisanal fishing units needed to land their catches as soon as their storage capacity was full at a landing site with the necessary infrastructures. Simulations 4, 5 and 6 show that even when fishing efficiencies were underestimated, linear relationships defining catches lead to the rapid collapse of fish stocks.

To promote the emergence of an overexploitation equilibrium as described theoretically in mathematical models (e.g. Auger et al., 2010), we applied a Holling type III functional response to define catches as a function of biomass. This reflects the fact that when fish abundance becomes low, their capture becomes increasingly difficult as they are difficult to locate. Considering a situation where the catchability, or fishing efficiency, would also be much lower (2 orders of magnitude) than initially estimated, we then obtain an equilibrium between exploitation and stock renewal with ~250,000 tons of fish caught per year in Senegal (simulation 8). Model calibration then seems satisfactory, with average artisanal fishing landings comparable to observed catches in the 2000s (Figure 17). But, based on the current structure of the artisanal fishery, *i.e.* the number of fishing units (estimated in 2014), their catch capacity, and current onshore processing capacity (estimated in 2020), we found that the fishery should collapses within a few years (simulation 8.1).

It is interesting to compare the dynamics of the model with historical data. Although no historical simulations were performed yet, the historical data inform of the fishing yields obtained for different number of artisanal fishers, which can be compared with the different level of fishing pressure tested in the model sensitivity experiment. These show that, after a sharp increase in the 1980s, since the 2000s Senegal's fish production, all fisheries combined, has plateaued at ~370,000 tons per year (~300,000 tons of pelagic and ~70,000 tons demersal species; Figure 17), although during this period the number of fishers increased from 10 000 to 22 000. These landings were largely dominated by artisanal fishing, but the total quantities of fish extracted actually include IUU catches, which may account for 50% according to Belhabib et al. (2014). Including these undeclared catches, total catches in Senegal in 2010 would thus be close to ~700,000 tons. However, in our configuration, when the catches approach one million tons per year, the fish biomass collapsed rapidly, in less than 5 years. Observations show that the fishery for the main pelagic target species, *Sardinella aurita*, has been showing signs of collapse since 2020 (Brochier et al., 2023). In practice, the Senegalese artisanal fishery still exist because it switched to other, less valuated target species. Consequently, and if the estimates of Belhabib et al. (2014) were correct, the biomass renewal capacities of the total fish populations, defined by the capacity of the environment and the reproductive rate of the populations of the different species, must then be higher than those parameterized in our configuration. These parameters could therefore be reviewed in the light of these results in future versions of the model. However, another explanation for the artisanal fishery to maintain would be the reduction of fishing effort as since 2022 an increasing number of fishers used their canoe to migrate to Europe through Canary archipelago.

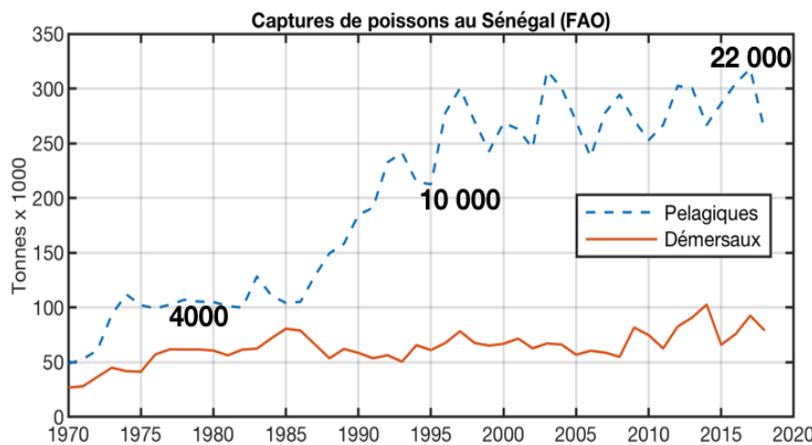

*Figure 17: Fish catch in Senegal. Red = demersal fish, blue = pelagic fish. The numbers in bold in the graph are the number of declared fishing units. Constructed from FAO data and governmental reports.*

The climate scenarios explored were defined by the range of sea surface temperature (SST) increases in the various IPCC climate projections for the region, reaching a maximum of +3°C by 2100 (see attached report "Analysis of IPCC Climate scenario for Sea Surface Temperature in North-West Africa" in Annex). To maintain the best possible representation of coastal dynamics, we have used a reanalysis of past climate (ERA5, the reference simulation, period 1979-2015) to which we added 1.5°C (scenario 1) or 3°C (scenario 2) uniformly to the SST. These scenarios have a major impact on the distribution of target species (Appendix Figure 4, Appendix Figure 5, Appendix Figure 6, Appendix Figure 7). The habitat of species with "Saharan affinities", whose habitat was restricted to the north of the system for 6 months of the year in the reference simulation was further restricted to the north for 9 months of the year in the +3°C scenario. Similarly, species with Guinean affinities, present all year round in the south of the system in the baseline simulation, were restricted to Mauritania or even Morocco for 6 months of the year. So, in general, under the effect of a uniform increase in SST, fish populations move towards the north of the system, and actually concentrated off Mauritania. Thus, it increased the accessibility of the fish biomass in the northern part of the system, where fishing infrastructures are more developed.

Despite the notable effect of climate scenarios on fish habitat distribution, we found that the effect of climate change was of 2ⁿᵈ order behind the effect of fishing pressure. Indeed, the application of the +1.5°C and + 3°C climate scenarios had little impact on the trajectories of Senegalese artisanal fishing. Also, considering a contrasted situation of a fishery in a sustainable equilibrium, the climate changes scenarios had virtually no impact on this equilibrium (simulations 8.1, 9.1, 10.1). These results, which may seem surprising, were in line with the feelings of the stakeholders as perceived during our field surveys. Climate change was rarely mentioned in focus groups, while over-exploitation was continuously denounced.

The impacts of the distribution and amplitude of processing capacity at landing sites as well as the number of fishing units were not explored in the preliminary simulations. However, it is probable that these parameters may be decisive in the mechanics of overexploitation. This was suggested by comparing the predicted distribution of *Sardinella aurita* with the distribution of fishmeal factories (Figure 18).

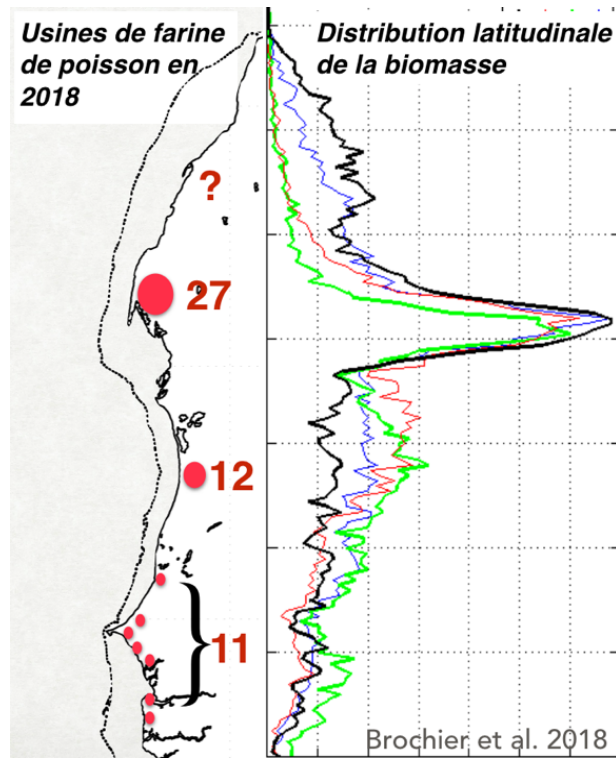

The kinematic of overexploitation as it emerged from the model was as follows: (1) fishing units rapidly deplete resources in the vicinity of Senegalese landing sites, (2) most fishing units carry out campaigns, i.e. migrations lasting several months, towards landing sites in surrounding countries (e.g. Figure 13).

*Figure 18: Position of the fishmeal factories (2018, left) and Sardinella aurita biomass seasonal distribution estimated by Evol-Deb model (Brochier et al., 2018).*

## 4.3 Limits of the Study

The present study did not consider the effects of climate change other than the increase in sea surface temperature, which was used as a proxy for fish habitat. However, other processes are likely to play a major role, such as reduced enrichment of surface waters as a result of weaker upwelling and acidification and deoxygenation of waters as a result of increased temperature and partial pressure of $CO_2$. These processes would cause a reduction in the carrying capacity of the environment due to the degradation of fish habitat, and therefore to a reduction in fish biomass and catch at equilibrium.

The present model cannot be used to explain the catch evolution of a particular species. Indeed, the four model-species used to represent fish biomass in the model stand for an agglomeration of target species. Thus, the estimation of the population parameters for these model-species were very uncertain, which may impact the predicted thresholds of fishing effort that lead to over-exploitation. Also, the landing sites processing capacities did not differentiate the processing of the different species, which may over-estimate the processing capacity for the demersal species. Actually, each species can have very different economic value and that actually correspond to different markets.

Finally, the behavior of fisher's might be more flexible in reality than it was implemented in the model. In particular, we considered a fixed number of fishers, while this number may be variable according to the profitability of the activity.

---

**Box: "Follow-up of this Work"**

***"AI to rapidly monitor the artisanal fishery"***

Informatic research in this project also bring interesting insights for an automated system to count fishing units on the landing site based on aerial images taken either from drone or satellite. It resulted in the creation of a large database of annotated images from satellite and drone, and the training of deep learning algorithms (Olaluwoye et al., accepted). Such system could help gather data at lower cost and could be used by the Senegalese institutions in charge of artisanal fishery monitoring. In particular, the counting from satellite images is promising to reconstruct past migration patterns.

***"A role-playing game to communicate the model to the stakeholders"***

An efficient way to communicate the results of a model is the construction of a role-playing game based on the model. The UMMISCO team has extensive experience of such participative method, historically for pastoralism models. This approach also allows the researcher to observe players logics. In the model, no entity could modify the socio-economic parameters, but the model user can play the role of this entity. The sustainable development managers objective would be to maximize the quantity of fish caught at equilibrium, while a short-term development manager objective would be to maximize the catch for the coming year for example. These objectives would dictate management measures such as landing sites infrastructure, fishing licenses and international fishing agreements. That is, a role-playing game may help actors to understand the long-term and short-term consequences of such management measures. In the continuation of this work a fishery role playing game illustrated by a Senegalese artist is under development at UMMISCO, directed toward to fishers, fish-traders and managers at the national and international levels.

---

## 4.4 Conclusions and Perspectives

This project intertwisted modelling work and field studies, crossing lines between sociology, ecology, oceanography and economy, allowing us to propose a conceptual model using mathematic and informatic frameworks. It resulted from an effective, constructive international research collaboration that contributed to form several students to the interdisciplinary approach of artisanal fisheries in Senegal. The research activities generated by the development of this theme at UMMISCO, spurred on by Habitable funding at UCAD and IRD, and by extension to ISRA partners (CRODT), also led to the involvement of researchers and teacher-researchers in different disciplines. Interviews and focus groups provided many information on Senegalese fisher's migrations, that we partly used to construct the current model but that may still be used in future research.

Preliminary results suggest that overexploitation is the main threat to artisanal fisheries. Although preliminary, these results are robust, mainly illustrating the fact that artisanal fishers, thanks to their migration capabilities, can adapt to the change in target fish habitat distribution predicted by the future climate scenarios tested. Although the estimate of fishing efficiency is a difficult task, based on realistic data of the artisanal fishing fleet and fish populations biomass, the model suggests that, with or without climate change, in its current configuration the Senegalese artisanal fishery is likely to collapse rapidly. A complete, rapid collapse was predicted in the model because the number of fishers was constant, but is unlikely to occur in real world because the fishing effort reduce as fishers leave the fishery. Indeed, fishers leave the fishery when the fish abundance reduce down to a threshold that cannot allow them to increase in fishing effort in order to maintain the catch minimum level. This is in line with observations since 2021 for one of the main specific small pelagic fish species fisheries (Brochier et al., 2023). Thus, fishers' migrations may continue to increase in the short term if no management measures are taken, regardless of climate change. This is a strong message to managers, because climate was often pointed as responsible for the current crisis, while bad management might be to blame. Reducing fishing effort and adapting to the effects of climate change on fish distribution could be achieved through the management of the fishery at the fish migration scale in the area, *i.e.* including stakeholders from Morocco to Guinea. Such management should include a monitoring of fish industrial and artisanal fishing effort, and processing capacities at landing sites, including fishmeal factories and exportation places. Further model explorations may allow to better understand the effect of coastal infrastructures distribution along the coast. In any case, the sub-regional management should ensure that no competition occur between different fisheries, as competition mechanically lead to overexploitation (Hieu et al., 2018).

Here the impact of climate change scenarios was considered through the change in fish species distribution. The results suggested that a good management of the fisheries at the regional scale could make artisanal fishery resilient to climate change. This result contrast with previous qualitative studies based on macroecological theory that predicted a drop in catch in West Africa due to climate change based on the assumption of a drop of primary productivity (Belhabib et al., 2016; Cheung et al., 2008). However, such drop in primary production was not yet observed and remain uncertain in North-West Africa climate scenarios. However, the effect of climate change on fish species carrying capacity through changes in primary productivity may be possible in a near future, as coupled hydrodynamical and biogeochemical simulation under climate change forcing (e.g. ROMS-PISCES) becomes available.

Artisanal fisheries in Senegal used to be a refuge sector for climate refugee, but its collapse due to overexploitation forces the reconversion of many fishers. In the current context, many people in North West Africa are candidate for emigration to Europe, and experimented fishers can valorize their knowledge of the sea for the transport of clandestine migrants to the Canary Archipelago (Spain). The specialization of Senegalese fishers in such activity may increase the attractivity of Senegal as a departure point for emigration. Reducing this phenomenon would be possible through sub-regional management to limit fishing capacity and processing capacity, and or international agreements to reduce the global consumption of fish extracted from the southern canary ecosystem, in particular fishmeal, either extracted by industrial or artisanal fisheries.

# 5 - References

Abernethy, K. E., Allison, E. H., Molloy, P. P., & Côté, I. M. (2007). Why do fishers fish where they fish? Using the ideal free distribution to understand the behaviour of artisanal reef fishers. *Canadian Journal of Fisheries and Aquatic Sciences*, *64*(11), 1595–1604. https://doi.org/10.1139/f07-125

Agnew, D. J., Pearce, J., Pramod, G., Peatman, T., Watson, R., Beddington, J. R., & Pitcher, T. J. (2009). Estimating the Worldwide Extent of Illegal Fishing. *PLoS ONE*, *4*(2), e4570. https://doi.org/10.1371/journal.pone.0004570

Atlas des pêches et pêcheurs artisans d'Afrique de l'Ouest: États membres de l'UEMOA : Bénin, Burkina Faso, Côte d'Ivoire, Guinée-Bissau, Mali, Niger, Sénégal, Togo. (2022). In P. Chavance & P. Morand (Eds.), *Atlas des pêches et pêcheurs artisans d'Afrique de l'Ouest: États* membres de l'UEMOA : Bénin, Burkina Faso, Côte d'Ivoire, Guinée-Bissau, Mali, Niger, Sénégal, Togo. IRD Éditions. https://doi.org/10.4000/books.irdeditions.43778

Auger, P., Mchich, R., Raïssi, N., & Kooi, B. W. (2010). Effects of market price on the dynamics of a spatial fishery model: Over-exploited fishery/traditional fishery. *Ecological Complexity*, *7*(1), 13–20.

Ba, A., Chaboud, C., Brehmer, P., & Schmidt, J. O. (2022). Are subsidies still relevant in West African artisanal small pelagic fishery? Insights from long run bioeconomic scenarios. *Marine Policy*, *146*, 105294. https://doi.org/10.1016/j.marpol.2022.105294

Ba, A., Schmidt, J., Dème, M., Lancker, K., Chaboud, C., Cury, P., Thiao, D., Diouf, M., & Brehmer, P. (2017). Profitability and economic drivers of small pelagic fisheries in West Africa: A twenty year perspective. *Marine Policy*, *76*, 152–158. https://doi.org/10.1016/j.marpol.2016.11.008

Ba, K., Thiaw, M., Fall, M., Thiam, N., Meissa, B., Jouffre, D., Thiaw, O. T., & Gascuel, D. (2018). Long-term fishing impact on the Senegalese coastal demersal resources: Diagnosing from stock assessment models. *Aquatic Living Resources*, *31*, 8. https://doi.org/10.1051/alr/2017046

Barry-Gérard, M. (n.d.). *Migrations des poissons le long du littoral sénégalais*. 20.

Bars, F. L., & Jaulin, L. (2013). *Robotic Sailing 2013: Proceedings of the 6th International Robotic Sailing Conference*. Springer Science & Business Media.

Belhabib, D., Koutob, V., Sall, A., Lam, V. W. Y., & Pauly, D. (2014). Fisheries catch misreporting and its implications: The case of Senegal. *Fisheries Research*, *151*, 1–11. https://doi.org/10.1016/j.fishres.2013.12.006

Binet, T., Failler, P., & Thorpe, A. (2012). Migration of Senegalese fishers: A case for regional approach to management. *Maritime Studies*, *11*(1), 1–14. https://doi.org/10.1186/2212-9790-11-1

Boely, T., Chabanne, J., Fréon, P., & Stéquert, B. (1982). Cycle sexuel et migrations de Sardinella aurita sur le plateau continental ouest-africain, des Iles Bissagos à la Mauritanie. *Rapport Proces Verbal de La Réunion Du Conseil International Pour l'Exploration de La Mer*, *180*, 350–355.

Boely, T., Chabanne, T., & Fréon, P. (1978). *SCHEMAS MIGRATOIRES, AIRES DE CONCENTRATIONS ET PERIODES DE REPRODUCTION DES PRINCIPALES ESPECES DE POISSONS PELAGIQUES COTIERS DANS LA ZONE SENEGALO-MAURITANIENNE* [Rapport du Groupe de Travail Ad Hoc sur les Poissons Pélagiques Côtiers Ouest-Africains de la Mauritanie au Libéria (26° N à 5° N)]. CRODT.

Boelyl, T., Chabanne, J., Fréon, P., & Stéquert, B. (1982). *Cycle sexuel et migrations de Sardinella aurita sur le plateau continental ouest-africain, des Iles Bissagos à la Mauritanie*. http://horizon.documentation.ird.fr/exl-doc/pleins_textes/pleins_textes_5/b_fdi_14-15/17549.pdf

Bourdieu, P. (1979). Les trois états du capital culturel. *Actes de la Recherche en Sciences Sociales*, *30*(1), 3–6. https://doi.org/10.3406/arss.1979.2654

Branch, T. A., Hilborn, R., Haynie, A. C., Fay, G., Flynn, L., Griffiths, J., Marshall, K. N., Randall, J. K., Scheuerell, J. M., Ward, E. J., & Young, M. (2006). Fleet dynamics and fishermen behavior: Lessons for fisheries managers. *Canadian Journal of Fisheries and Aquatic Sciences*, *63*(7), 1647–1668. https://doi.org/10.1139/f06-072

Brochier, T., Auger, P. A., Pecquerie, L., Machu, E., Capet, X., Thiaw, M., Mbaye, B. C., Braham, C. B., Ettahiri, O., Charouki, N., Sene, O. N., Werner, F., & Brehmer, P. (2018). Complex small pelagic fish population patterns arising from individual behavioral responses to their environment. *Progress in Oceanography*, *164*, 12–27. https://doi.org/10.1016/j.pocean.2018.03.011

Brochier, T., Auger, P., Thiam, N., Sow, M., Diouf, S., Sloterdijk, H., & Brehmer, P. (2015). Implementation of artificial habitats: Inside or outside the marine protected areas? Insights from a mathematical approach. *Ecological Modelling*, *297*, 98–106. https://doi.org/10.1016/j.ecolmodel.2014.10.034

Brochier, T., Auger, P., Thiao, D., Bah, A., Ly, S., Nguyen-Huu, T., & Brehmer, P. (2018). Can overexploited fisheries recover by self-organization ? Reallocation of fishing effort as an emergent form of governance. *Marine Policy*, *95*, 46–56. https://doi.org/10.1016/j.marpol.2018.06.009

Brochier, T., & Bah, A. (2021). *FisherMob: Un modèle bioéconomique de la mobilité des pêcheurs*. https://hal.archives-ouvertes.fr/hal-03183811

Brochier, T., Brehmer, P., Mbaye, A., Diop, M., Watanuki, N., Terashima, H., Kaplan, D., & Auger, P. (2021). Successful artificial reefs depend on getting the context right due to complex socio-bio-economic interactions. *Scientific Reports*, *11*(1), 16698. https://doi.org/10.1038/s41598-021-95454-0

Brochier, T., Colas, F., Lett, C., Echevin, V., Cubillos, L. A., Tam, J., Chlaida, M., Mullon, C., & Fréon, P. (2009). Small pelagic fish reproductive strategies in upwelling systems: A natal homing evolutionary model to study environmental constraints. *Progress in Oceanography*, *83*(1–4), 261–269.

Brochier, T., Echevin, V., Tam, J., Chaigneau, A., Goubanova, K., & Bertrand, A. (2013). Climate change scenarios experiments predict a future reduction in small pelagic fish recruitment in the Humboldt Current system. *Global Change Biology*, *19*(6), 1841–1853. https://doi.org/10.1111/gcb.12184

Brochier, T., Ecoutin, J. M., de Morais, L. T., Kaplan, D. M., & Lae, R. (2013). A multi-agent ecosystem model for studying changes in a tropical estuarine fish assemblage within a marine protected area. *Aquatic Living Resources*, *26*(02), 147–158. https://doi.org/10.1051/alr/2012028

Brochier, T., Mason, E., Moyano, M., Berraho, A., Colas, F., Sangrà, P., Hernández-León, S., Ettahiri, O., & Lett, C. (2011). Ichthyoplankton transport from the African coast to the Canary Islands. *Journal of Marine Systems*, *87*(2), 109–122. https://doi.org/10.1016/j.jmarsys.2011.02.025

Brochier, T., Ramzi, A., Lett, C., Machu, E., Berraho, A., Fréon, P., & Hernández-León, S. (2008). Modelling sardine and anchovy ichthyoplankton transport in the Canary Current System. *Journal of Plankton Research*, *30*(10), 1133–1146.

Castillo-Jordán, C., Cubillos, L. A., & Paramo, J. (2007). The spawning spatial structure of two co-occurring small pelagic fish off central southern Chile in 2005. *Aquatic Living Resources*, *20*(1), 77–84. https://doi.org/10.1051/alr:2007018

Charles, A. T., Mazany, R. L., & Cross, M. L. (1999). The Economics of Illegal Fishing: A Behavioral Model. *Marine Resource Economics*, *14*(2), 95–110.

Chauveau, J.-P. (1989). Histoire de la pêche industrielle au Sénégal et politiques d'industrialisation: 1ère partie : cinq siècles de pêche européenne (du 15ème siècle au milieu des années 1950). *Cahiers des Sciences Humaines*, *25*(1–2), 237–258.

Chauveau, J.-P., Jul-Larsen, E., & Chaboud, Christian. (2000). *Les pêches piroguières en Afrique de l'Ouest—Pouvoirs, mobilités, marchés* (IRD Éditions/Karthala). https://www.editions.ird.fr/produit/157/9782845860711/les-peches-piroguieres-en-afrique-de-l-ouest

Chérel, G., Cottineau, C., & Reuillon, R. (2015). Beyond Corroboration: Strengthening Model Validation by Looking for Unexpected Patterns. *PLOS ONE*, *10*(9), e0138212. https://doi.org/10.1371/journal.pone.0138212

Cormier Salem, M.-C., & Dahou, T. (Eds.). (2009). Gouverner la mer: Etats, pirates, sociétés. *Politique Africaine*, *116*. http://www.documentation.ird.fr/hor/fdi:010048588

Corten, A., Braham, C.-B., & Sadegh, A. S. (2017). The development of a fishmeal industry in Mauritania and its impact on the regional stocks of sardinella and other small pelagics in Northwest Africa. *Fisheries Research*, *186, Part 1*, 328–336. https://doi.org/10.1016/j.fishres.2016.10.009

Corten, A., & Sadegh, A. S. (2014). *The development of a fish meal industry in Mauritania and its impact on the regional stocks of sardinella and other small pelagics* (p. 10) [FAO Working Group on small Pelagic Fish in West Africa].

Cury, P., & Roy, C. (1988). Migration saisonnière du thiof (Epinephelus aeneus) au Sénégal: Influence des upwellings. *Oceanologica Acta*, *11*(1), 25–36.

Dahou, T. (2009). La politique des espaces maritimes en Afrique: Louvoyer entre local et global. *Politique Africaine*, *116*, 5–22.

Dahou, T. (2010). Gérer l'espace sans gouverner les hommes: Le dilemme des Aires marines protégées (Saloum, Sénégal). *Anthropologie et Sociétés*, *34*(1), 75–93. https://doi.org/10.7202/044197ar

Deme, E. H. B., Failler, P., & Deme, M. (2021). Migration of Senegalese artisanal fishermen in West Africa: Patterns and impacts. *African Identities*, *19*(3), 253–265. https://doi.org/10.1080/14725843.2021.1937049

Deme, E. hadj B., Deme, M., & Failler, P. (2022). Small pelagic fish in Senegal: A multi-usage resource. *Marine Policy*, *141*, 105083. https://doi.org/10.1016/j.marpol.2022.105083

Deme, E. hadj B., Failler, P., & Touron-Gardic, G. (2021). La gouvernance des aires marines protégées au Sénégal: Difficulté de la gestion participative et immobilisme des comités de gestion. *VertigO - la revue électronique en sciences de l'environnement*, *Volume 21 numéro 1*, Article Volume 21 numéro 1. https://doi.org/10.4000/vertigo.30880

Deme, E. J. B., Brehmer, P. P., Bâ, A., & Failler, P. (2021). Résilience et réactivité des pêcheurs artisans sénégalais: La crise écologique comme moteur d'innovations. *Mondes En Développement*, *49*(193), 109–127. https://doi.org/10.3917/med.193.0113

Diankha, O., Ba, A., Brehmer, P., Brochier, T., Sow, B. A., Thiaw, M., Gaye, A. T., Ngom, F., & Demarcq, H. (2018). Contrasted optimal environmental windows for both sardinella species in Senegalese waters. *Fisheries Oceanography*, *27*(4), 351–365. https://doi.org/10.1111/fog.12257

Diankha, O., Thiaw, M., Sow, B. A., Brochier, T., Gaye, A. T., & Brehmer, P. (2015). Round sardinella (Sardinella aurita) and anchovy (Engraulis encrasicolus) abundance as related to temperature in the senegalese waters. *Thalassas*, *31*(2), 9–17.

Diallo, M. (2021). *Analyse des pratiques de mobilité chez les pêcheurs artisanaux du Sénégal dans une perspective de modélisation: Cas des pêcheurs de Joal, Mbour, Saint-Louis et Soumbédioune* (p. 120) [Mémoire de fin d'étude]. Institut Universitaire de pêche et d'aquaculture (IUPA).

Essington, T. E., Moriarty, P. E., Froehlich, H. E., Hodgson, E. E., Koehn, L. E., Oken, K. L., Siple, M. C., & Stawitz, C. C. (2015). Fishing amplifies forage fish population collapses. *Proceedings of the National Academy of Sciences*, *112*(21), 6648–6652. https://doi.org/10.1073/pnas.1422020112

Fadiaba, O., Ndao, S., Seck, S., Signate, M., & Guitton, J. (2022). Sénégal. In P. Chavance & P. Morand (Eds.), *Atlas des pêches et pêcheurs artisans d'Afrique de l'Ouest: États membres de l'UEMOA : Bénin, Burkina Faso, Côte d'Ivoire, Guinée-Bissau, Mali, Niger, Sénégal, Togo* (pp. 131–136). IRD Éditions. https://doi.org/10.4000/books.irdeditions.44091

Failler, P. (2014). Climate variability and food security in Africa: The case of small pelagic fish in West Africa. *Journal of Fisheries & Livestock Production*, *2014*. http://www.esciencecentral.org/journals/climate-variability-and-food-security-in-africa-the-case-of-small-pelagic-fish-in-west-africa-2332-2608.1000122.php?aid=32593

Failler, P., & Binet, T. (2010). Sénégal. Les pêcheurs migrants: Réfugiés climatiques et écologiques. *Hommes et migrations. Revue française de référence sur les dynamiques migratoires*, *1284*, 98–111. https://doi.org/10.4000/hommesmigrations.1250

Failler, P., Binet, T., Dème, E. hadj B., & Deme, M. (2020). Importance de la pêche migrante ouest- africaine au début du XXIe siècle. *Revue Africaine des Migrations Internationales*, *1*(1), Article 1. https://revues.imist.ma/index.php/RAMI/article/view/21436

Fall, T. (2020). *Les déterminants de la mobilité des pêcheurs artisanaux: Cas des pêcheurs de Saint- Louis, Yarakh et Joal* (p. 137) [Mémoire de fin d'étude]. Institut Universitaire de pêche et d'aquaculture (IUPA).

FAO. (2007). *The state of world fisheries and aquaculture 2006, Fisheries and Aquaculture Department, Food and Agriculture Organization of the United Nations*.

Garcia, S., Atlantic, P. for the D. of F. in the E. C., Food, Atlantic, A. O. of the U. N. F. C. for the E. C., Food, Nations, A. O. of the U., & Programme, U. N. D. (1982). *Distribution, migration and spawning of the main fish resources in the northern CECAF area*. Food and Agriculture Organization of the United Nations.

Garcia, S. M. (2010). Governance, science and society: The ecosystem approach to fisheries. *Handbook of Marine Fisheries Conservation and Management*, 87–98.

Garcia, S. M., & Cochrane, K. L. (2005). Ecosystem approach to fisheries: A review of implementation guidelines. *ICES Journal of Marine Science: Journal Du Conseil*, *62*(3), 311–318.

Gillis, D. M. (2003). Ideal free distributions in fleet dynamics: A behavioral perspective on vessel movement in fisheries analysis. *Canadian Journal of Zoology*, *81*(2), 177–187. https://doi.org/10.1139/z02-240

Gillis, D. M., van der Lee, A., & Walters, C. (2012). Advancing the application of the ideal free distribution to spatial models of fishing effort: The isodar approach. *Canadian Journal of Fisheries and Aquatic Sciences*, *69*(10), 1610–1620. https://doi.org/10.1139/f2012-091

Gorgues, T., Ménage, O., Terre, T., & Gaillard, F. (2011). An innovative approach of the surface layer sampling. *Journal Des Sciences Halieutique et Aquatique*, *4*, 105–109.

Goudiaby, M., Seye, M., & Mbaye, M. D. L. (n.d.). *Résultats énéraux des pêches maritimes 2018—Bureau des statistique* (p. 98). MINISTERE DES PECHES ET DE L'ECONOMIE MARITIME - DIRECTION DES PECHES MARITIMES.

Grimm, V., Berger, U., DeAngelis, D. L., Polhill, J. G., Giske, J., & Railsback, S. F. (2010). The ODD protocol: A review and first update. *Ecological Modelling*, *221*(23), 2760–2768. https://doi.org/10.1016/j.ecolmodel.2010.08.019

Grimm, V., Revilla, E., Berger, U., Jeltsch, F., Mooij, W. M., Railsback, S. F., Thulke, H. H., Weiner, J., Wiegand, T., & DeAngelis, D. L. (2005). Pattern-oriented modeling of agent-based complex systems: Lessons from ecology. *Science*, *310*(5750), 987–991.

Guénette, S., Meissa, B., & Gascuel, D. (2014). Assessing the contribution of marine protected areas to the trophic functioning of ecosystems: A model for the Banc d'Arguin and the Mauritanian shelf. *PloS One*, *9*(4), e94742.

Holland, D. S. (2000). A bioeconomic model of marine sanctuaries on Georges Bank. *Canadian Journal of Fisheries and Aquatic Sciences*, *57*(6), 1307–1319. https://doi.org/10.1139/f00-061

Holland, D. S., & Sutinen, J. G. (1999). An empirical model of fleet dynamics in New England trawl fisheries. *Canadian Journal of Fisheries and Aquatic Sciences*, *56*(2), 253–264. https://doi.org/10.1139/f98-169

Jaulin, L., & Bars, F. L. (2013). An Interval Approach for Stability Analysis: Application to Sailboat Robotics. *IEEE Transactions on Robotics*, *29*(1), 282–287. https://doi.org/10.1109/TRO.2012.2217794

Laurans, M., Gascuel, D., Chassot, E., & Thiam, D. (2004). Changes in the trophic structure of fish demersal communities in West Africa in the three last decades. *Aquatic Living Resources*, *17*(2), 163–174.

Machu, É., Brochier, T., Capet, X., Ndoya, S., Sidiki Ba, I., & Descroix, L. (2023). *Chapitre 2. Pollutions dans un monde liquide*. IRD Éditions. https://books.openedition.org/irdeditions/ https://books.openedition.org/irdeditions/44554

Machu, E., Capet, X., Estrade, P. A., Ndoye, S., Brajard, J., Baurand, F., Auger, P.-A., Lazar, A., & Brehmer, P. (2019). First Evidence of Anoxia and Nitrogen Loss in the Southern Canary Upwelling System. *Geophysical Research Letters*, *46*(5), 2619–2627. https://doi.org/10.1029/2018GL079622

Mazé, C., Dahou, T., Ragueneau, O., Danto, A., Mariat-Roy, E., Raimonet, M., & Weisbein, J. (2017). Knowledge and power in integrated coastal management. For a political anthropology of the sea combined with the sciences of the marine environment. *Comptes Rendus Geoscience*, *349*(6), 359–368. https://doi.org/10.1016/j.crte.2017.09.008

MAZEAUD, A., Barbier, R., Blondiaux, L., Chateauraynaud, F., Fourniau, J.-M., Lefebvre, R., Neveu, C., & Salles, D. (2013). *Dictionnaire critique et interdisciplinaire de la participation*. http://www.dicopart.fr/fr/dico/citoyenelutechnicien

Mbaye, A., Cormier-Salem, M.-C., Schmidt, J. O., & Brehmer, P. (2021). Senegalese artisanal fishers in the apprehension of changes of the marine environment: A universal knowledge? *Social Sciences & Humanities Open*, *4*(1), 100231. https://doi.org/10.1016/j.ssaho.2021.100231

Mbaye, A., Thiam, N., & Fall, M. (2018). Les zones de pêche protégées au Sénégal: Entre terroir du pêcheur et parcours du poisson. Quelle(s) échelle(s) de gestion ? *Développement durable et territoires. Économie, géographie, politique, droit, sociologie, Vol. 9, n°1*. https://doi.org/10.4000/developpementdurable.11999

Mbaye, B. C., Brochier, T., Echevin, V., Lazar, A., Lévy, M., Mason, E., Gaye, A. T., & Machu, E. (2015). Do Sardinella aurita spawning seasons match local retention patterns in the Senegalese–Mauritanian upwelling region? *Fisheries Oceanography*, *24*(1), 69–89. https://doi.org/10.1111/fog.12094

Mchich, R., Brochier, T., Auger, P., & Brehmer, P. (2016). Interactions Between the Cross-Shore Structure of Small Pelagic Fish Population, Offshore Industrial Fisheries and Near Shore Artisanal Fisheries: A Mathematical Approach. *Acta Biotheoretica*, *64*(4), 479–493. https://doi.org/10.1007/s10441-016-9299-7

Mcowen, C. J., Cheung, W. W. L., Rykaczewski, R. R., Watson, R. A., & Wood, L. J. (2015). Is fisheries production within Large Marine Ecosystems determined by bottom-up or top-down forcing? *Fish and Fisheries*, *16*(4), 623–632. https://doi.org/10.1111/faf.12082

Mellado, T., Brochier, T., Timor, J., & Vitancurt, J. (2014). Use of local knowledge in marine protected area management. *Marine Policy*, *44*, 390–396. https://doi.org/10.1016/j.marpol.2013.10.004

Menerault, P. (1999). Ces réseaux qui nous gouvernent ? (Sous la direction de Michel Marié et Michel Gariépy). *FLUX Cahiers scientifiques internationaux Réseaux et Territoires*, *15*(36), 69–73.

Ndiaye, M. (2023). *Cartographie participative de la distribution spatiale et saisonnière des principales espèces cibles de la pêche artisanale sénégalaise* (p. 76) [Mémoire de fin d'étude]. Institut Universitaire de pêche et d'aquaculture (IUPA).

Ndoye, S., Capet, X., Estrade, P., Sow, B., Machu, E., Brochier, T., Döring, J., & Brehmer, P. (2017). Dynamics of a "low-enrichment high-retention" upwelling center over the southern Senegal shelf. *Geophysical Research Letters*, *44*(10), 5034–5043.

Nguyen, T. H., Brochier, T., Auger, P., Trinh, V. D., & Brehmer, P. (2018). Competition or cooperation in transboundary fish stocks management: Insight from a dynamical model. *Journal of Theoretical Biology*, *447*, 1–11. https://doi.org/10.1016/j.jtbi.2018.03.017

Olaluwoye, O. (2024). *Enumeration of Pirogues Using Google Earth Images* (p. 47) [Internship report for the degree of Master of Mathematical Science]. ESP-UCAD / AIMS.

Pala, C. (2013). Detective work uncovers under-reported overfishing. *Nature*, *496*(7443), 18–18.

Pauly, D., Christensen, V., Dalsgaard, J., Froese, R., & Torres Jr, F. (1998). Fishing down marine food webs. *Science*, *279*(5352), 860.

Song, A. M., Scholtens, J., Stephen, J., Bavinck, M., & Chuenpagdee, R. (2017). Transboundary research in fisheries. *Marine Policy*, *76*, 8–18. https://doi.org/10.1016/j.marpol.2016.10.023

Sultan, B., Roudier, P., Quirion, P., Alhassane, A., Muller, B., Dingkuhn, M., Ciais, P., Guimberteau, M., Traore, S., & Baron, C. (2013). Assessing climate change impacts on sorghum and millet yields in the Sudanian and Sahelian savannas of West Africa. *Environmental Research Letters*, *8*(1), 014040. https://doi.org/10.1088/1748-9326/8/1/014040

Sumaila, U. R., Alder, J., & Keith, H. (2006). Global scope and economics of illegal fishing. *Marine Policy*, *30*(6), 696–703. https://doi.org/10.1016/j.marpol.2005.11.001

Sylla, A., Mignot, J., Capet, X., & Gaye, A. T. (2019). Weakening of the Senegalo–Mauritanian upwelling system under climate change. *Climate Dynamics*, *53*(7), 4447–4473. https://doi.org/10.1007/s00382-019-04797-y

Tacon, A. G. J. (2004). Use of fish meal and fish oil in aquaculture: A global perspective. *Aquatic Resources, Culture and Development*, *1*(1), 3–14.

Thiao, D., & Bunting, S. W. (2022). *Socio-economic and biological impacts of the fish-based feed industry for sub-Saharan Africa* (1236; Fisheries and Aquaculture Circular). FAO, Worldfish and University of Greenwich, Natural Resources Institute. https://doi.org/10.4060/cb7990en

Thiao, D., Chaboud, C., Samba, A., Laloë, F., & Cury, P. (2012). Economic dimension of the collapse of the false cod Epinephelus aeneus in a context of ineffective management of the small-scale fisheries in Senegal. *African Journal of Marine Science*, *34*(3), 305–311. https://doi.org/10.2989/1814232X.2012.725278

Thiaw, M., Auger, P.-A., Ngom, F., Brochier, T., Faye, S., Diankha, O., & Brehmer, P. (2017). Effect of environmental conditions on the seasonal and inter-annual variability of small pelagic fish abundance off North-West Africa: The case of both Senegalese sardinella. *Fisheries Oceanography*, *26*(5), 583–601.

Touron-Gardic, G., Hermansen, Ø., Failler, P., Dia, A. D., Tarbia, M. O. L., Brahim, K., Thorpe, A., Bara Dème, E. H., Beibou, E., Kane, E. A., Bouzouma, M., & Arias-Hansen, J. (2022). The small pelagics value chain in Mauritania – Recent changes and food security impacts. *Marine Policy*, *143*, 105190. https://doi.org/10.1016/j.marpol.2022.105190

Travers, M., Shin, Y.-J., Jennings, S., & Cury, P. (2007). Towards end-to-end models for investigating the effects of climate and fishing in marine ecosystems. *Progress In Oceanography*, *75*(4), 751–770. https://doi.org/10.1016/j.pocean.2007.08.001

Weigel, J.-Y., Montbrison, D. de, Giron, Y., Fossi, A., & Diop, H. (2014). *Etat de l'art sur la co-gestion des pêches: Synthèse mondiale, expérience, enseignements*. 30 p. multigr. https://halshs.archives-ouvertes.fr/halshs-01425953

Zickgraf, C. (2018). "The Fish Migrate And So Must We": The Relationship Between International And Internal Environmental Mobility In a Senegalese Fishing Community. *Medzinarodne Vztahy (Journal of International Relations)*, *16*(1), 5–21.

# 6 – Appendix

## Appendix A: Configuration files and parameters

Configuration files are here: https://dropsu.sorbonne-universite.fr/s/ij7FRyork6sBqJk

## Appendix B: Saltelli sensitivity analysis results

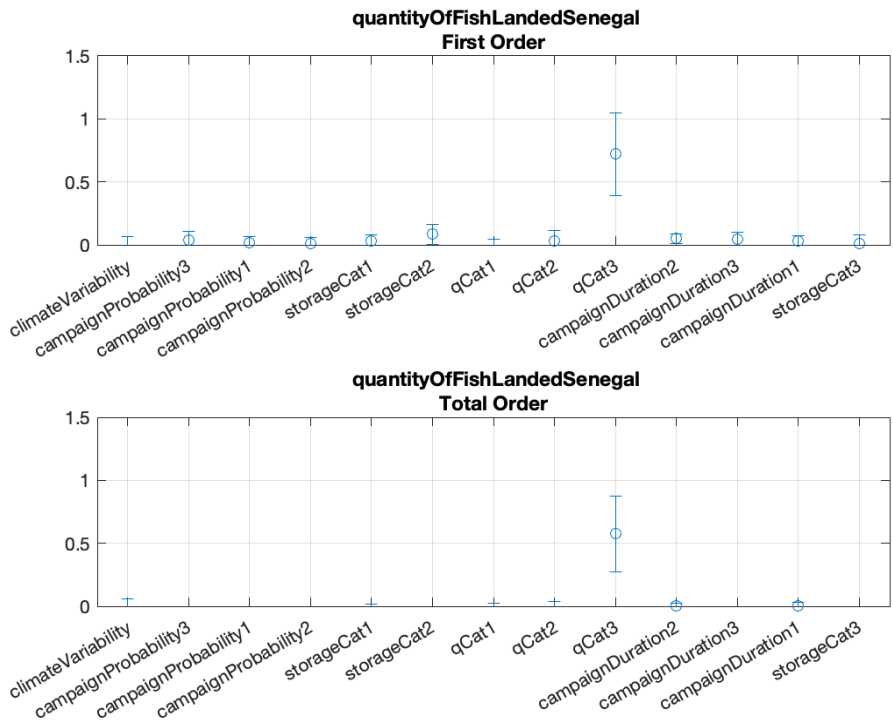

*Appendix Figure 1: Saltelli analysis of the effect of model parameters on cumulative catch in Senegal*

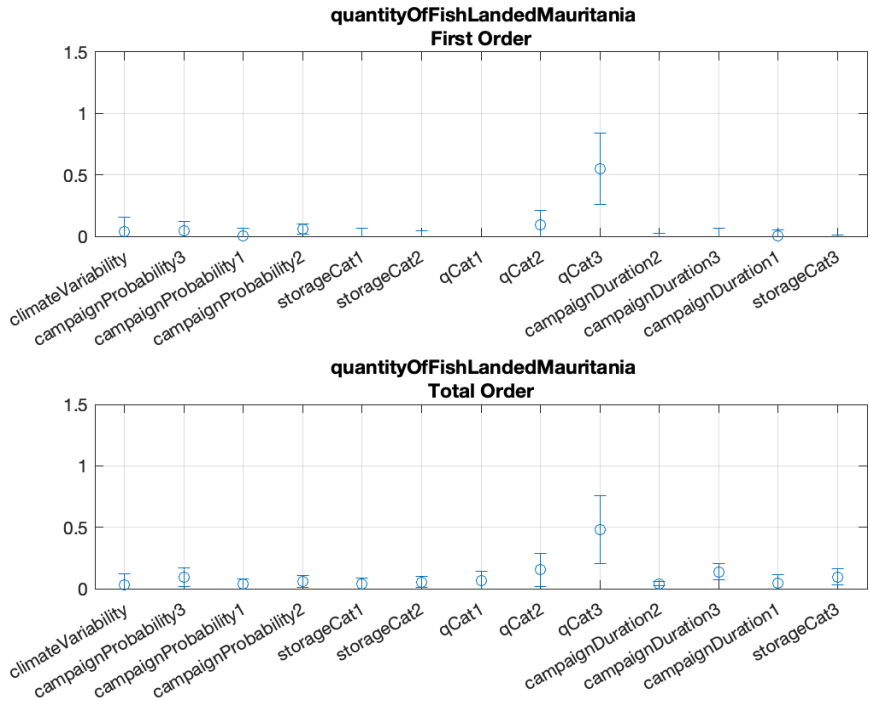

*Appendix Figure 2: Saltelli analysis of the effect of model parameters on cumulative catch in Mauritania*

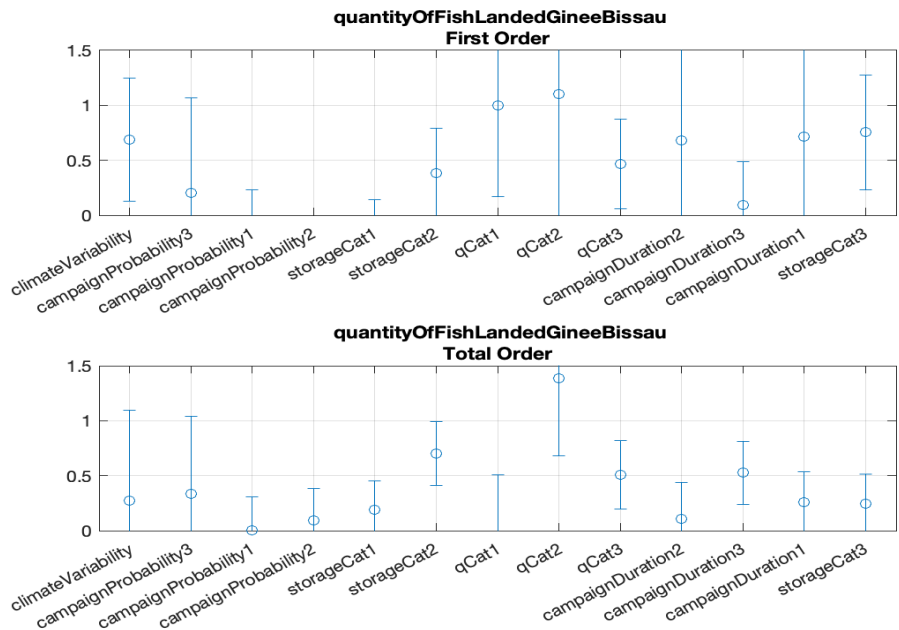

*Appendix Figure 3: Saltelli analysis of the effect of model parameters on the catch in Guinee Bissau*

## Appendix C: Fish Habitat Distribution in climate Scenarios

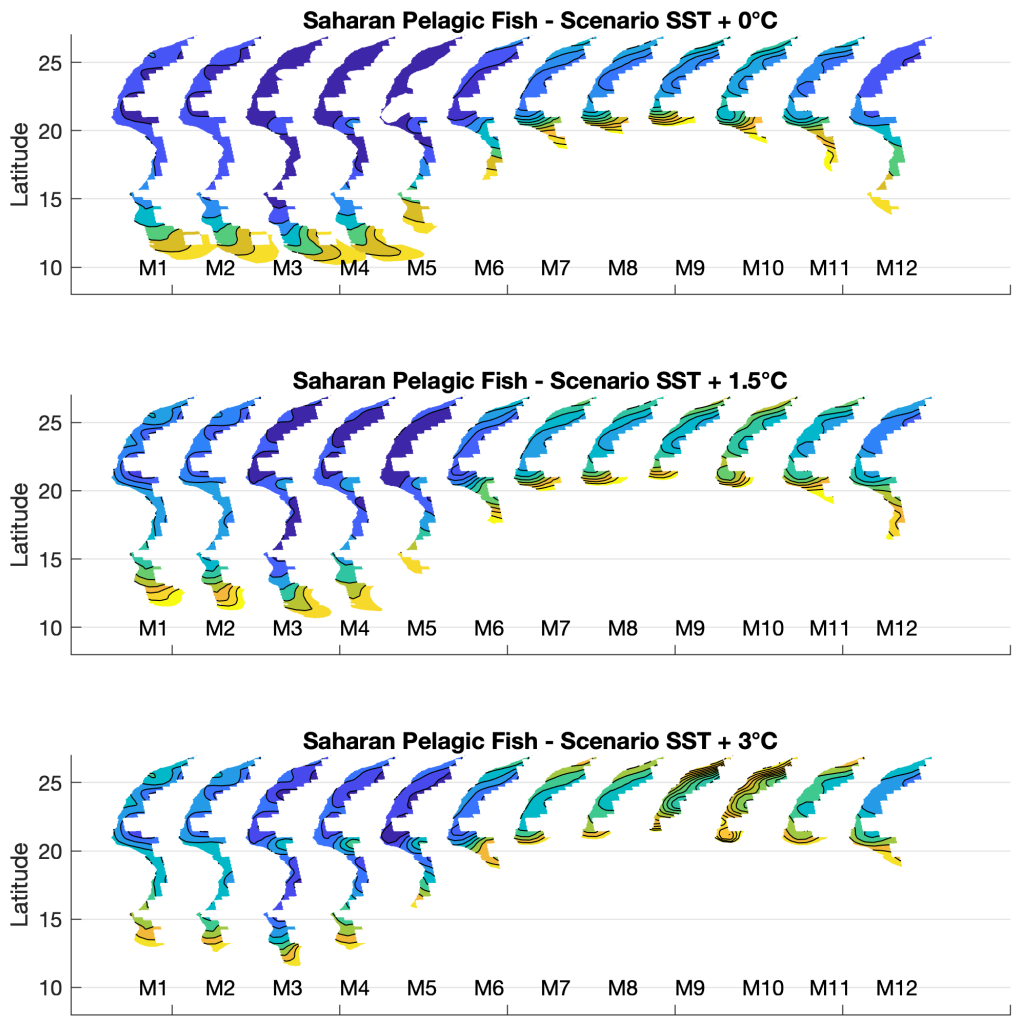

*Appendix Figure 4: Monthly distribution of Saharan pelagic fish model under present climate (top), mild climate change (+1,5C, middle) and extreme climate change (+3°C, bottom). For each month (M1-12), the represented part of the model grid correspond to the fish habitat, and color correspond to the SST.*

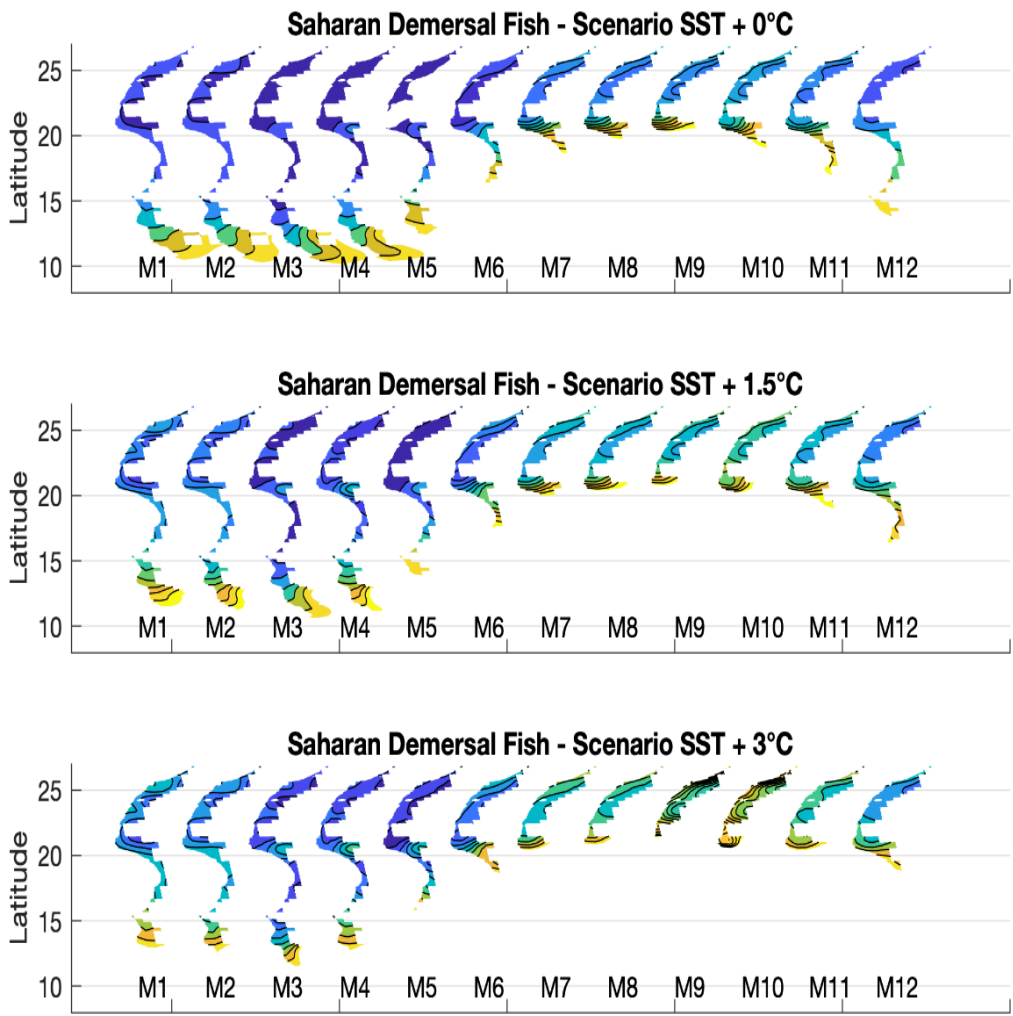

*Appendix Figure 5: Same as figure 22 but for Saharan Demersal fish.*

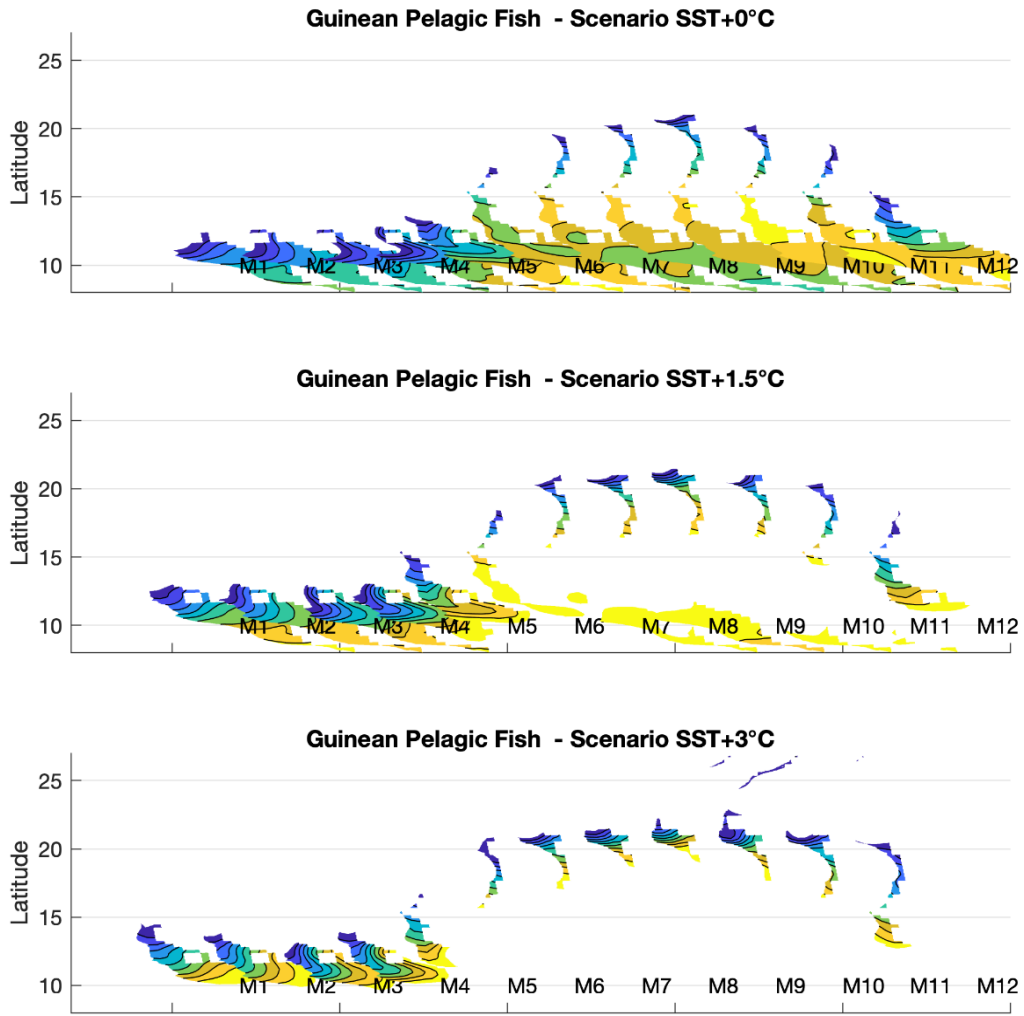

*Appendix Figure 6: Same as figure 22 but for Guinean pelagic fish.*

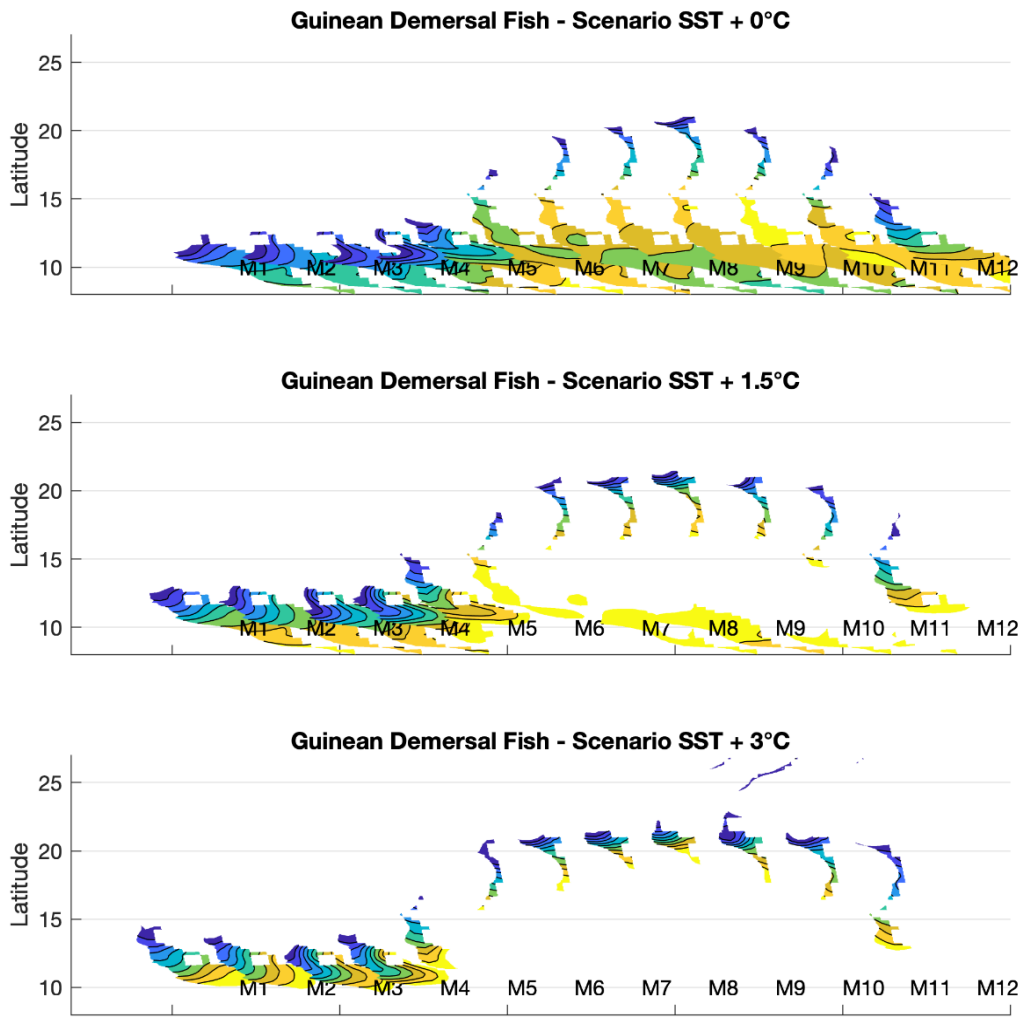

*Appendix Figure 7: Same as figure 22 but for Guinean Demersal fish.*

## Appendix D: Output figures

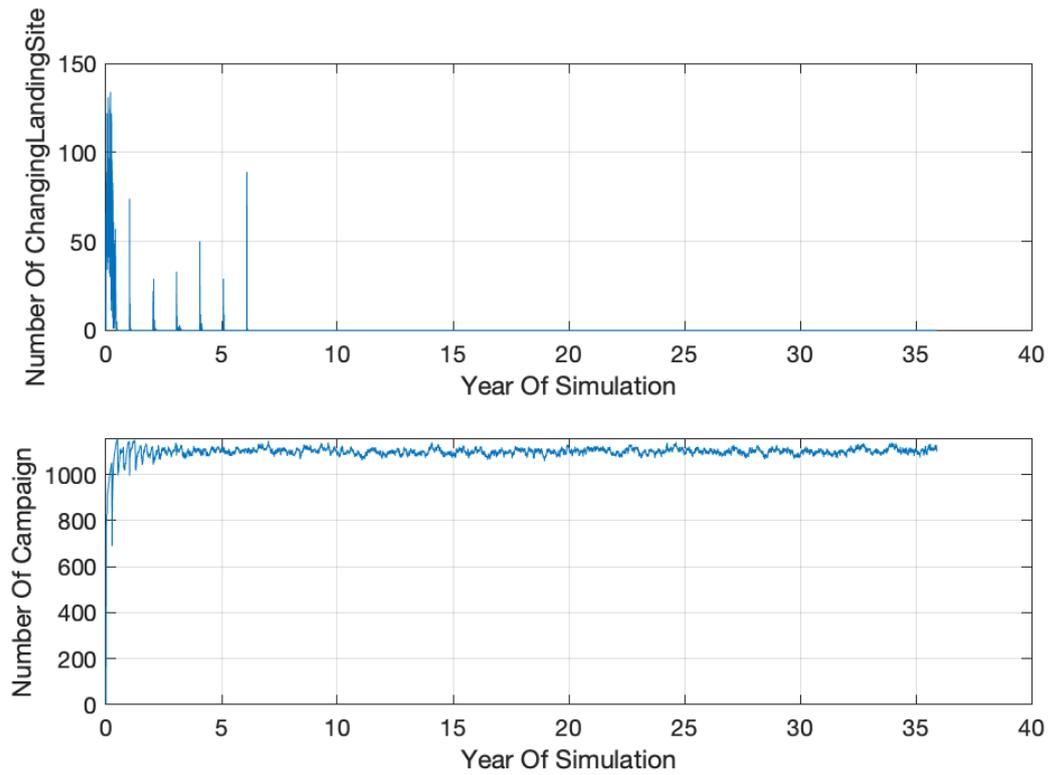

*Appendix Figure 8 : Evolution of the number of short-term (up) and long-term (down) migrations of the fishing units (simulation 1)*

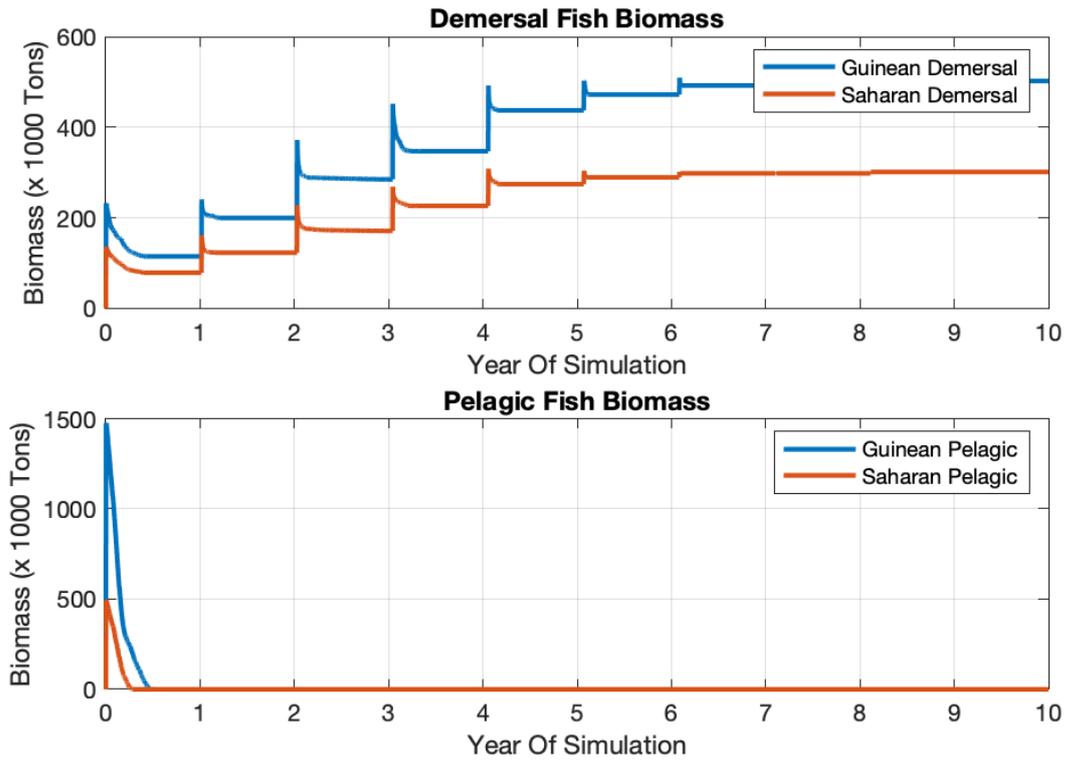

*Appendix Figure 9: Evolution of fish biomass for each model species in simulation 1. Top: demersal species, Bottom: pelagic species*

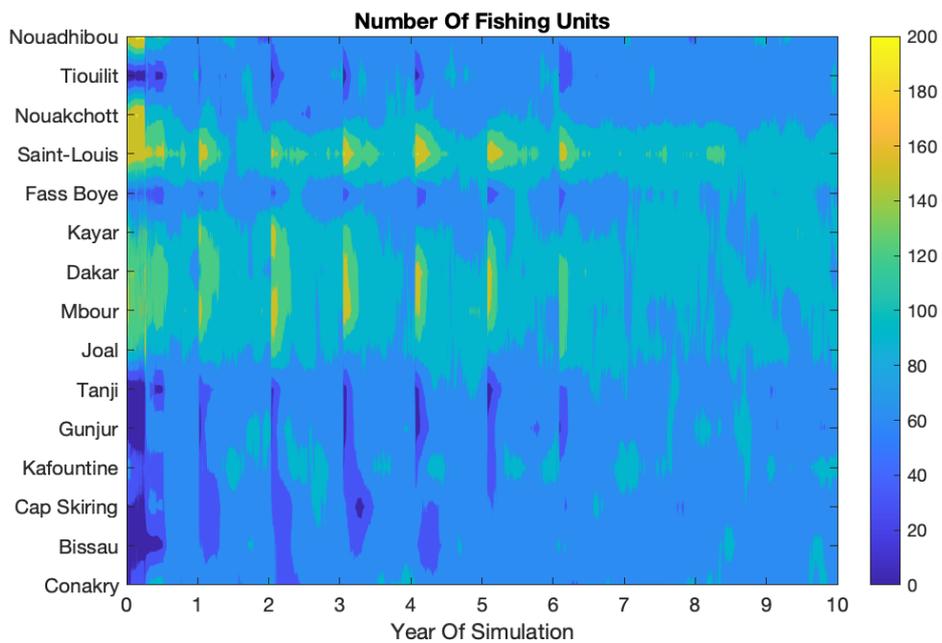

*Appendix Figure 10: Evolution of the number of pirogues in landing sites in simulation 1.*

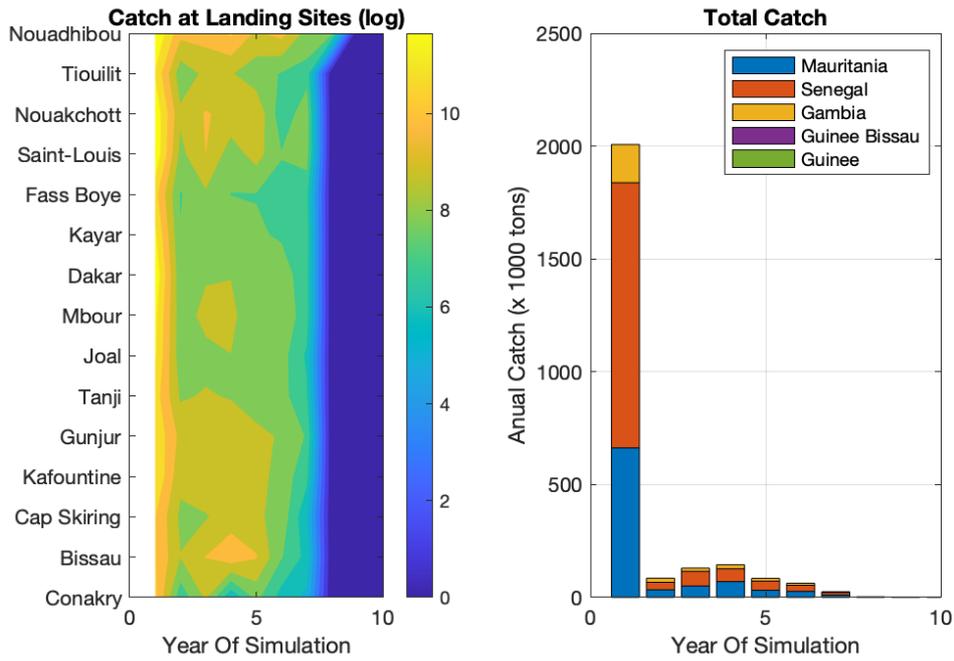

*Appendix Figure 11: Evolution of the catch in simulation 1. left: catch per landing site. rigth: total scattered per country.*

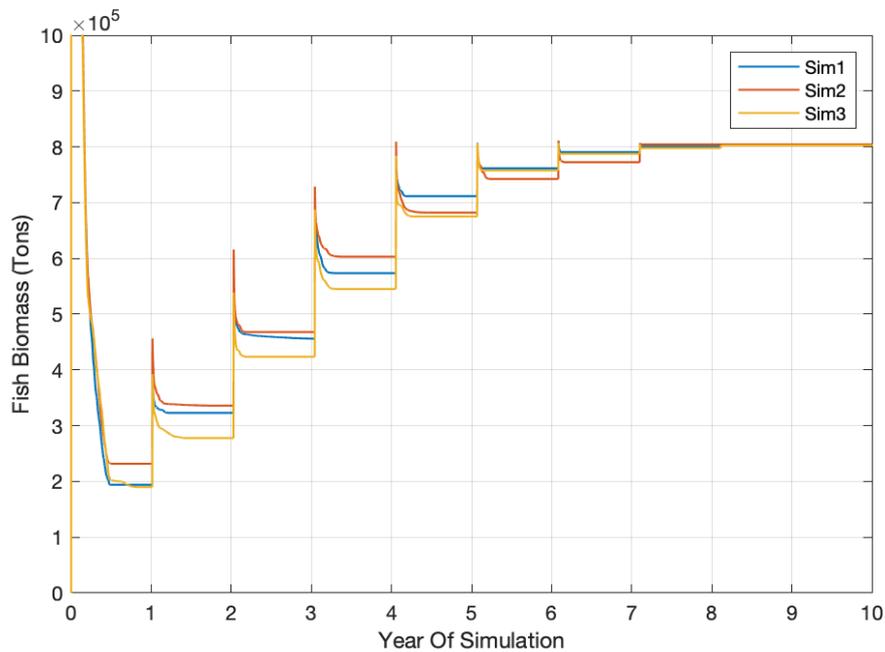

*Appendix Figure 12: Evolution of the catch in Senegal for simulations 1, 2 and 3*

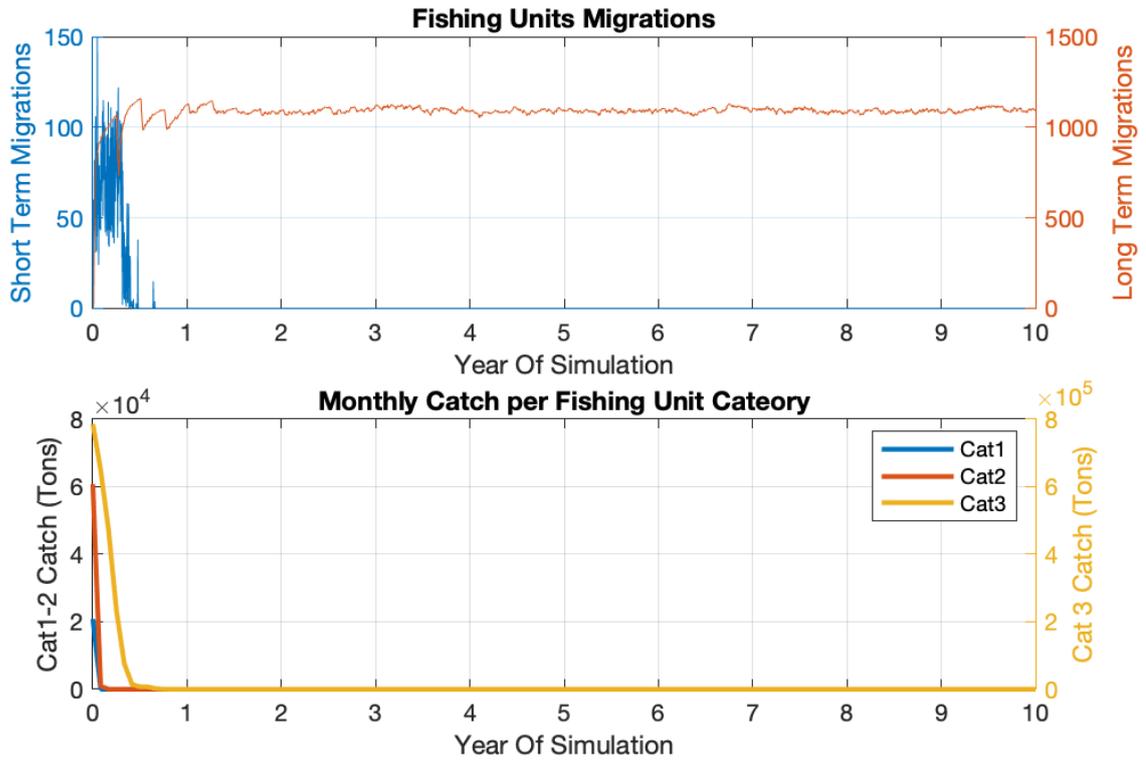

*Appendix Figure 13: Evolution of the catch in simulation 4*

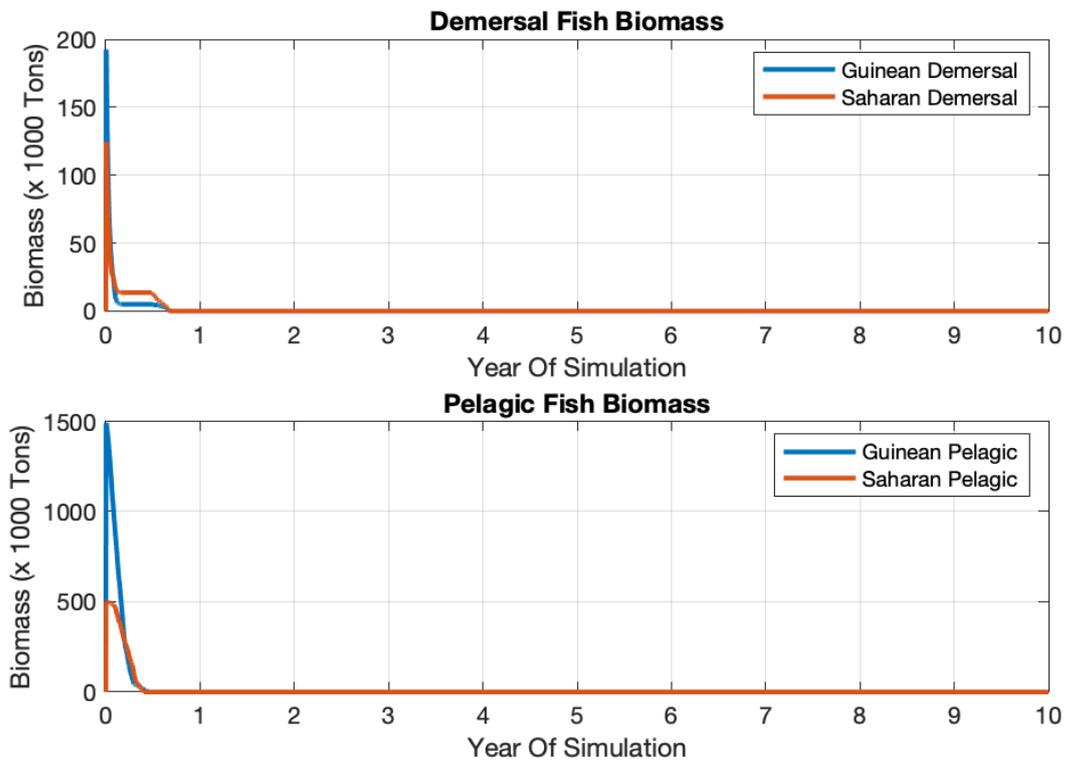

*Appendix Figure 14: Evolution of Biomass in simulation 4*

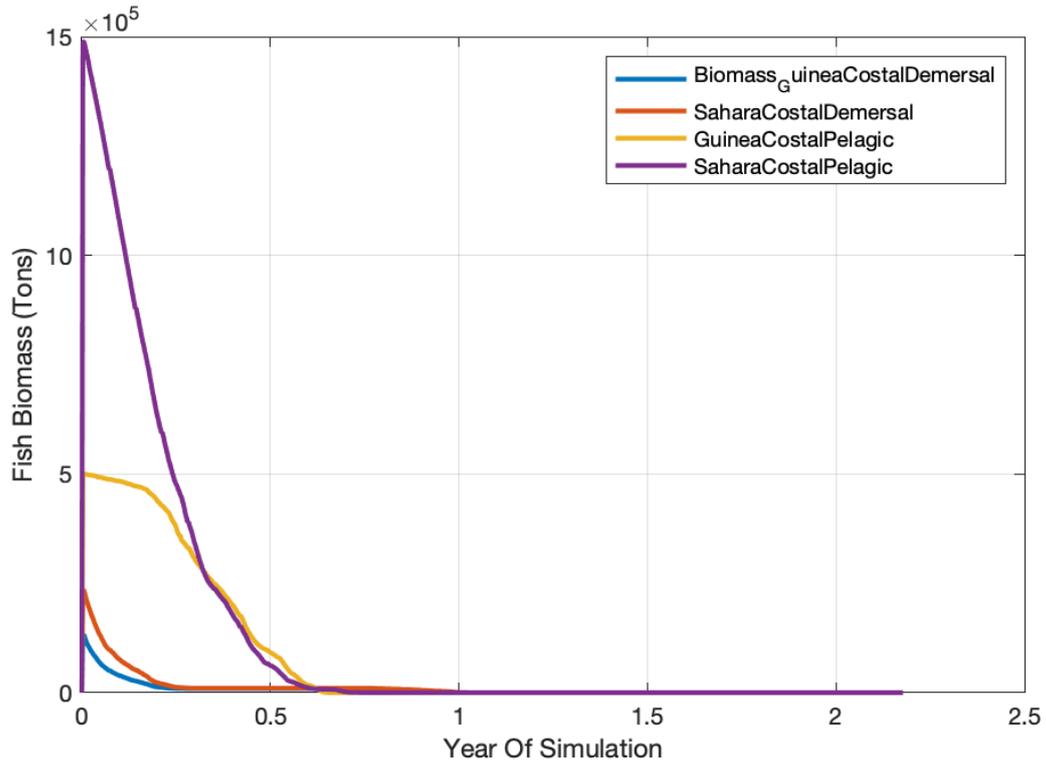

*Appendix Figure 15: Evolution of Fish Biomass in Simulation 5*

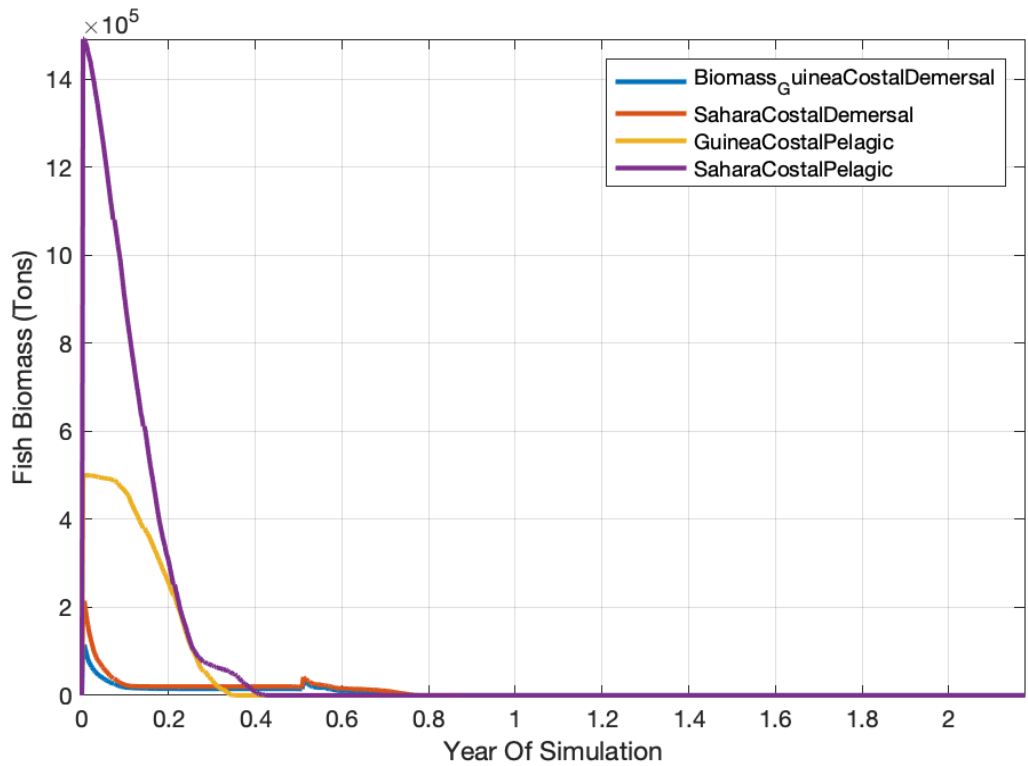

*Appendix Figure 16: Evolution of fish biomass in simulation 6*

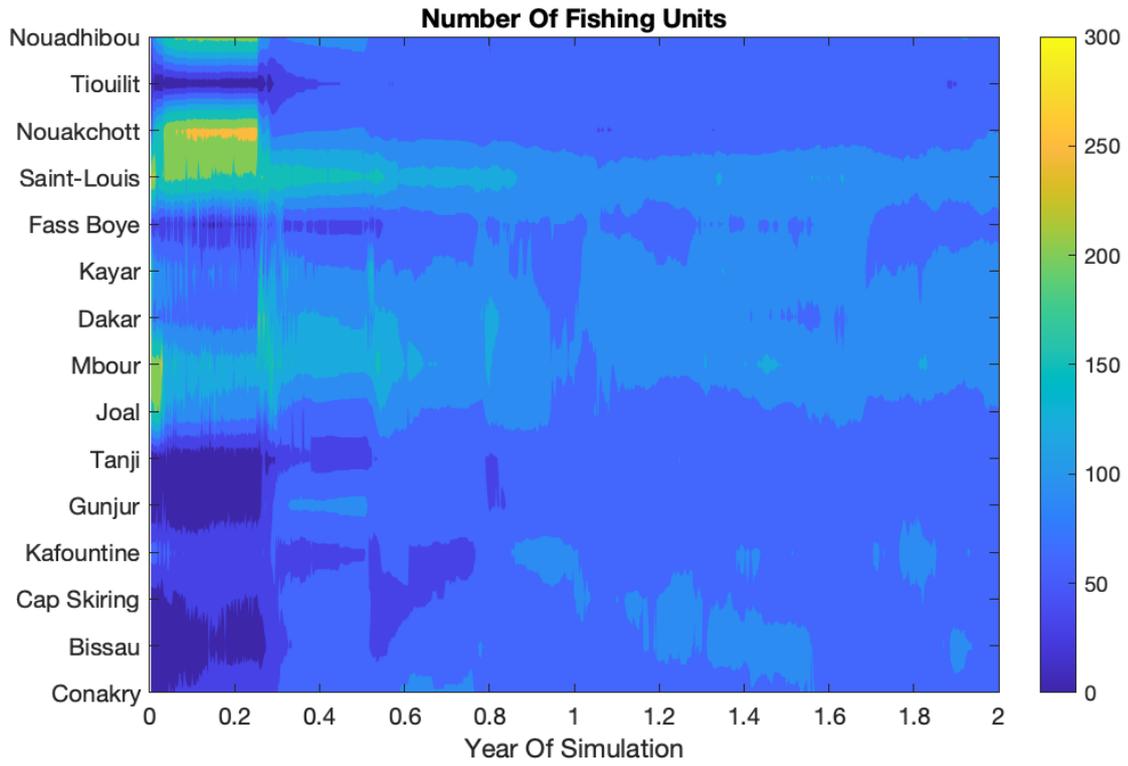

*Appendix Figure 17: Evolution of the distribution of fishing units in simulation 6.*

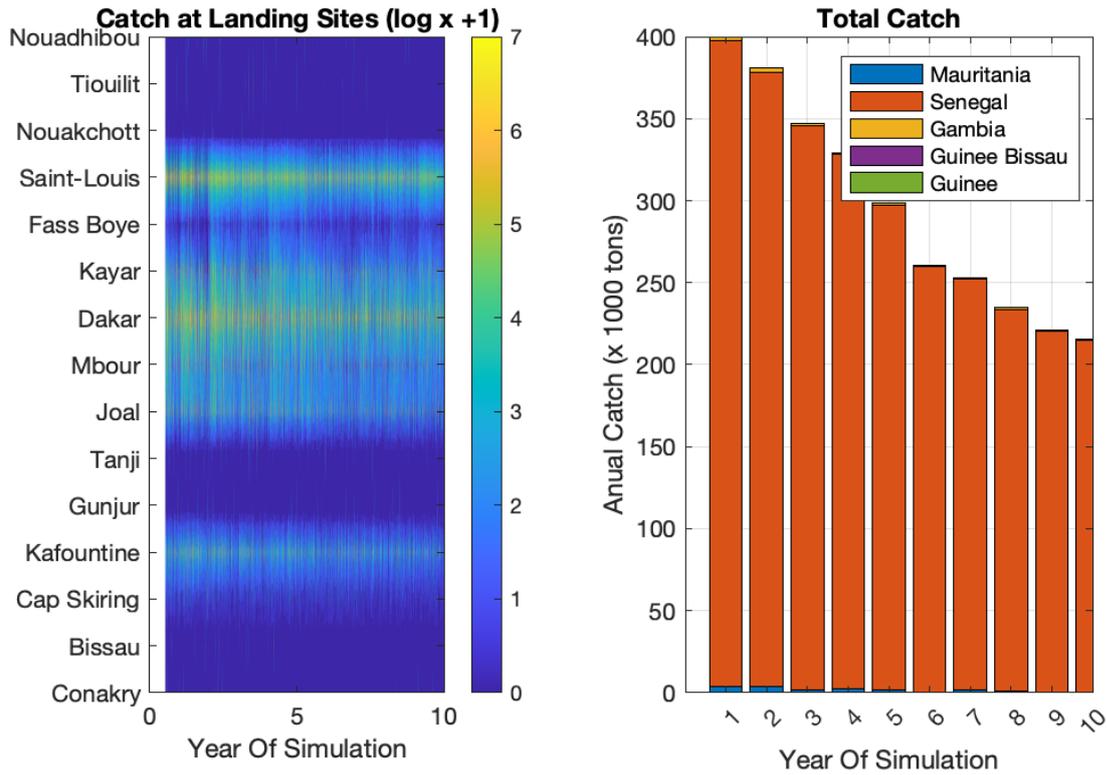

*Appendix Figure 18: Evolution of catches in simulation 7. Left: distribution of catch among landing site; right: total catch scattered per country.*

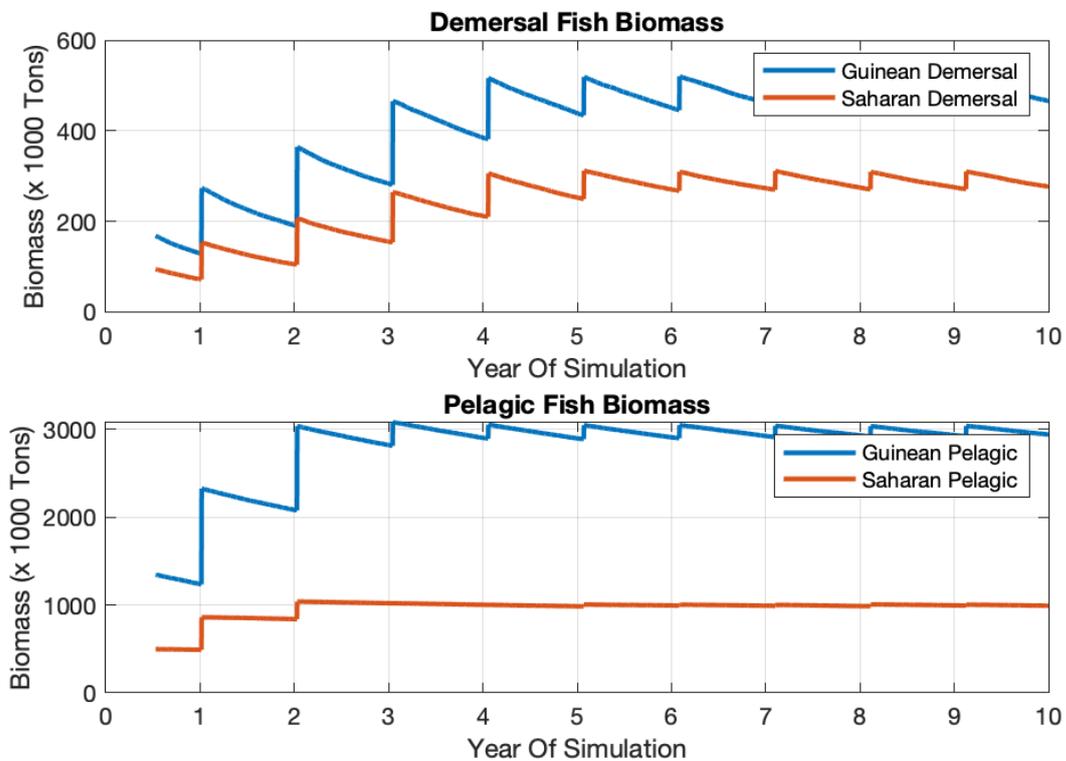

*Appendix Figure 19: Evolution of fish Biomass in Simulation 7. Top: Demersal species. Bottom: pelagic species*

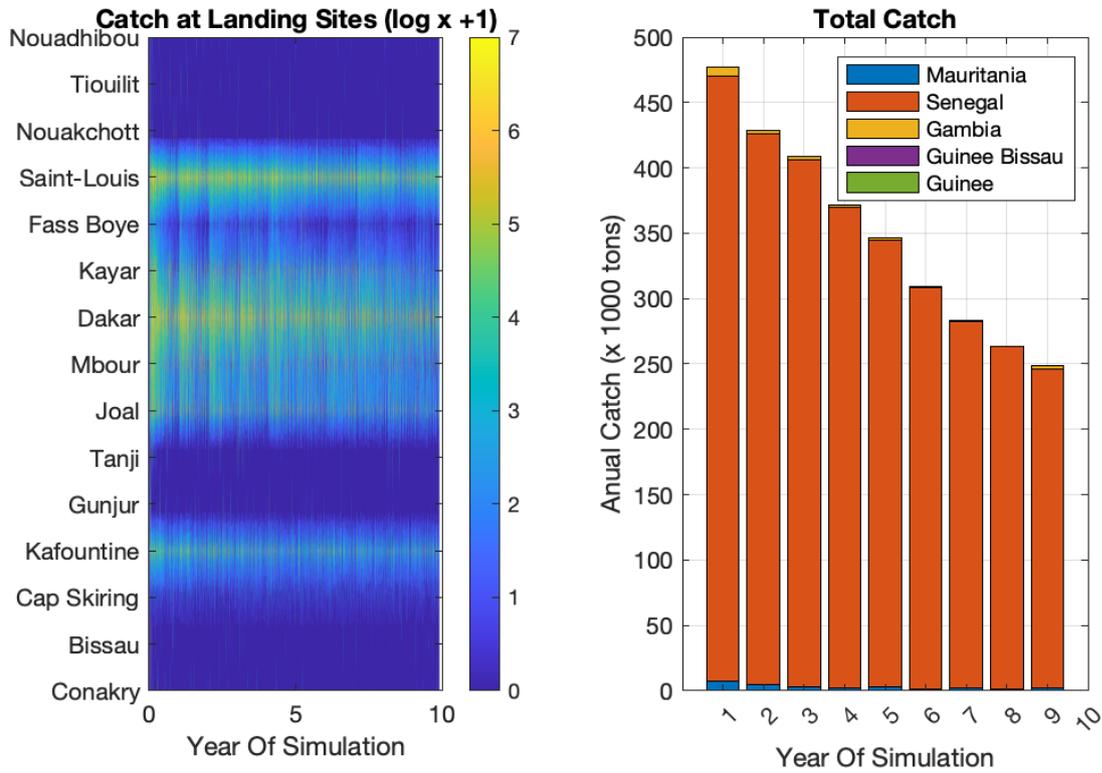

*Appendix Figure 20: Evolution of fish catch in simulation 8. Left: distribution of catch among landing site; right: total catch scattered per country.*

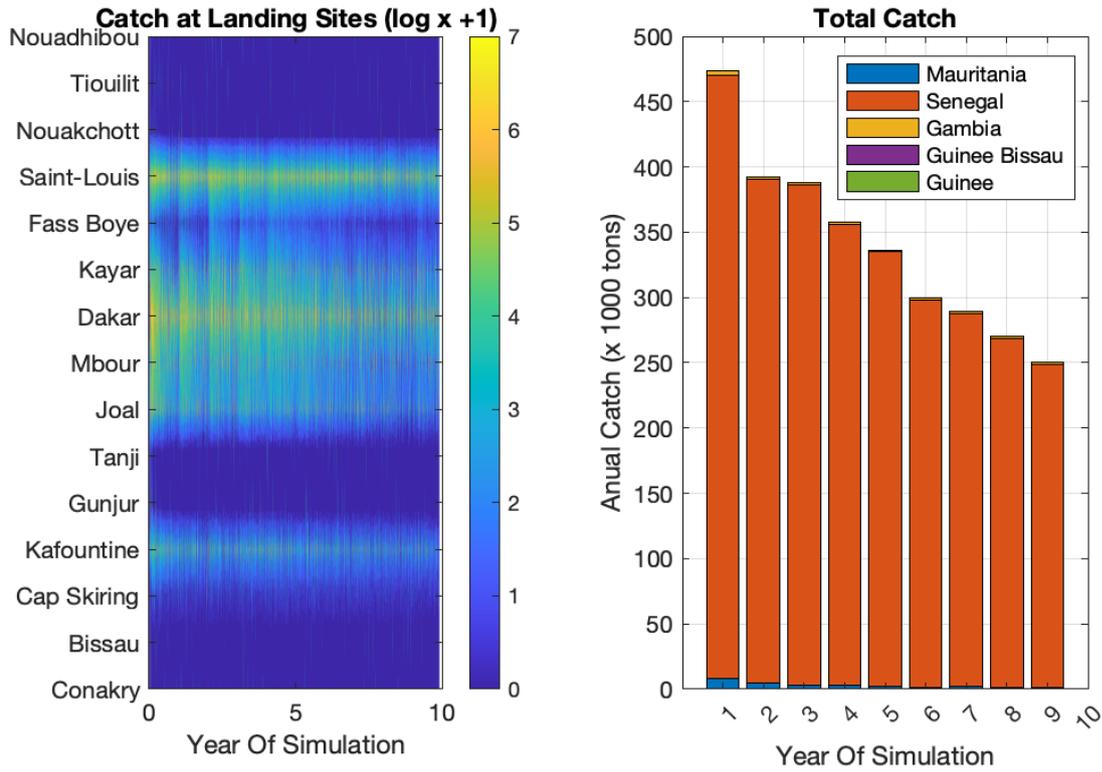

*Appendix Figure 21: Evolution of fish catch in simulation 9. Left: distribution of catch among landing site; right: total catch scattered per country.*

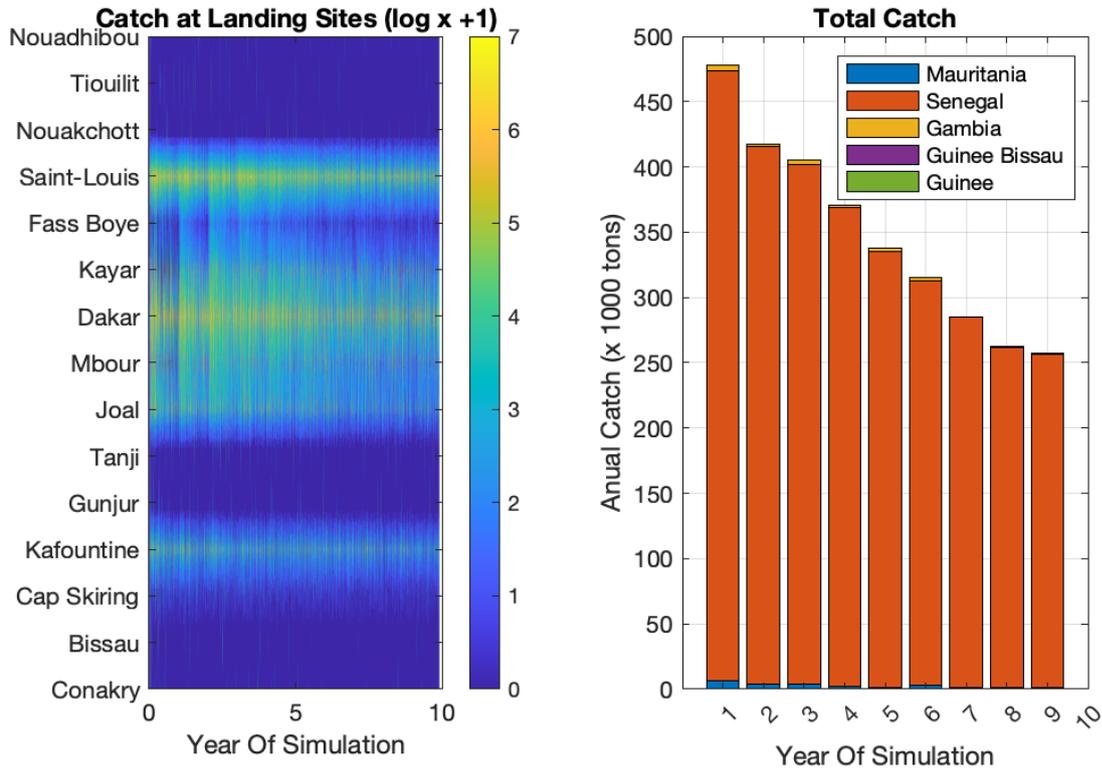

*Appendix Figure 22: Evolution of fish catch in simulation 10. Left: distribution of catch among landing site; right: total catch scattered per country.*

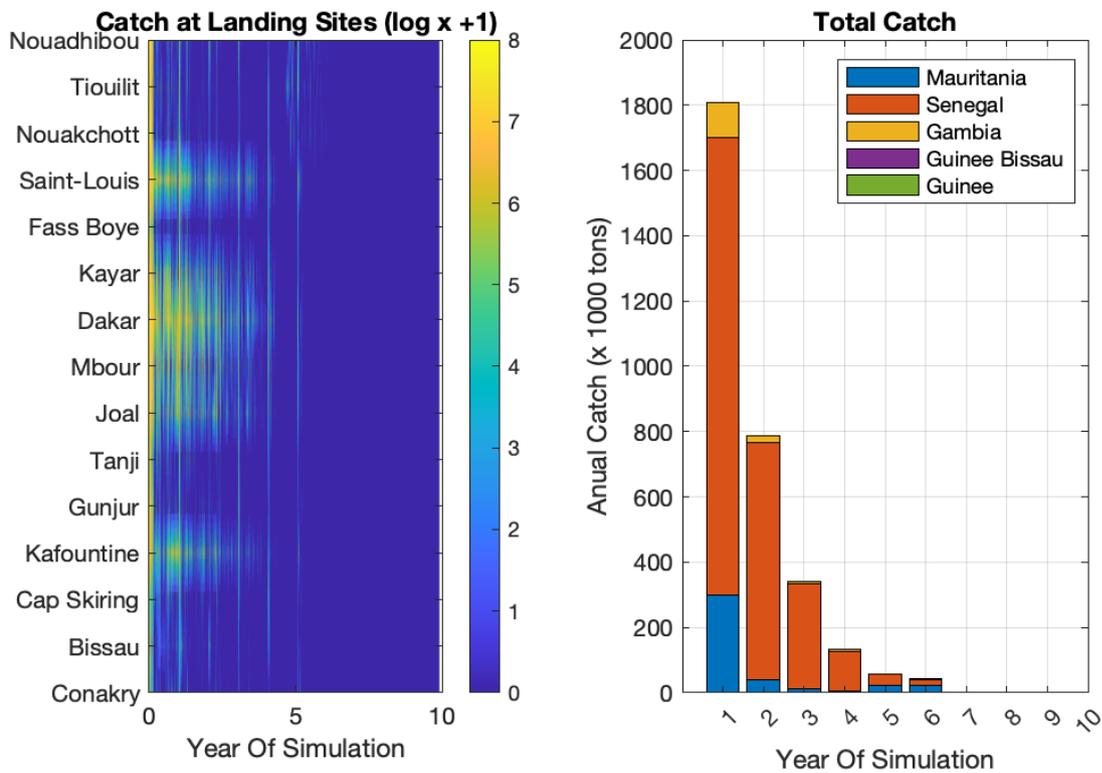

*Appendix Figure 23: Evolution of fish catch in simulation 8.1. Left: distribution of catch among landing site; right: total catch scattered per country.*

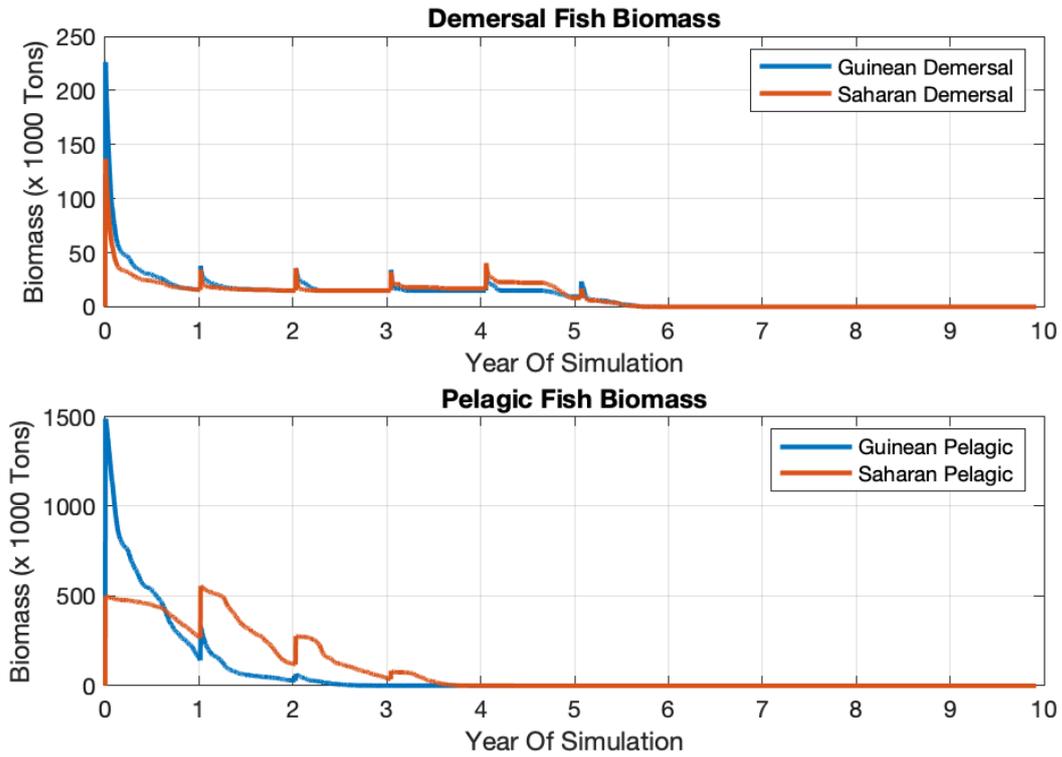

*Appendix Figure 24: Evolution of fish biomass in simulation 8.1.*

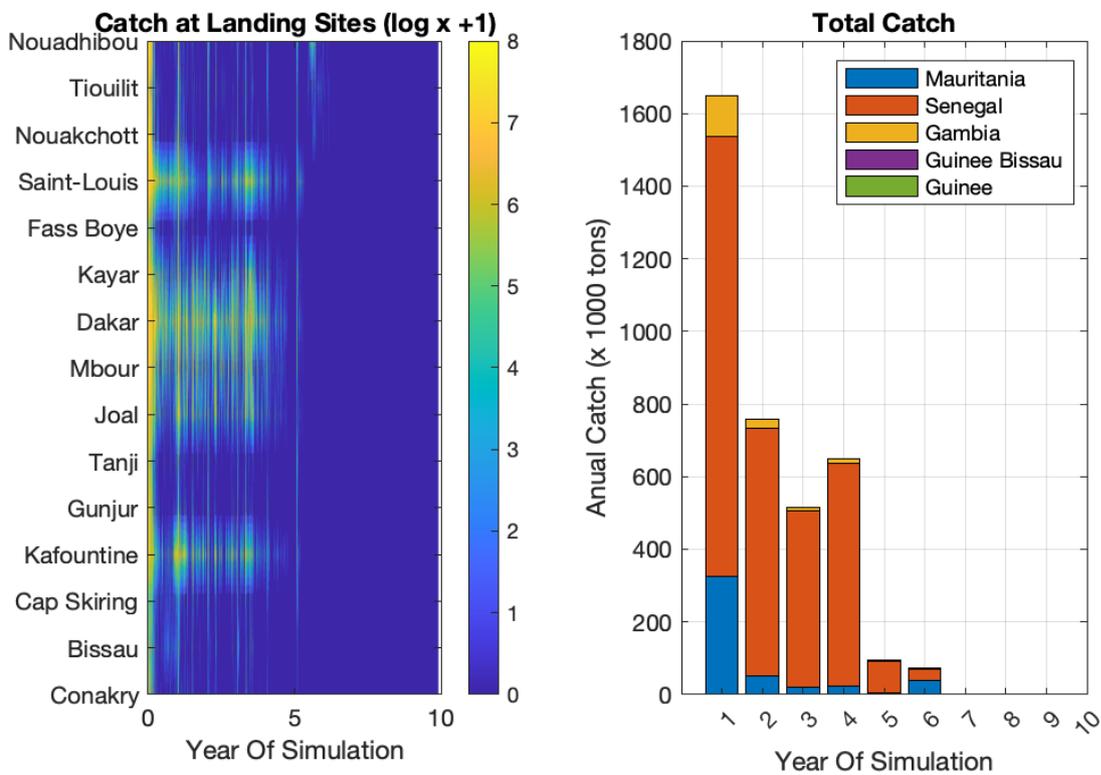

*Appendix Figure 25: Evolution of fish catch in simulation 9.1. Left: distribution of catch among landing site; right: total catch scattered per country.*

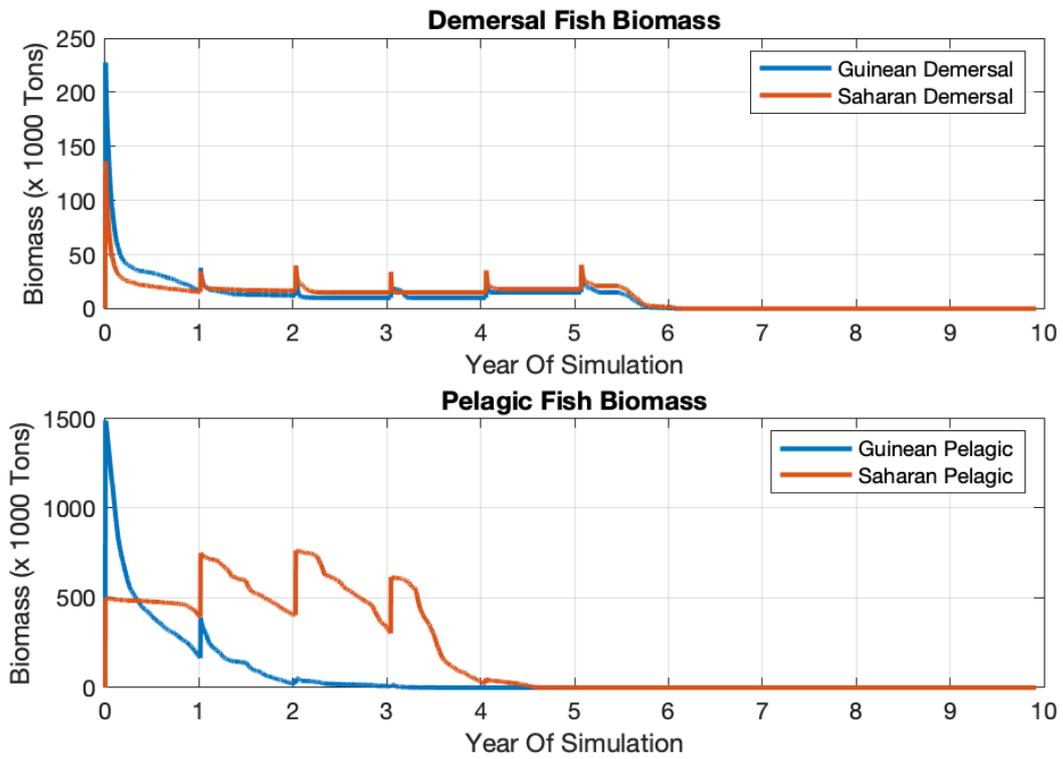

*Appendix Figure 26: Evolution of fish biomass in simulation 9.1*

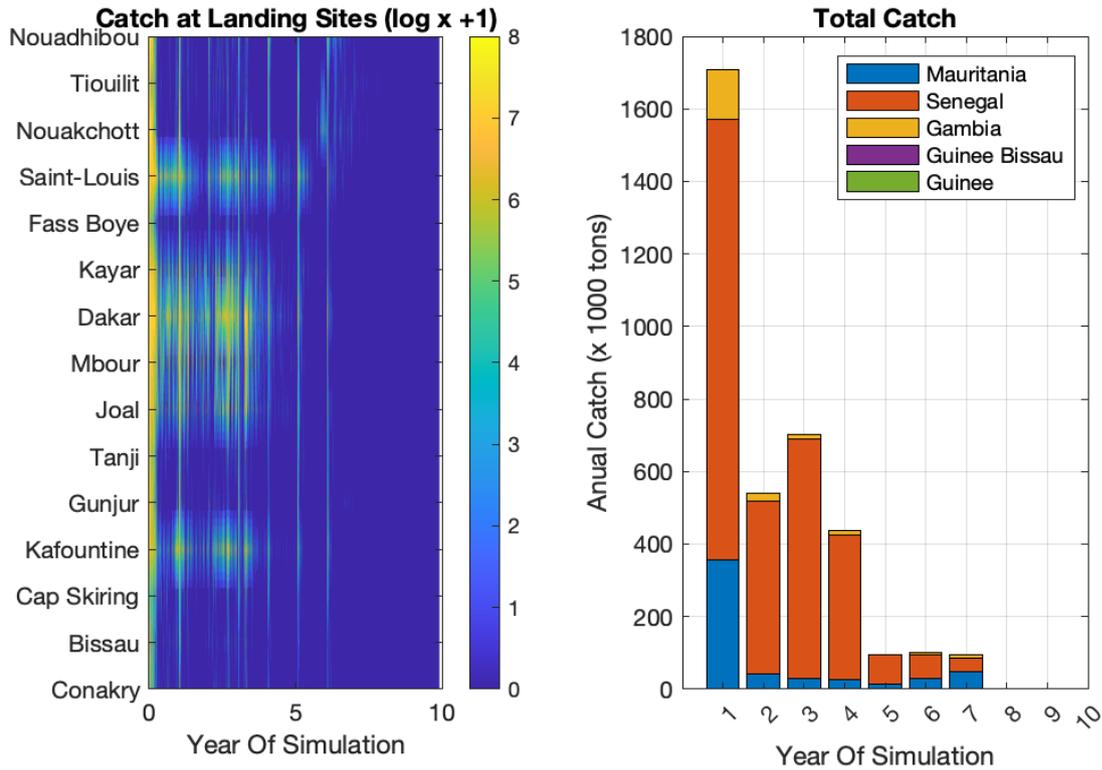

*Appendix Figure 27: Evolution of catch in simulation 10.1. Left: distribution of catch among landing site; right: total catch scattered per country.*

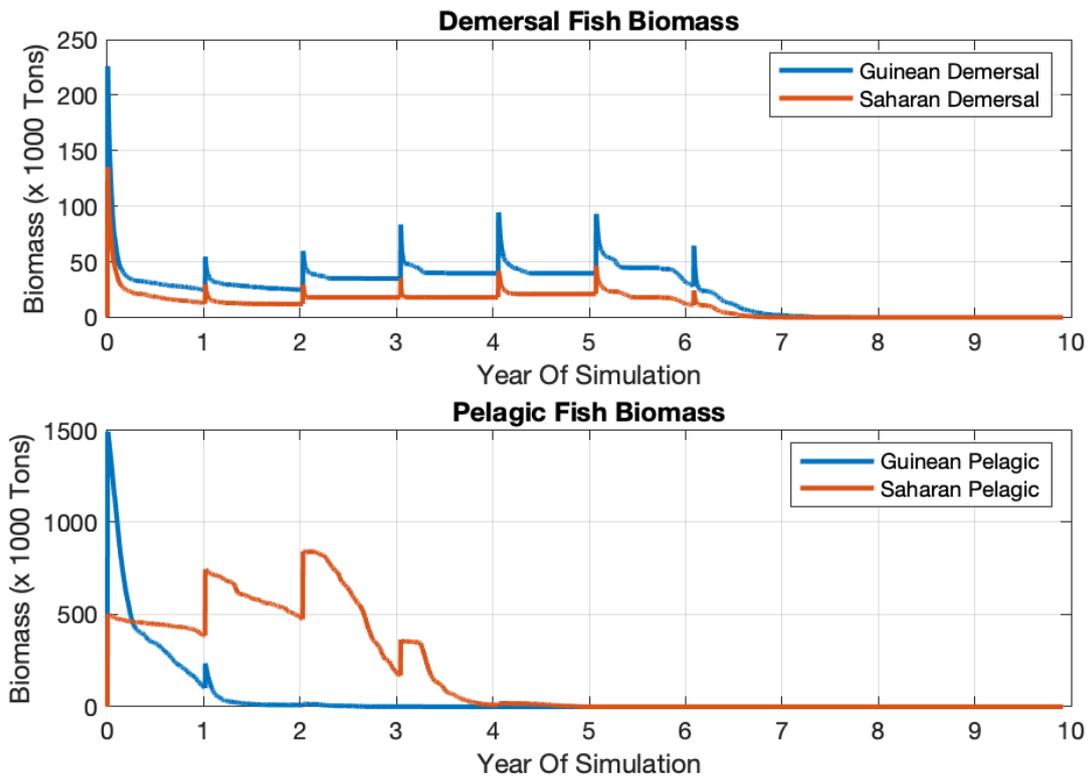

*Appendix Figure 28: Evolution of fish biomass in simu 10.1.*